\documentclass[twocolumn, times]{aastex631}
\usepackage{amsmath, latexsym, color, verbatim}
\usepackage{natbib}

\shorttitle{pySTARBURST99}
\shortauthors{Hawcroft et al. 2025}

\newcommand{\Msun}{\ensuremath{\rm M_\odot}}
\newcommand{\mdot}{\ensuremath{\rm M_\odot\,yr^{-1}}}

\newcommand{\logL}{$\log (L_{*}/L_{\odot})$}

\newcommand{\cmfgen}{\textsc{cmfgen}}
\newcommand{\fastwind}{\textsc{Fastwind}}
\newcommand{\genec}{\textsc{Genec}}
\newcommand{\powr}{\textsc{powr}}
\newcommand{\wmbasic}{\textsc{WMbasic}}
\newcommand{\galaxev}{\textsc{Galaxev}}
\newcommand{\cigale}{\textsc{cigale}}
\newcommand{\fsps}{\textsc{fsps}}
\newcommand{\galev}{\textsc{galev}}
\newcommand{\miles}{\textsc{miles}}
\newcommand{\pegase}{\textsc{Pegase}}
\newcommand{\galsevn}{\textsc{Galsevn}}
\newcommand{\slug}{\textsc{Slug}}
\newcommand{\hrpypopstar}{\textsc{HR-pyPopStar}}
\newcommand{\starburst}{\textsc{Starburst99}}
\newcommand{\nstarburst}{py\textsc{Starburst99}}
\newcommand{\bpass}{\textsc{bpass}}
\newcommand{\fortran}{\textsc{Fortran}}

\newcommand{\kms}{$\rm{km}\,\rm{s}^{-1}$}

\begin{document}

\title{pySTARBURST99: The Next Generation of STARBURST99}


\author{Calum Hawcroft} 
\affiliation{Space Telescope Science Institute, 3700 San Martin Drive, Baltimore, MD 21218, USA \\ email: chawcroft@stsci.edu or calum.hawcroft@gmail.com}

\author{Claus Leitherer} 
\affiliation{Space Telescope Science Institute, 3700 San Martin Drive, Baltimore, MD 21218, USA \\ email: chawcroft@stsci.edu or calum.hawcroft@gmail.com}

\author{Oskar Arangur\'e} 
\affiliation{Instituto de Astronom\'ia, Universidad Nacional Aut\'onoma de M\'exico, Unidad Acad\'emica en Ensenada, Km 103 Carr. Tijuana-Ensenada, Ensenada, B.C., C.P. 22860, M\'exico} 

\author{John Chisholm}
\affiliation{Department of Astronomy, The University of Texas at Austin, 2515 Speedway, Stop C1400, Austin, TX 78712, USA}

\author{Sylvia Ekstr\"om}
\affiliation{Observatoire de Genève, Chemin Pegasi 51, 1290 Versoix, Switzerland}

\author{S\'ebastien Martinet}
\affiliation{Institut d’Astronomie et d’Astrophysique, Université Libre de Bruxelles (ULB), CP 226, B-1050 Brussels, Belgium}

\author{Lucimara P. Martins}
\affiliation{NAT - Universidade Cidade de São Paulo, Rua Galvão Bueno, 868, São Paulo, Brazil}

\author{Georges Meynet}
\affiliation{Observatoire de Genève, Chemin Pegasi 51, 1290 Versoix, Switzerland}

\author{Christophe Morisset}
\affiliation{Instituto de Astronom\'ia, Universidad Nacional Aut\'onoma de M\'exico, Unidad Acad\'emica en Ensenada, Km 103 Carr. Tijuana-Ensenada, Ensenada, B.C., C.P. 22860, M\'exico} 
\affiliation{Instituto de Ciencias Físicas, Universidad Nacional Autónoma de México, Av. Universidad s/n, 62210 Cuernavaca, Mor., México}

\author{Andreas A. C. Sander}
\affiliation{Zentrum f{\"u}r Astronomie der Universit{\"a}t Heidelberg, Astronomisches Rechen-Institut, M{\"o}nchhofstr. 12-14, 69120 Heidelberg, Germany}

\author{Aida Wofford}
\affiliation{Instituto de Astronom\'ia, Universidad Nacional Aut\'onoma de M\'exico, Unidad Acad\'emica en Ensenada, Km 103 Carr. Tijuana-Ensenada, Ensenada, B.C., C.P. 22860, M\'exico} 
\affiliation{Department of Astronomy and Astrophysics, University of California, San Diego, 9500 Gilman Drive, La Jolla, CA 92093, USA}

\begin{abstract}

\starburst\, is a population synthesis code tailored to predict the integrated properties or observational characteristics of star-forming galaxies.
Here we present an update to \starburst\, where we port the code to python, include new evolutionary tracks both rotating and non-rotating at a range of low metallicity environments. We complement these tracks with a corresponding grid of new synthetic SEDs. Additionally we include both evolutionary and spectral models of stars up to 300-500\Msun. Synthesis models made with the python version of the code and new input stellar models are labelled \nstarburst.
We make new predictions for many properties, such as ionising flux, SED, bolometric luminosity, wind power, hydrogen line equivalent widths and the UV $\beta$-slope. These properties are all assessed over wider coverage in metallicity, mass and resolution than in previous versions of \starburst.
A notable finding from these updates is an increase in \ion{H}{1} ionising flux of 0.3 dex in the first 2Myr when increasing the upper mass limit from 120 to 300\Msun. Changing metallicity has little impact on \ion{H}{1} in the first 2Myr (range of 0.015 dex from $Z=0.02$ to $0.0$) but lower metallicities have higher \ion{H}{1} by 1 dex (comparing $Z=0.02$ to $0.0004$) at later times, with $Z=0.0$ having even higher \ion{H}{1} at later times. Rotating models have significantly higher \ion{H}{1} than their equivalent non-rotating models at any time after 2Myr. Similar trends are found for \ion{He}{1} and \ion{He}{2}, bolometric luminosity and wind momentum, with more complex relations found for hydrogen line equivalent widths and UV $\beta$-slopes. 

\end{abstract}

\section{Introduction} \label{sec: introduction}

Young stellar populations are characterised by the intense UV flux from the most massive stars (e.g. \citealp{Leitherer2020}). Therefore, key observational diagnostics in star-forming galaxies come from the extended atmospheres of stars which are hot and luminous enough to drive stellar winds, through radiation-pressure acting on metal ions in the photosphere. Beside being responsible for a significant amount of the ionising flux produced in starburst galaxies, due to their intrinsically hot temperatures, the strong winds lead to a significant environmental enrichment of momentum, energy and heavy elements (e.g. \citealp{Geen2023}).
With this in mind, synthetic stellar populations are an essential tool, to make predictions of both directly measurable and only inferrable integrated properties of galaxies with ongoing star formation, both as probes of the stellar content and as a fundamental element in the assessment of nebular properties.

However, synthetic stellar populations crucially depend on the capabilities of the underlying stellar models they are built upon (e.g. \citealp{Conroy2009}). Consequently, the inherent uncertainties in the composition of the stellar population or the evolutionary pathways directly limit the predictive power of the population models.
While fundamental questions about the physical processes in massive stars remain, population synthesis models need to keep up with the progress made in the field, for example related to mass-loss, angular momentum transport, and the impact of stellar interactions in binary and multiple systems (see e.g. \citealp{Langer2012,Eldridge2022,Marchant2024}). Moreover, there are further fundamental limitations imposed by the scope of available grids of underlying stellar models. In particular we focus on two quantities, the metallicity and the upper mass limit of the IMF, which are key to understand unresolved populations throughout the Universe. 

The initial stellar mass is the primary driver of the evolutionary pathway. Massive stars stars lose a significant fraction of their mass through their stellar winds, which are inherently metallicity dependent, therefore it is crucial that models at various metallicities are available. Especially given our increased understanding of local populations with sub-solar metallicity, e.g., through large observing programmes like ULLYSES \citep{RomanDuval2020} and its optical complement XShootU \citep{Vink2023} on the level of individual stars as well as CLASSY \citep{Berg2022} and the rapidly expanding list of observed galaxies in the early Universe with JWST on the level of star-forming galaxies. There are a wide range of available stellar evolutionary models with massive stars (\bpass, \citealp{Eldridge2017}; \cigale, \citealp{Boquien2019}; \fsps, \citealp{conroy2010}; \galev, \citealp{Kotulla2009}; \galaxev, \citealp{Bruzual2003, Plat2019}; \galsevn, \citealp{Lecroq2024}; \miles, \citealp{Vazdekis2016}; \pegase, \citealp{LeBorgne2004}; \slug, \citealp{Krumholz2015}; \hrpypopstar, \citealp{Millan-Irigoyen2021}; \starburst, \citealp{Leitherer1999, Leitherer2014}). \pegase\, and \miles\, are based on empirical spectral libraries, and \galev\, uses the \cite{lejeune1997} theoretical library. The other codes (\bpass, \cigale, \fsps, \galaxev, \slug, \starburst) are more tailored for spectral synthesis including massive stars. All of these codes utilise \wmbasic\, \citep{Pauldrach2001} atmosphere models for OB stars to produce synthetic spectra, apart from \hrpypopstar\, which implements the \powr\, model grid \citep{Hainich2019} covering $Z=0.014, 0.006, 0.002$\footnote{$Z$ is defined as the fractional abundance of metals, such that H + He + Z = 1.}. The individual stellar parameters in each \wmbasic\, grid vary but the metallicites are the same, with the range $Z=0.04, 0.02, 0.008, 0.004, 0.001$ in \cigale, \galaxev, \slug\, and \starburst\,. \bpass\, adds models at $Z=0.006$ and $0.002$ and \fsps\, adds models at $Z=0.0001, 0.003, 0.01, 0.014, 0.03$. However, the various stellar evolutionary tracks (discussed further in Sect. \ref{sec: evolution}) used by \bpass, \cigale, \galaxev, \slug, \starburst), offer many more metallicites (see e.g. Table 10 in \citealp{Sanchez2022}, Table 1 in \citealp{Eldridge2017} and Table 1 in \citealp{Leitherer2014}). This means that the majority of population synthesis models rely on interpolation in metallicity to produce spectra, especially at low metallicity. 

A growing area of interest is the impact of stars with (initial) masses above 100\,$M_{\odot{}}$, also known as very massive stars (VMS). Such stars are exceedingly rare (from an IMF perspective) but highly impactful, and their presence in stellar populations is inferred in a number of observations of distant galaxies \citep{Senchyna2021, Mestric2023, Upadhyaya2024}. Resolved VMS have been directly detected in Galactic clusters (NGC3603; \citealp{Schnurr2008, Crowther2010}, Arches; \citealp{Figer2002, Najarro2004, Martins2008, Lohr2018}) with the most massive stars found in the LMC cluster R136 \citep[see, e.g.,][]{Crowther2010, Crowther2016, Bestenlehner2020, Kalari2022, Brands2022}. Tentative detections of these objects in relatively nearby star-forming regions has become increasingly common \citep[e.g.,][]{Wofford2014, Smith2016, Wofford2023, Smith2023}. This could enable some verification of VMS predictions. Coupling this progress with the argument that VMS form more readily in low metallicity environments; the motivation to properly account for VMS in stellar population models grows as they help inform our understanding of the earliest generations of stars. Some population synthesis codes have taken steps to account for the presence of VMS. The \galaxev\, and BPASS codes now include stellar evolutionary predictions for stars up to initial masses of 300\,$M_{\odot{}}$, albeit without dedicated updates to the VMS spectral libraries, although \galaxev\ do include PoWR spectra for WNL stars as a close proxy to the spectra of VMS. New studies are building upon these evolutionary models with tailored complementary stellar atmosphere models in attempts to reproduce galactic spectra \citep{Martins2022, Schaerer2024, Martins2025}. There have also been empirical approaches with works such as \cite{Crowther2024} creating composite stellar populations using dedicated UV spectroscopy as well as the ULLYSES library of massive star spectra to recreate composite observations of R136, which can inform our interpretation of spectra of highly star-forming regions at larger distances. An additional observational complication is the difficulty in distinguishing between VMS and Wolf-Rayet (WR) stars in composite rest-frame UV spectra \citep{Martins2023, Rivera-Thorsen2024, Berg2024}, with these two populations of stars having a number of similar spectral signatures while providing significantly different contributions to integrated population properties. 


Significant progress has been made in the field of stellar evolution in the past decade. With respect to \starburst, of particular interest is the tremendous extension of the \genec\ suite of stellar tracks as the underlying mapping of stellar evolution in the population synthesis modelling. Available \genec\ evolutionary models for initial masses from from 1 to 120\,$M_{\odot}$ now cover a considerable range of metallicities, representing the conditions in a wide selection of environments, from the metal enhanced Galactic Nucleus through solar composition to the metal-poor Large and Small Magellanic Clouds, as well as very low metallicity galaxies such as IZw18 and completely metal-free regions, which would harbour Pop III objects. 
The \genec\ grids have further been extended in the mass domain with models available for initial masses up to 500 $M_{\odot}$ at $Z=0.02$ and $0.014$, and up to 300 $M_{\odot}$ at $Z=0.006$ and $0.0$. Following the success found in better reproducing the distribution of stars across the HRD by including a treatment of rotation in previous evolutionary grids (\citealp{Levesque2012, Leitherer2014}), models including rotation are available for all the aforementioned \genec\ extensions as well.
Yet, to fully realise the predictive power of new evolutionary model grids, they must be paired with complementary synthetic model spectra from stellar atmosphere models which are representative of their parameter space coverage. There has also been substantial progress in the development of NLTE atmosphere models since the release of the commonly used \wmbasic\ grid (in codes such as \cmfgen\, \citep{Hillier1998}, \powr\, \citep{Grafener2002,Sander2015} and \fastwind\, \citealp{Santolaya-Rey1997a, Puls2005}, discussed further in Sect. \ref{sec: atmospheres}), providing a range of options to update stellar atmospheres for spectral synthesis modelling of starburst populations. 

This tremendous progress in the modelling of stellar evolution and atmospheres has yet to be combined and implemented in a population synthesis framework to explain unresolved stellar populations.
In this work, we aim to utilise these advancements in stellar models to update the capabilities of the population synthesis code \starburst, with a focus on the impact of metallicity and upper mass limit. We present a new version of \starburst, in which we implement the latest generation of \genec\ releases (covering $Z=0.02$, $0.014$, $0.006$, $0.002$, $0.0004$ and $0.0$) with rotation and include VMS (500\,\Msun\ at 0.014 and 300\,\Msun\ at lower metallicity), along with a new grid of complementary synthetic spectra from \fastwind. 

Sect. \ref{sec: updates} discusses the implementation of the aforementioned stellar evolutionary models in \starburst, along with new tailored stellar atmosphere models produced to complement the tracks and our efforts to translate the \starburst\, code from \fortran\, to python. Sect. \ref{sec: results} makes new predictions on a few key properties of the composite populations, such as their spectral energy distribution (SED), especially the ionising spectrum, as well as wind power and UV slope. Finally, we discuss the impact of these new predictions and highlight upcoming future work in Sect. \ref{sec: conclusions}.

\section{Updates} \label{sec: updates}


In this section, we will outline the major updates made to \starburst\, focusing on the developments made in three key areas: (i) the included evolutionary models, (ii) the applied stellar atmosphere models, and (iii) the framework of the software itself being ported to python from \fortran.

\subsection{Stellar evolution} \label{sec: evolution}

Stellar evolutionary predictions have developed significantly in the past decade, with various libraries available which focus on massive stars produced with different codes (and each tailored to a specific parameter space). These include MIST (MESA models; \citealp{Paxton2013, Dotter2016, Choi2016}) and BONN tracks tailored to the Magellanic Clouds and their extension through BoOST (\citealp{Brott2011, Kohler2015, Szecsi2022}), both of which offer a wide range of masses and metallicities for single star evolutionary scenarios and include rotation. Many other options are available for more specific purposes such as investigating post-MS binary evolution or very massive stars (\citealp{Marchant2017, Stevenson2017, Pauli2022b, Broekgaarden2021, Martins2022, Sabhahit2022, Fragos2023}). For those used in population spectral synthesis, PARSEC models in GALEXEV (\citealp{Bruzual2003, Chen2015}) which have high sampling in rotation and are adapted for binaries with \galsevn, STARS including binaries for BPASS \citealp{Eldridge2004, Eldridge2017}, and the \genec\, models in \starburst\, (\citealp{Maeder1994, Leitherer1999, Levesque2012, Leitherer2014, Ekstrom2012, Georgy2013}). We also note that caution should be taken when comparing evolutionary tracks, as there may be differences in predictions for similar stars as discussed in e.g. \cite{Agrawal2021}. For specific physical effects such as rotation, there can also be differences in implementation between codes \citep{Nandal2024}. 

In principle, any or all of the aforementioned tracks could be incorporated within the framework of \starburst.
For the immediate purpose of this work to extend \starburst\ models to lower metallicity and higher mass, and the straight-forward comparison with the currently established \starburst\, models, we implement the \genec\, evolutionary tracks \citep{Ekstrom2012, Georgy2013, Groh2019, Eggenberger2021, Murphy2021, Yusof2022, Martinet2023}. 
A further extension to allow data from different evolutionary codes to be processed is envisioned, but is beyond the scope of the current work.

The implementation of solar and SMC metallicity tracks, defined relative to the solar metallicity of $Z=0.014$ \citep{Asplund2009}, from \cite{Ekstrom2012} and \cite{Georgy2013} are described in \cite{Leitherer2014}. We follow the same routine to include the LMC, IZw18 and Z=0 (Z0) tracks. The LMC and IZw18 tracks follow the same physical recipes as described in \cite{Ekstrom2012} and \cite{Georgy2013}. The \cite{Vink2001} mass-loss rates are implemented on the main sequence (MS), \cite{deJager1988} is used in the blue supergiant (BSG) phase, a combination of \cite{Sylvester1998} and \cite{vanLoon1999} in the red supergiant (RSG) phase and either \cite{NugisLamers2000} or \cite{Grafener2008} in the WR regime. Additionally, the correction factor from \cite{Maeder2000} is applied to the mass-loss rates in rotating models. The applied mass-loss descriptions are consistent within one set of tracks. Between different sets, there are only slight differences in mass-loss scaling with metallicity as $\dot{M}(Z) = (Z/Z_{\odot})^{\alpha}\dot{M}(Z_{\odot})$ (see Table \ref{tbl:Evolution}). We discuss the mass-loss rates in the context of \starburst\, outputs further in Sect. \ref{sec: winds}. For the Z0 tracks, there is no mass loss unless the star reaches a critical rotation threshold, at which point an average mass-loss rate of $10^{-5}\mdot$ is applied until the star passes back under the rotation threshold. This mass loss is invoked only for parts of the evolution in the most massive stellar models ($>85\Msun$). 
For stars with $M > 40\Msun$, as described in \cite{Ekstrom2012} and \cite{Georgy2013}, a density-scale based mixing length for including turbulent pressure and acoustic flux in the envelope is used. 
All grids offer two rotation options (non-rotating and an initial velocity of 40\% critical rotation), and none include the effects of magnetic fields. Therefore the models are differentially rotating.

We also note that the mass ranges and sampling differ between some of the grids. Mainly there is a slight decrease in range and sampling below SMC metallicity and an increase in sampling in the MW grid. For tests of the impact of mass resolution in the MW, comparing isochrones generated with either 24 or 33 evolutionary tracks, we find that there is very little impact on the isochrones during the main sequence. There are changes at higher masses when the post-MS evolution is more sensitive to smaller changes in mass. There are therefore significant differences at times when WR stars are present in the population, likely due to the addition of a $50\Msun$ track which reduces interpolation distance between the $40\Msun$ and $60\Msun$ tracks. The impact of these additional tracks is discussed in Sect. \ref{sec: seds}.

Initial heavy element mixtures at all metallicites follow \cite{Ekstrom2012} and are tuned down, essentially meaning they are solar-scaled abundances. This may not be fully representative of the suggested regions as $\alpha$-element ratios may vary in low metallicity environments as evidenced in nearby low-Z star forming regions \cite{Bouret2015, Schosser2025} and galaxies close to Cosmic Noon and at high redshift \cite{Steidel2016, Cullen2021, Strom2022, Cameron2023, Welch2025}. Recent works have begun to investigate and quantify the impact of non-solar abundance patterns on predictions of stellar populations \cite{Pietrinferni2021, Grasha2021, Byrne2025}.

For the VMS models from \cite{Martinet2023}, the input physics is the same as for other \genec\, models, apart from an increase in overshooting to 20\% of the pressure scale height at the \textit{Ledoux} boundary and the inclusion of electron-positron pair production in the equation of state. The mass loss-metallicity scaling follows \cite{Georgy2013}. We note the evolution models do not contain a general increase in mass-loss rate for VMS on the main sequence, which would be physically motivated by proximity to the Eddington limit \citep{Grafener2008,Vink2011,Bestenlehner2014,Sabhahit2022}. The initial masses added are 180, 250 and 300 $\Msun$ at MW, LMC and Z0. This means there are no VMS tracks to match the exact metallicities of the SMC or IZw18. There is further a $500\,\Msun$ non-rotating track at solar metallicity. 

\begin{deluxetable*}{cccccccccccc}
\label{tbl:Evolution}
\tablecolumns{12}
\tablewidth{0pc}
\tablecaption{Properties of the stellar evolutionary models.}
\tablehead{
\colhead{Region} & \colhead{Ref} & \colhead{H} & \colhead{He} & \colhead{Z} & \colhead{$\alpha$} & \colhead{$\alpha$} & \colhead{$\alpha$} & \colhead{$\alpha$} & \colhead{M} & \colhead{M} & \colhead{No. tracks}
\\
\colhead{} & \colhead{} & \colhead{} & \colhead{} & \colhead{} & \colhead{$\dot{M}_\mathrm{O}$} & \colhead{$\dot{M}_\mathrm{BSG}$} & \colhead{$\dot{M}_\mathrm{WR}$} & \colhead{$\dot{M}_\mathrm{R(S)G}$} & \colhead{$v_{r0}$} & \colhead{$v_{r4}$} & \colhead{}
}
\startdata
GalC &	Yus22 & 0.7064 & 0.2735 &	0.02   & 0.85   & 0.5   & 0.66   & 1 & 0.8-500 & 0.8-300 & 33\\
MW	 &	Eks12	& 0.72   & 0.266  &	0.014  & 1    & 1   & 1    & 1 & 0.8-500 & 0.8-300 & 33\\
LMC  &	Egg21	& 0.738	 & 0.256  & 0.006  & 0.7  & 0.7 & 0.7  & 1 & 0.8-300 & 0.8-300 &24\\
SMC	 &	Geo13	& 0.747	 & 0.251  & 0.002  & 0.85 & 0.5 & 0.66 & 1 & 0.8-120 & 0.8-120 &24\\
IZw18&  Gro19	& 0.7507 & 0.2489 &	0.0004 & 0.85 & 0.5 & 0.66 & 1 & 1.7-120 & 0.8-120 &17\\
Z0	 &	Mur21 & 0.7516 & 0.2484 &	0.0	   & -    & -   & -    & - & 1.7-300 & 0.8-300 &16\\
\hline
VMS &	Mar23 & 0.7516 & 0.2484 &	0.00001	& 0.7 & 0.7 & 0.7 & 1 & 180-300 & 180-300 &3\\
\enddata
\tablenotetext{}{The columns are as follows: Label/Name of region with representative metallicity of the models, label for reference to evolutionary tracks, hydrogen abundance, helium abundance, metallicity, mass-loss rate metallicity relation exponent for O-type star evolutionary phase, exponent for blue supergiant phase, exponent for Wolf-Rayet phase, exponenet for red supergiant phase, mass range of non-rotating models, mass range of rotating (at 40\% critical) models, number of tracks within the mass range.}
\end{deluxetable*}

\begin{figure}[t!]
    \includegraphics[width=\columnwidth]{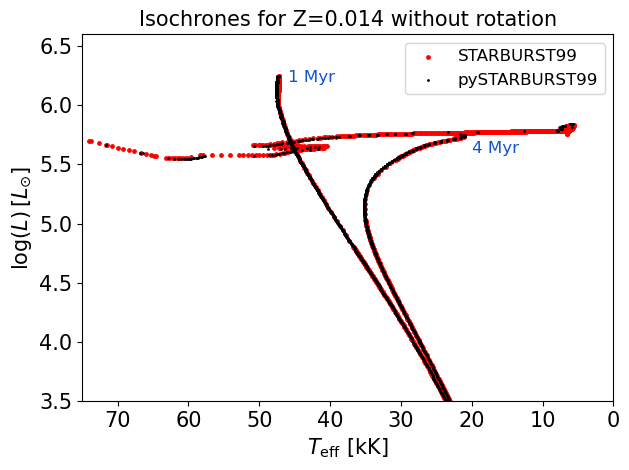}
    \caption{Isochrones computed with $Z=0.014$ \genec\, tracks without rotation from 1-4 Myr. The tracks in red are produced with \starburst\, and the tracks in black are produced with \nstarburst, showing good agreement between the two code versions.}
    \label{fig: isochrones}
\end{figure}

\subsection{Stellar spectral library} \label{sec: atmospheres}

Two different synthetic spectral libraries for MS OB stars are used in \starburst. The first is a grid of 33 models per metallicity, linearly sampled in effective temperature and surface gravity, which have large wavelength coverage ($90-16\times10^{5}$\AA) at relatively low (and wavelength dependent) resolution. These spectra are used to generate SEDs. The second is a set of 86 models per metallicity, with base stellar parameters tailored to the evolutionary tracks. These models are higher resolution UV spectra ($900-3000$\AA) used to generate spectra with synthetic line profiles. Both of these libraries were produced with \wmbasic. Other spectral libraries are available, e.g. for the post-MS, at optical wavelengths or from empirical spectra. The \starburst\, spectral libraries are discussed further in e.g \cite{Leitherer2010, Leitherer2014}. To fully realise the impact of the new suite of evolutionary models we generate an entirely new grid of stellar atmosphere models with \fastwind\, (\citealp{Santolaya-Rey1997a}) to replace the current \wmbasic\, OB library and to fill gaps in the newly established parameter space. In this work we create models to replace only the low resolution synthetic spectral library used to produce SEDs, with plans to update the high resolution spectral library in a future work. Both \wmbasic\, and \fastwind\, are designed for unified stellar atmosphere and wind modelling of hot (OBA) stars with steady-state, spherically symmetric radiatively driven winds in NLTE (non-local-thermodynamic-equilibirium). 
One unique aspect of \wmbasic\ is the option to solve the hydrodynamical equations with a set of line force parameters to inherently obtain the mass-loss rate and velocity profile rather than using a prescription. For the \wmbasic-models applied in \starburst\, \citet{Leitherer2010}, the line force parameters were not iterated self-consistently, but pre-specified based on the metallicity-dependence for the wind-momentum luminosity relation derived by \citet{Kudritzki2000}. In the \fastwind\ models applied in this work, we instead use the common, numerically less costly approach to prescribe the hydrodynamic structure with a quasi-hydrostatic photosphere connected to a $\beta$-type velocity law. This is a standard treatment to create grids of synthetic stellar spectra \citep[see, e.g., the recent method overview in][]{Sander2024}. Specifically, our calculated \fastwind\ models employ a smooth transition at 10\% of the isothermal gas sound speed. This results in additional input parameters relating to the description of the stellar wind, the total mass-loss rate $\dot{M}$, the aforementioned $\beta$ and the terminal wind velocity $v_{\infty}$. We do not make use of the v11-branch of \fastwind\ \citep{Puls2020} to compute the radiative transfer of all line transitions in the comoving frame (CMF), but instead use the CMF only for specified 'foreground' elements while approximating the rest through an approximated line-blanketing formalism. This enables a substantial speed-up in computation time compared to other codes, such as \cmfgen\, or \powr. Nonetheless, the new \fastwind\ models contain improvements in the implementation of physical processes such as line-blanketing and pressure broadening, as well as atomic data. An extensive comparison between \wmbasic\, and \fastwind\, was made in \cite{Puls2005}, finding generally very good agreement in the UV between the two codes with a few caveats in the wavelength range below 400\AA. Since this comparison was made, there have been two full further releases of \fastwind, consisting of a number of important updates, such as the treatments of X-rays and wind clumping \citep{Carneiro2016, Sundqvist2018}. While these have a significant effect on individual line profiles, they should not have too much impact on the overall SED.

\begin{figure}[t!]
    \includegraphics[width=\columnwidth]{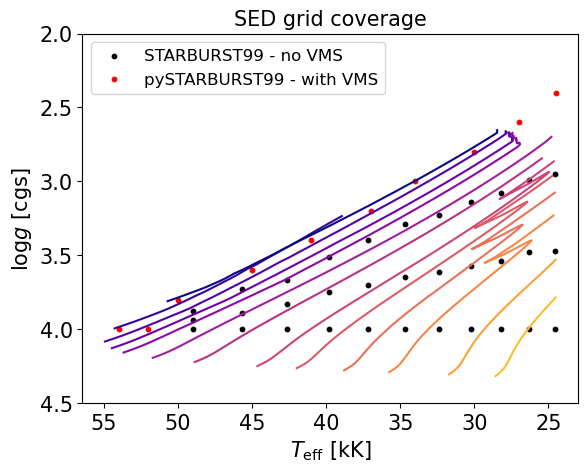}
    \caption{Example of grid coverage for SED predictions with extension for VMS at $Z=0.014$. \genec\, evolutionary tracks are overplotted at regular intervals from 9-500\Msun. This coverage is available for all metallicities. A similar level of coverage is available at zero-metallicity where the tracks are much hotter than any others.} 
    \label{fig: grid_coverage}
\end{figure}

The current \wmbasic\, model grid consists of 33 models per metallicity, at 5 metallicites ($Z=0.001$, $0.004$, $0.008$, $0.02$, and $0.04$). We replicate the parameter space coverage with \fastwind, but adjust the input metallicities for consistency with the \genec\, models ($Z=0.0$, $0.0004$, $0.002$, $0.006$, $0.014$ and $0.02$)\footnote{For numerical reasons the input metallicity to \fastwind\ cannot be $0.0$, instead these extremely low metallicity models are generated with $Z=10^{-6}$.}. 
Additionally, we extend the parameter space coverage to account for evolutionary predictions of VMS. This is achieved through a 33\% increase in the size of the model grid, illustrated in Figure \ref{fig: grid_coverage}. 

For this set of low-resolution stellar SEDs we do not include the effects of X-rays or wind clumping. Additionally we do not tailor mass-loss rates. All of these physical properties will need to be included to reproduce high resolution individual line profiles but have little impact on the overall SED.


\subsection{Python} \label{sec: python}

In translating the code to python from \fortran, our primary goal is to increase the portability and usability of \starburst. With the widespread use of python in the community we hope the new version will facilitate the use of \nstarburst\ with other astrophysical python packages and make the code more accessible. For the present work, we do not reproduce the full capabilities of \starburst\, but focus, for now, on a few key modules: spectral synthesis at low resolution in order to produce SEDs and the ability to make predictions on the ionising flux, wind power and $\beta$-slope. At the foundation of the framework of \starburst\, is the translation of stellar evolutionary tracks into stellar isochrones, in order to interpolate between tracks and make predictions at arbitrary resolution in mass within the range covered by the tracks. One of the often cited drawbacks for a python programme is the speed relative to \fortran, although much of this can be alleviated with the use of packages which act as wrappers around compiled code languages including \fortran\ and C, such as NumPy and SciPy. The bulk of the computational time occupied by \starburst\ is the isochrone synthesis. Thus, the computation time primarily scales with the resolution in mass of the isochrones, which in turn affects the speed of any dependent routines, i.e., higher resolution isochrones will be inherently more expensive in assigning synthetic spectra. To ensure similar runtimes we implement SciPy and NumPy functions for intensive computations in \nstarburst. This results in typical \nstarburst\ runtimes ranging between 15 to 60 seconds, depending on the input options, which is comparable to \starburst. In an attempt to further speed-up the code we also re-establish the resolution as a flexible input parameter and test which resolution is sufficient to produce statistically indistinguishable SEDs to make a recommendation for the input resolution in order to minimise computational load. For instance, it may be optimal to reduce resolution when running exploratory models or applying a flexible fitting routine and increase it when generating a grid or verifying a best-fit solution. Alternately, we introduce the option to increase the resolution within a specified mass range. However, as the interpolation is computed track-to-track, the specificity of the mass range is limited by the ZAMS mass sampling of the input evolutionary tracks. The interpolation scheme is the most tangible variation between \starburst\ and \nstarburst\ with \starburst\ implementing methods from numerical recipes (McCracken and Dorn) and \nstarburst\ using Scipy functions (based on Direckx). However this has fairly little impact on the results of the interpolation and production of isochrones, as shown in Fig. \ref{fig: isochrones}. For the resolution implemented as default, there is essentially no difference between \fortran\, and the python versions. 

We hope that the flexibility of the new language will make the code more portable and user-friendly, enabling more users to make predictions for starburst galaxies themselves and to tailor the code to their interests. However, to avoid blocking functionality of existing user-built frameworks, we also include the new range of metallicity in evolutionary tracks and spectral models in the standard \fortran\, version of \starburst, available on request or via the \starburst\ website (\href{https://massivestars.stsci.edu/starburst99/docs/default.htm}{\starburst}), although we do not include VMS in the fortran version. The full functionalities of \starburst\, will be built into the python version over time and released alongside further improved builds. The current version of \nstarburst\ is available to download at GitHub (\href{https://github.com/CalumHawcroft/Starburst}{GitHub - \nstarburst}) or to run in browser with Colab (\href{https://colab.research.google.com/drive/1MN4P8Q47jshh3bw2sEODb3eJDNxgAbLc?usp=sharing}{Colab - \nstarburst}). User feedback is encouraged by email to pystarburst99@gmail.com.

\section{Results} \label{sec: results}

\subsection{Spectral energy distributions} \label{sec: seds}

\begin{figure}[t!]
    \includegraphics[width=\columnwidth]{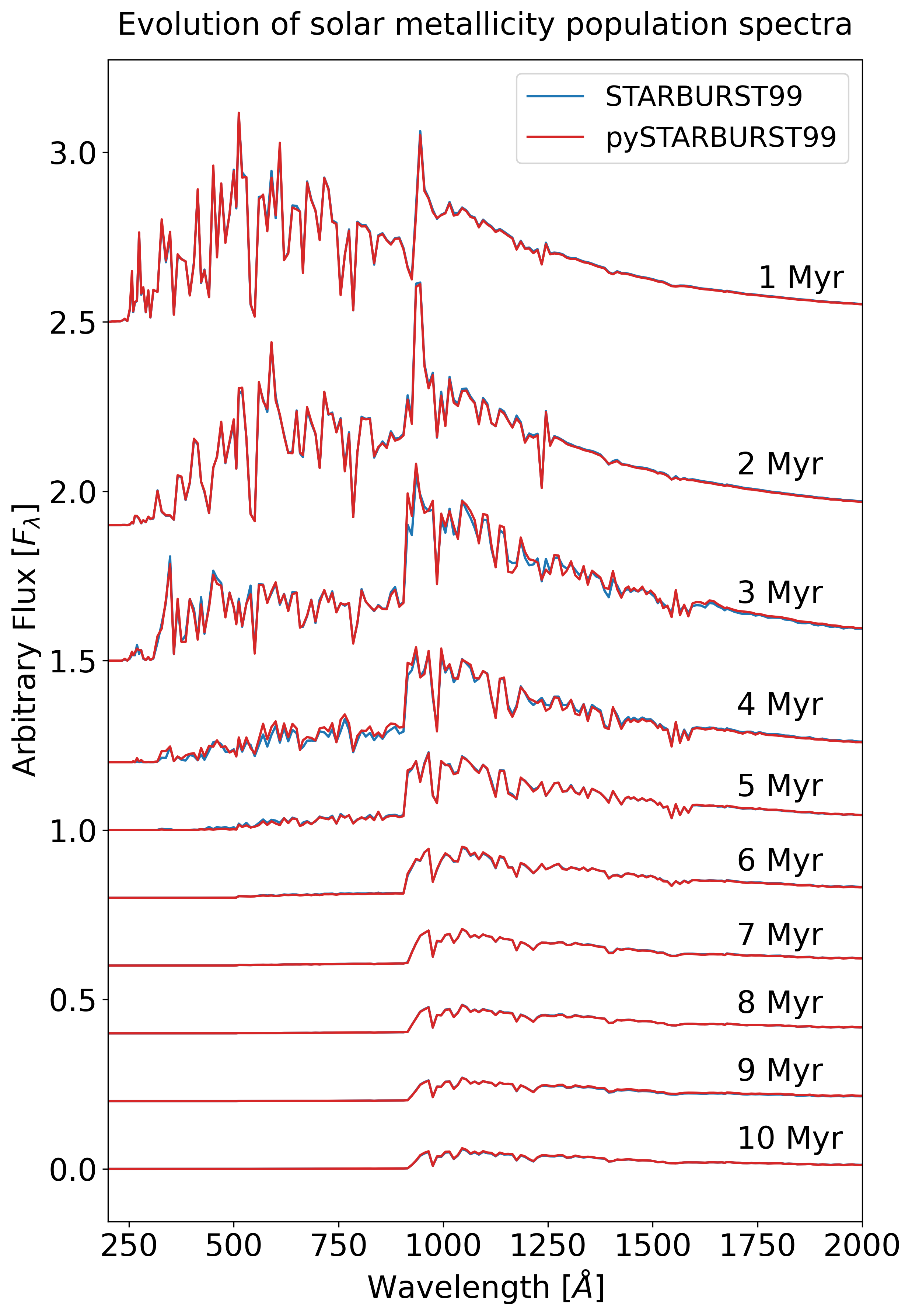}
    \caption{Synthetic FUV SEDs from \starburst\, compared to those produced with \nstarburst\, in both cases utilising the \cite{Ekstrom2012} stellar evolutionary models at Z=0.014 and \wmbasic\, spectra at Z=0.02. The evolution with time is shown from 1 Myr to 10 Myr at intervals of 1 Myr.}
    \label{fig: py_fort_comp}
\end{figure}

With the updates described in Sect.\,\ref{sec: updates}, there are a number of comparisons we can make with respect to computing SEDs. The first is to assess the quality of the SEDs produced with the python version of \starburst\, in comparison to the \fortran\, version. For this task, we include only the well established stellar models used in previous versions of \starburst, i.e., with the \genec\, and \wmbasic\, models included in earlier studies \citep[e.g.,][]{Levesque2012} and using a Kroupa IMF \citep{Kroupa2002} with an upper mass limit of 120$M_{\odot{}}$. In Fig. \ref{fig: py_fort_comp} we show an example for such a population at various times and for non-rotating evolutionary models and include the comparison with rotation in Fig. \ref{fig: py_fort_comp_rot}. The agreement between both versions of the code is very good with any significant variations in flux being associated with differences in the output time of the stellar population, due to the precision of the time increment, which in turn affects the output at a given time. For example, an output spectrum from SB99 at 1.01Myr can have flux differences up to 300\% at short wavelengths ($<1000$\AA\,) over a range of $50\,$\AA\ compared to the 1.01Myr \nstarburst\ spectrum, whereas the \nstarburst\ output at 1.04Myr is within 5\% of the SB99 1.01Myr spectrum with small spikes of 25\% in the residuals on the scale of only 1 resolution element in wavelength\footnote{the resolution of SB99 SEDs can vary across the spectrum, with the highest resolution element of 2\AA\ and a lowest of 10\AA.}. Such small scale discrepancies can be attributed to slight differences in the numerical solutions of the interpolation between stellar evolutionary tracks in the two versions of the code. 
The inclusion of additional masses in the evolutionary grids can also have an impact on the SEDs due to a resulting difference in the isochrones in either code version. The agreement between output SEDs from SB99 and \nstarburst\ has been tested and verified for solar, LMC, and SMC models as defined in the current version of \starburst, with more comparisons shown in Appendix \ref{sec: app-plots}. 

\begin{figure}[t!]
    \includegraphics[width=\columnwidth]{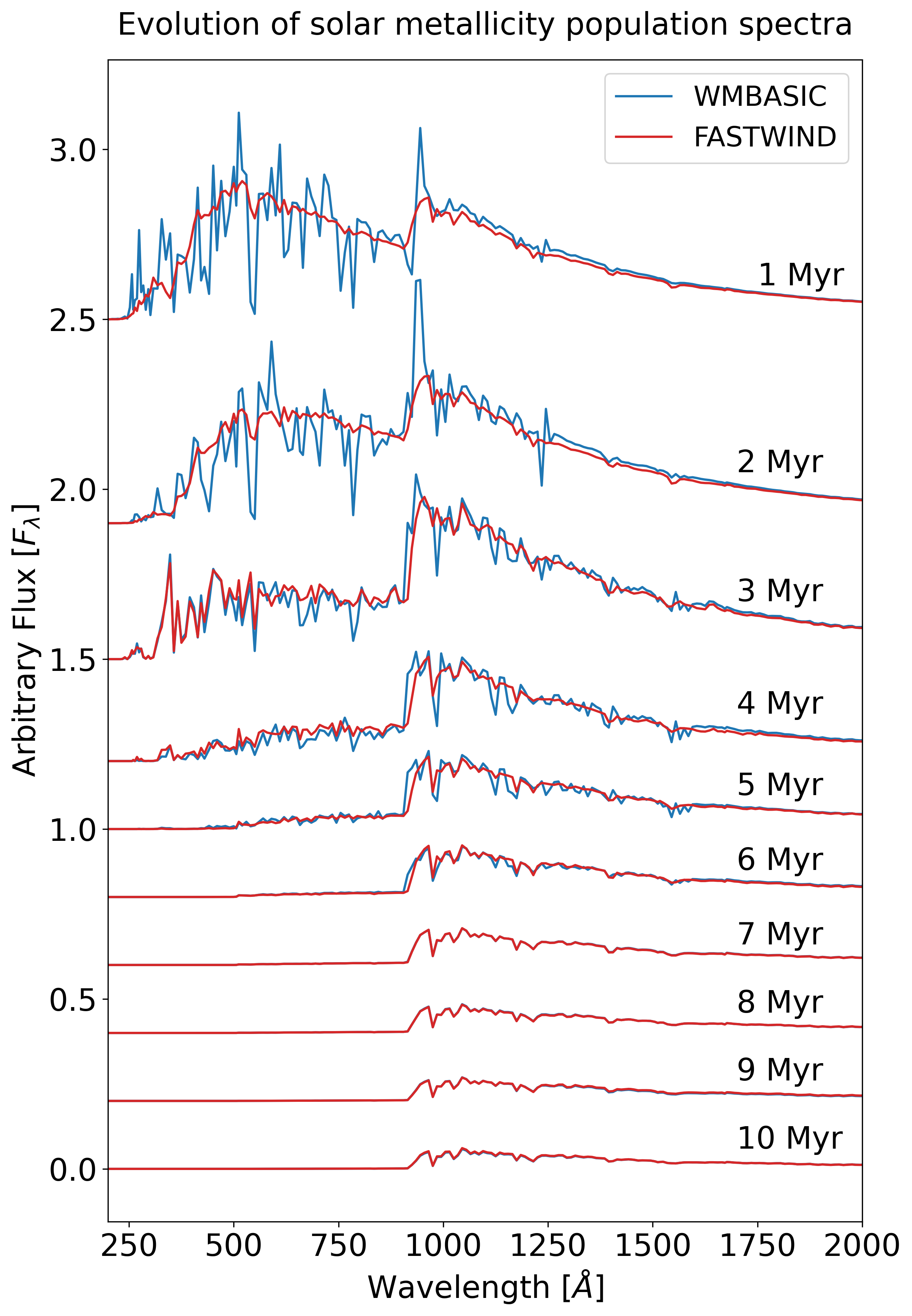}
    \caption{Synthetic FUV SEDs, utilising the \cite{Ekstrom2012} stellar evolutionary models at $Z=0.014$ for both, with \wmbasic\, spectra at $Z=0.02$ compared with \fastwind\, spectra at $Z=0.014$. The evolution with time is shown from 1 Myr to 10 Myr at intervals of 1 Myr.}
    \label{fig: wm_fw_comp}
\end{figure}

With the verification of \nstarburst, our next step is to make a comparison between SEDs produced with \fastwind\, models and those with \wmbasic\, models, keeping the evolutionary models consistent. It is difficult to provide a quantitative assessment of the difference between SEDs produced with \nstarburst\ using \wmbasic\, and \fastwind\, spectra as there are significant differences in the fluxes on scales of $\sim$10\AA\, in the UV (as shown in Fig. \ref{fig: wm_fw_comp}), despite all input spectra being resampled to same wavelength grid. Qualitatively, it appears as though \fastwind\, offers a much smoother SED below $\sim$1500\AA\, which may reduce the uncertainties in SED fitting to observed spectra. There are also specific limitations to certain aspects of the \wmbasic\ spectra, for example the peak at 930\AA is a known artifact due to the lack of pressure broadening in the models. For a few example SEDs in the model grid we also compare the \fastwind\, and \wmbasic\, models with \powr\, models, resampled to the same wavelength grid, and find good agreement between \powr\, and \fastwind.

\begin{figure}[t!]
    \includegraphics[width=\columnwidth]{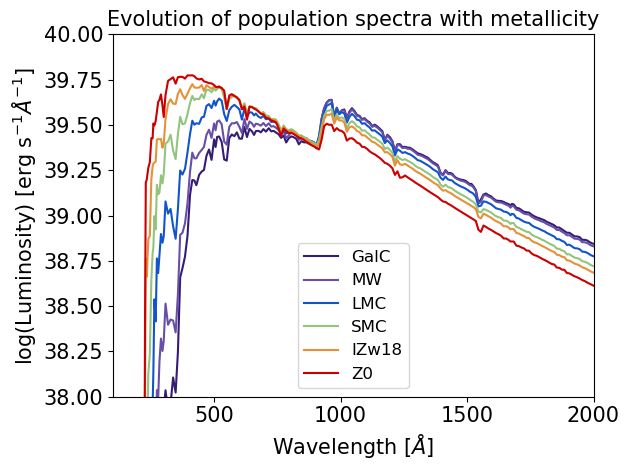}
    \caption{Low resolution synthetic SEDs with metallicities representative of populations in the Galactic centre to zero metallicity. These are produced at 2Myr using \nstarburst\ with the \genec\, tracks, new \fastwind\, model grid, an upper mass limit of 120$\Msun$ and no rotation.}
    \label{fig: sed_metallicity}
\end{figure}

Using the \fastwind\ spectra as well as atmosphere models consistent with the metallicity of the employed \genec\ stellar evolution models, we can produce new population SEDs for all the newly available metallicities. In Fig.\,\ref{fig: sed_metallicity}, we can see a clear trend in the peak of the SED flux shifting to shorter wavelengths at lower metallicity. The bolometric luminosity is held constant for this comparison, showing the effect of lower metallicity resulting in overall higher temperatures for the stellar population at a given time. 

\begin{figure}[t!]
    \includegraphics[width=\columnwidth]{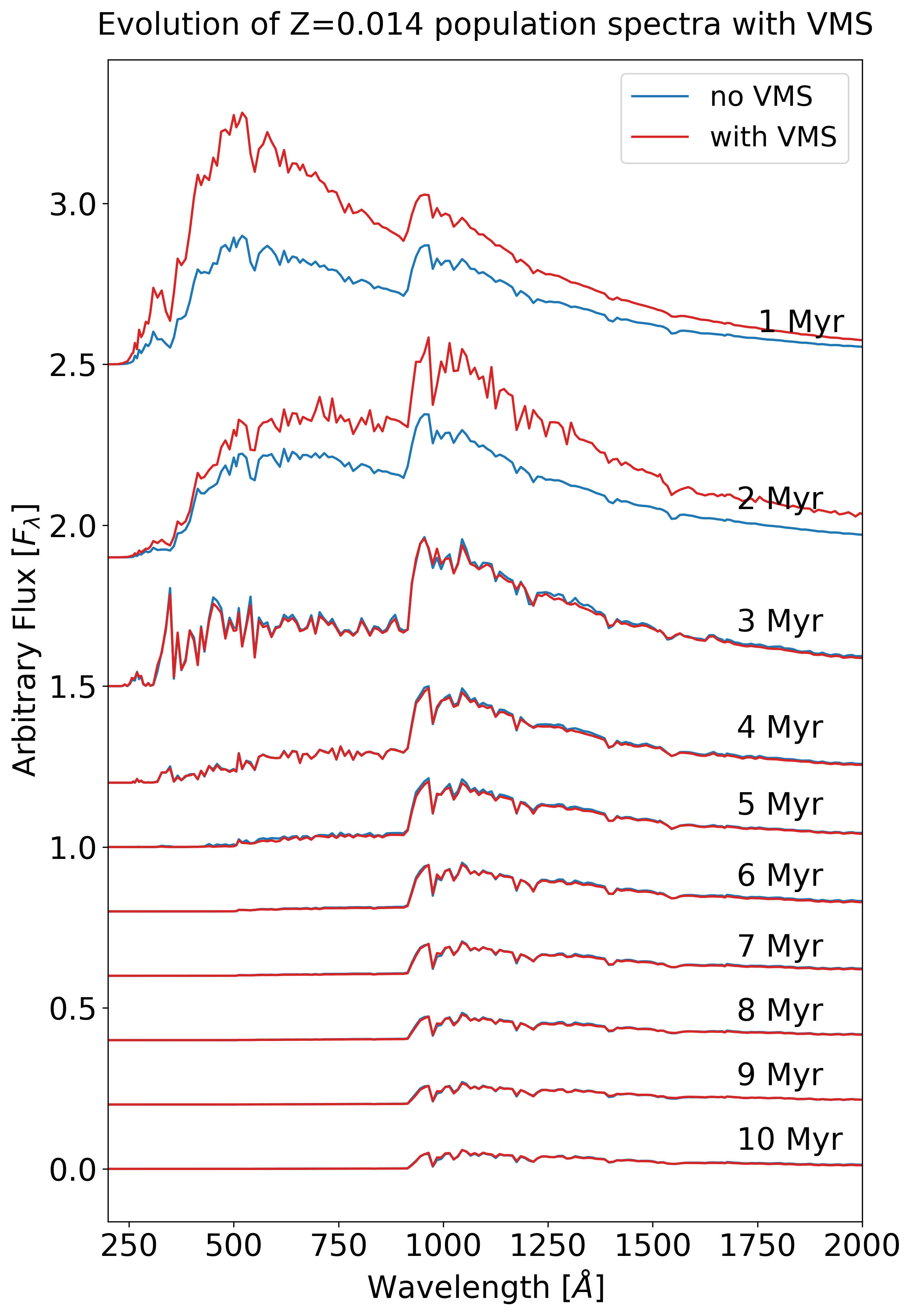}
    \caption{Synthetic FUV SEDs from \nstarburst, utilising the \cite{Ekstrom2012} stellar evolutionary models at Z=0.014 and \fastwind\, spectra at Z=0.014. This is compared with the addition of VMS evolutionary tracks up to $300M_{\odot}$ from \cite{Martinet2023} and \fastwind\, models tailored to match the extended parameter space coverage from VMS evolutionary tracks. The evolution with time is shown from 1 Myr to 10 Myr at intervals of 1 Myr.}
    \label{fig: sed_VMS}
\end{figure}

Finally, we can investigate the impact of the inclusion of VMS in the SEDs. This is shown in Fig.\,\ref{fig: sed_VMS}, where we do not modify the slope of the upper IMF and simply extend the upper mass limit. There is a significant increase in flux which peaks in the FUV at early times and shifts to longer UV wavelengths before all the VMS have reached the end of core carbon burning by 3 Myr. 

\subsection{Ionising fluxes} \label{sec: ionflux}

\begin{figure}[t!]
    \includegraphics[width=\columnwidth]{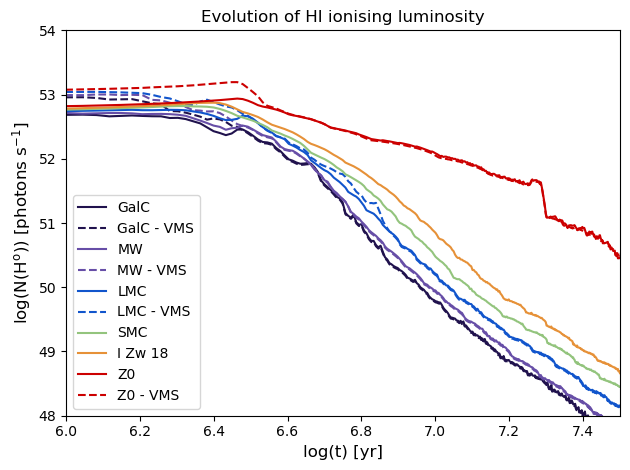}
    \includegraphics[width=\columnwidth]{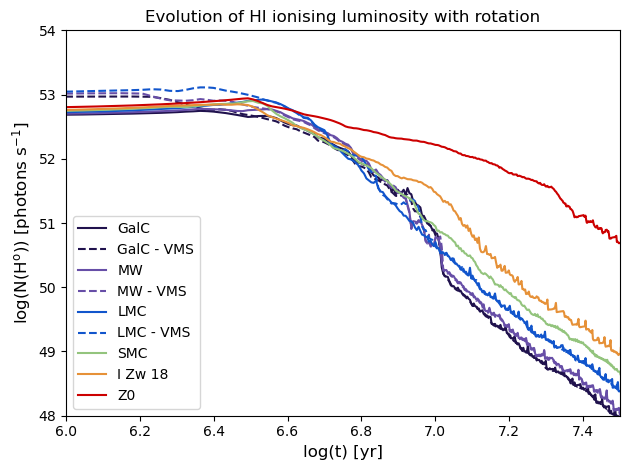}
    \caption{\ion{H}{1} ionising fluxes with metallicity from solar to zero metallicity. The solid lines show \ion{H}{1} ionising fluxes made with \fastwind\, models and upper mass limit of 120\Msun, without rotation. The dashed lines are as the solid lines but with an upper mass limit of 300\Msun.}
    \label{fig: ionflux_metallicity}
\end{figure}

\begin{figure}[t!]
    \includegraphics[width=\columnwidth]{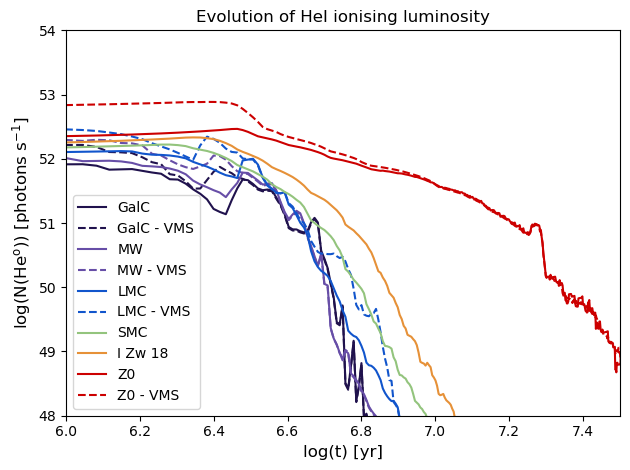}
    \includegraphics[width=\columnwidth]{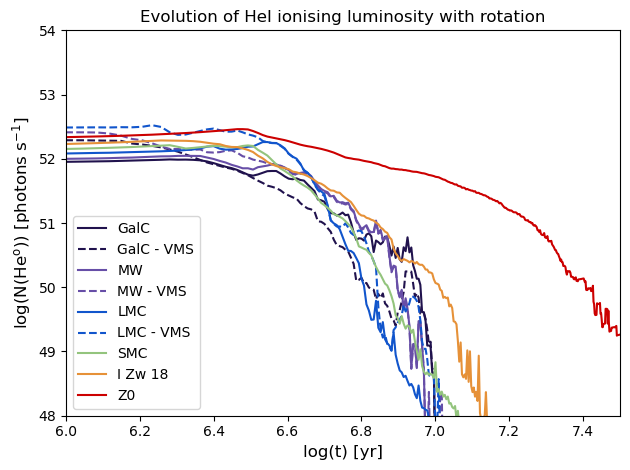}
    \caption{\ion{He}{1} ionising fluxes as in Fig. \ref{fig: ionflux_metallicity}.}
    \label{fig: HeIionflux_metallicity}
\end{figure}

\begin{figure}[t!]
    \includegraphics[width=\columnwidth]{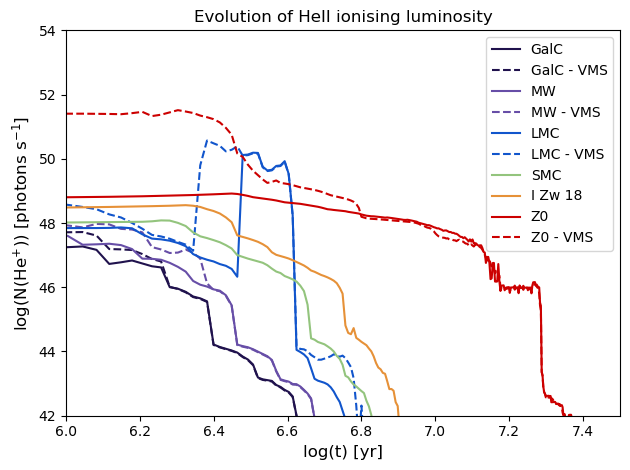}
    \includegraphics[width=\columnwidth]{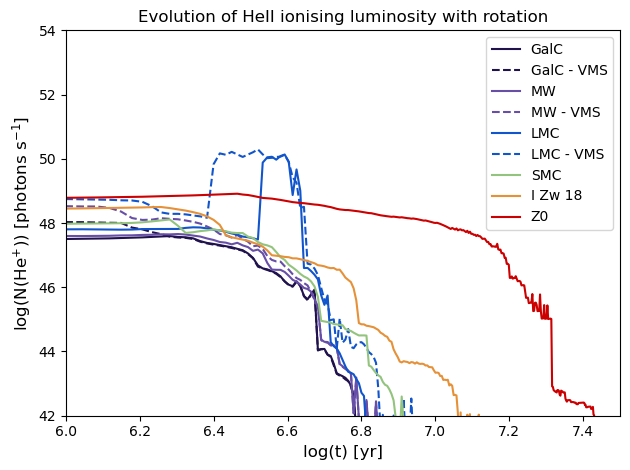}
    \caption{\ion{He}{2} ionising fluxes as in Fig. \ref{fig: ionflux_metallicity}.}
    \label{fig: HeIIionflux_metallicity}
\end{figure}

Of particular interest in the study of starburst galaxies is the nature of the ionising spectrum, which can be greatly impacted by the intrinsic metallicity, as well as assumptions about the composition and evolution of the synthetic stellar population. Significant changes can be caused by the accounting and treatment of rotation \citep{Levesque2012, Leitherer2014}, binarity \citep{Gotberg2020}, abundances \citep{Grasha2021}, and the presumed upper mass limit \citep{Schaerer2024}. The resulting uncertainty in the predicted ionising fluxes propagates into all simulations building on these results, such as photoionisation models and large-scale hydrodynamical simulations. The ionising photon production becomes even more important in the early Universe when Lyman continuum emitting galaxies contributed significantly to the reionisation of the Universe, particularly the reionisation of hydrogen. 
Yet, also the escape fraction must be taken into consideration. While this is beyond the scope of the present work, where we focus on the resulting intrinsic spectra of stellar populations with the updated version of \starburst, escape fractions for massive star populations are an active field of theoretical and observational research and have for example recently been discussed in \citet{Izotov2016,Chisholm2018,Gotberg2020,Pahl2020,Kimm2022,Marques-Chaves2022,Flury2022,Roy2024}.

The behaviour of ionising fluxes with time (shown in Fig. \ref{fig: ionflux_metallicity}) correlates well with the evolution of the stellar population. For the first $\sim$$2\,$Myr, the \ion{H}{1} ionising flux is essentially constant as the stars at the upper end of the IMF expand while the maximum luminosity of the most massive stars increases. There is then a down-turn as the maximum temperature dips below 40\,kK and the most massive stars expand more rapidly, until around 2.5\,Myr when there is an uptick in \ion{H}{1} ionising flux when the most massive stars enter post-MS evolution and are predicted to produce hot WR stars which have lost a considerable part of their hydrogen envelope. The \ion{H}{1} ionising flux reaches a secondary peak around 3.1 Myr, which is immediately followed by the first supernova and a steady decrease in \ion{H}{1} ionising flux as stars with decreasing initial masses reach the end of their evolution. This general trend is true for stellar populations at all metallicities, with the exception of the lack of the WR bump once the metallicity, and therefore wind strength, becomes too low for non-rotating single stars to enter the WR phase. 

When comparing with the original \starburst\, outputs, care has to be taken to be consistent with respect to the metallicity of the evolution models and employed spectral atmosphere models. For example, one could observe a slight increase in ionising flux at solar metallicity in \nstarburst\ compared to \starburst. This is due to the stellar library from \wmbasic\, being computed at $Z=0.02$ while the new \fastwind\ grid uses $Z=0.014$. 

At all metallicities, the initial \ion{H}{1} ionising flux is roughly constant, with a slightly higher ionising flux at early times for lower metallicities (differing by $0.015\,$dex between the most extreme metallicities). At later times, the ionising flux steadily decreases, but for any given time after $\sim$3 Myr, the ionising flux will be much higher at lower metallicity. This trend is a result of the difference in stellar evolution with metallicity. Notably, the input metallicity has essentially no impact on the hydrogen ionising flux of an individual stellar atmosphere model if all other stellar parameters are held constant. While there is less line-blanketing in the atmospheres of lower metallicity stars, photons which experience more scattering and absorption are eventually emitted at similar energies to their initial energy \citep{Mokiem2004}. This even holds for WR stars \citep{SanderVink2020}, but is no longer true for helium ionising photons. The physical process resulting in harder ionising hydrogen spectra at low metallicity comes from the fact that stars are in general hotter and more luminous when the metallicity decreases. Due to the dearth of the CNO cycle the stars must be denser and hotter to reach thermal equilibrium. 
There is an also a secondary opacity effect, in that lowered opacity increases the luminosity to mass ratio. Overall, the low metallicity star is hotter at similar luminosity \citep{Mowlavi1998}. Additionally, the zero metallicity population produces much higher ionising flux at late times compared to stars with any initial metal content. These stars do not become significantly more luminous, but metal-free stars must contract more than higher metallicity stars, until sufficient He is burnt in the core to trigger the CNO cycle. The onset of the CNO cycle is accompanied by a change of slope in the HRD: while contraction goes up-bluewards, when the CNO ignites, the star starts evolving up-redwards. Above 30\,\Msun, this happens already before the ZAMS and the evolution is directly redwards from the ZAMS on. This extreme contraction results in much higher temperatures, for example, a $20M_{\odot}$ star reaches 65\,kK at the ZAMS. This is 20\,kK hotter than a $20M_{\odot}$ star with SMC metallicity. These stars also remain hotter for the full duration of their evolution: A $30M_{\odot}$ star never drops below 25\,kK at zero metallicity, meaning that the ionising flux output remains comparably high throughout the full lifetime. 


The initial ionising flux output only increases with the upper mass limit. If we increase the upper mass limit to the maximum covered by the \genec\, tracks ($500\,M_{\odot}$ at Z=0.014 and $300\,M_{\odot}$ at Z=0.006 and 0.0), there is a corresponding increase in \ion{H}{1} ionising flux by $\sim0.4$\,dex for $500\,M_{\odot}$ and $\sim0.3$\,dex for $300\,M_{\odot}$. The \ion{H}{1} ionising flux quickly returns to that of a traditional population (upper mass limit of $120\,M_{\odot}$) once the VMS have disappeared within the first 3 Myrs. 

With the addition of rotation to any specific population we see that the ionising fluxes stay higher for longer, due to a combination of two effects, first the increase in stellar lifetimes which comes with the rejuvenation of hydrogen in the core, and secondly the increase in temperature and luminosity which results from the larger convective core and lower surface opacity. 

We observe very similar trends for the \ion{He}{1} and \ion{He}{2} ionising fluxes as we do for \ion{H}{1}, with the boosts in ionising flux with rotation and VMS being more pronounced for He than H. We note the presence of a significant enhancement in \ion{He}{2} ionising flux at LMC metallicity during the WR lifetime. This WR bump is not present at lower metallicity due to the lack of WR stars predicted by single-star evolutionary tracks at low metallicity. However the WR bump is also not observed at higher metallicity were WR stars are predicted. There are two contributing factors to this, the main reason being that the WR synthetic spectral grids with $Z\geq0.014$ show significantly less flux below 300\AA\, than lower metallicity models for the same spectral types. The secondary factor is that the LMC isochrones predict much more luminous (by $\sim0.3$dex) WR stars than higher metallicity tracks. Therefore if the LMC WR spectra were not scaled to the luminosity of the isochrones and simply assigned, maintaining the input luminosity of closest stellar atmosphere model the WR bump would disappear.

\begin{figure}[t!]
    \includegraphics[width=\columnwidth]{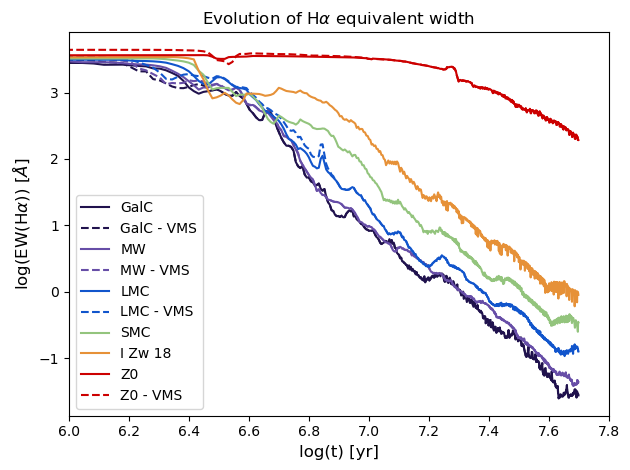}
    \includegraphics[width=\columnwidth]{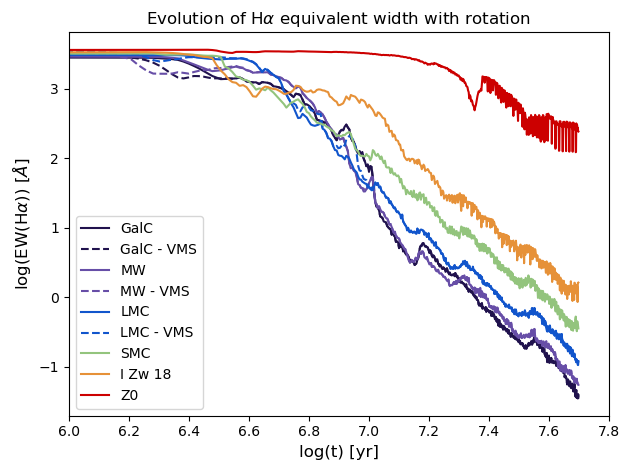}
    \caption{H$\alpha$ equivalent widths as in Fig. \ref{fig: ionflux_metallicity}. }
    \label{fig: ha_ews}
\end{figure}

We can also make predictions of the equivalent widths (EW) of key diagnostics at longer wavelengths (e.g. H$\alpha$, H$\beta$, Pa$\beta$, Br$\gamma$). The EW is defined as the ratio of the luminosity of these lines, which are estimated from the \ion{H}{1} ionising flux, to the continuum flux at the wavelength of the transition. In Fig. \ref{fig: ha_ews} the general trend of H$\alpha$ EW is similar to the \ion{H}{1} ionising flux in time and metallicity, but the behaviour differs with the addition of VMS. There is a similar increase in H$\alpha$ EW with the addition of VMS initially, but from 1.6 Myr to 2.2 Myr the H$\alpha$ EW actually decreases when VMS are included. This is caused by a large jump in the temperatures of the isochrones, with the VMS decreasing in temperature by 20\,kK in a single timestep. This results in the sudden switch to a population of luminous (\logL$> 6.5$) but relatively cool ($\sim20$\,kK) stars, which provide high continuum fluxes with hardly any ionising flux. The trend in H$\alpha$ EW returns to normal after 2.2 Myr when WR stars are predicted.

\begin{figure}[t!]
    \includegraphics[width=\columnwidth]{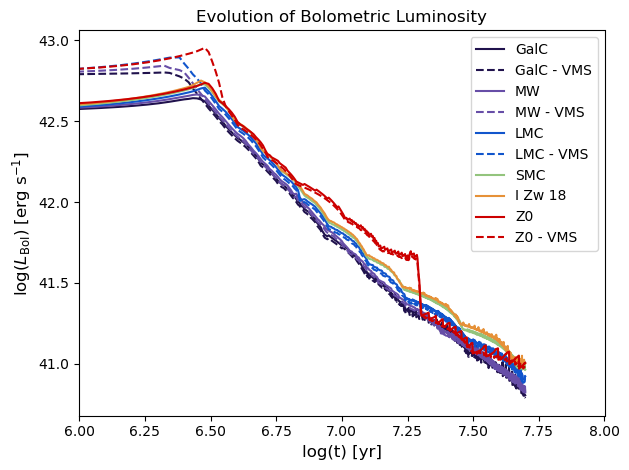}
    \includegraphics[width=\columnwidth]{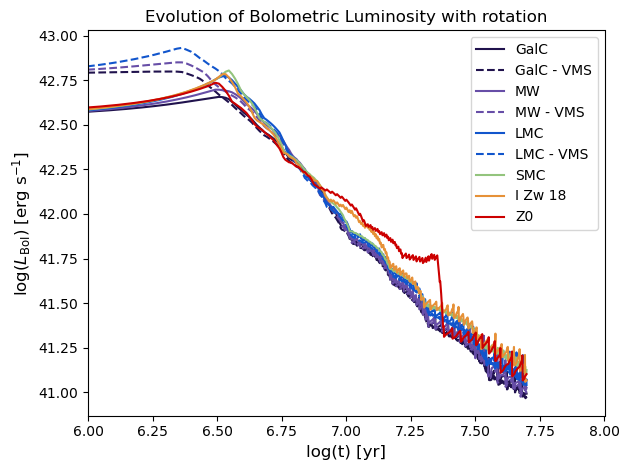}
    \caption{As described in Fig. \ref{fig: ionflux_metallicity} but showing bolometric luminosity.}
    \label{fig: luminosity_metallicity}
\end{figure}

\subsection{Wind power} \label{sec: winds}

Along with ionising processes, the mechanical output of massive stars is an important feedback mechanism, ejecting stellar masses worth of material at speeds on the order of thousands of $\rm{km\,s^{-1}}$. Such high energy outflows affect a number of Galaxy-scale processes, ranging from galaxy and star formation to driving galactic outflows. In addition, significant quantities of metals are deposited into the ISM.

\begin{figure}[t!]
    \includegraphics[width=\columnwidth]{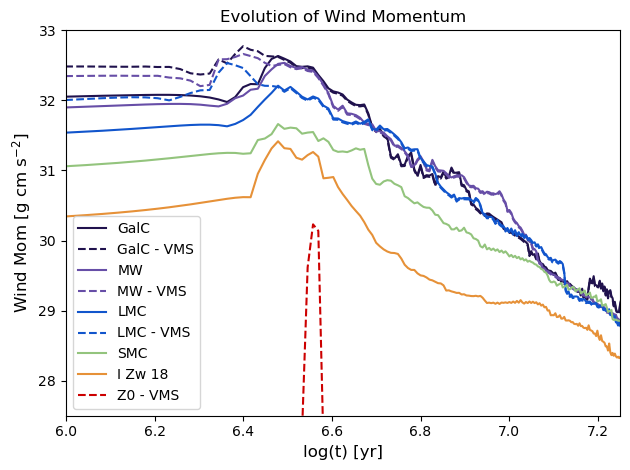}   
    \includegraphics[width=\columnwidth]{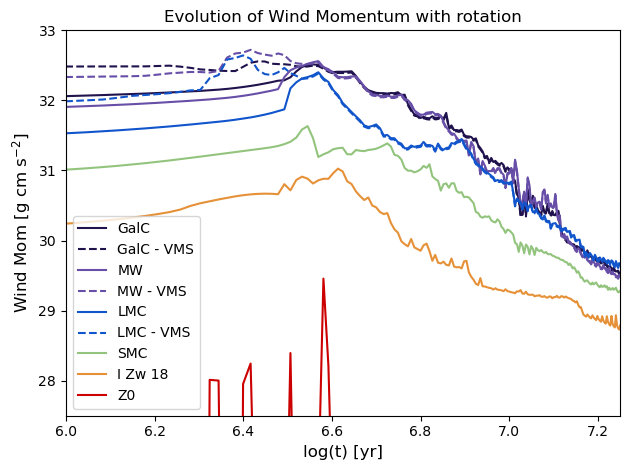} 
    \caption{As described in Fig. \ref{fig: ionflux_metallicity} but showing wind momentum.}
    \label{fig: windmom_VMS}
\end{figure}

The wind power computed with \starburst\ is a combination of the mass-loss rate and terminal wind speed summed over the full population. In Fig.\,\ref{fig: windmom_VMS}, we show the evolution of the wind momentum ($\dot{M}$$v_{\infty}$) with age and metallicity. These predictions are computed using the inputs of the \genec\, models themselves, which are the theoretical predictions of $\dot{M}$ from \cite{Vink2001} and $v_{\infty}$ from \cite{Leitherer1992} on the main sequence. In contrast to the predictions for the ionising flux, we see a decrease in wind power with decreasing metallicity. Such a trend is well established (\citealp{Mokiem2007, Garcia2014, Bouret2015, Marcolino2022, Brands2022, Hawcroft2023, Hawcroft2024, Backs2024, Telford2024}) and expected due to the reduction in line-driving that comes with the reduced number of metal ions in a stellar atmosphere (e.g. \citealp{Vink2001, Krticka2017, Bjorklund2021, VinkSander2021}). At any given metallicity, the inclusion of VMS increases the wind power. With an upper mass limit of $300\,M_{\odot}$, we find an increase in wind momentum of $0.43$ and $0.47\,$dex for the predictions at $Z=0.014$ and $0.006$ respectively. The general trend in wind momentum with time is consistent across all metallicities and corresponds directly to changes in the mass-loss rate prescription used in the \genec\, models. A notable deviation from the general trend are the predictions at zero metallicity. In this case, no mass is lost from the surface of the star in the \genec\ models unless the star reaches either the Eddington limit or critical rotation, at which point the mass-loss rate is increased significantly and is reflected in the non-rotating VMS models or typical masses with rotation. There are currently no estimates of $v_{\infty}$ for zero metallicity models and so to obtain an estimate for the wind power, we implement a metallicity of $Z=10^{-5}$, resulting in maximum $v_{\infty}$ values around 400 \kms\, for the most extreme cases. 

\begin{figure}[t!]
    \includegraphics[width=\columnwidth]{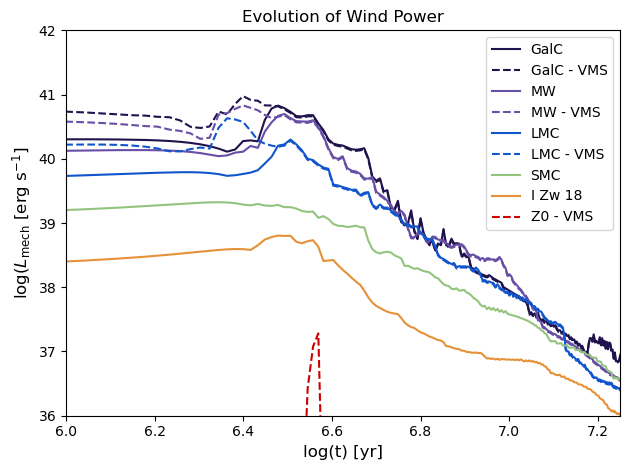}   
    \caption{As described in Fig. \ref{fig: ionflux_metallicity} but showing wind power.}
    \label{fig: windpower_VMS}
\end{figure}

We can also make predictions for the wind power with new theoretical predictions and empirical calibrations for the wind strengths of massive stars as these can be applied directly to pre-computed isochrones. Although the alternate predictions have not been included in the evolutionary models and thus will not be be entirely consistent with the predicted HRD positions, applying them to the isochrones computed using the \cite{Vink2001} predictions gives us a first order of magnitude estimate of the impact of these updated prescriptions with respect to mechanical feedback. 
In the future, we aim to recompute these and other outputs with new evolutionary tracks tailored to our improved understanding of the wind properties of massive stars, ideally covering both the main sequence and post-MS stages. 

For solar metallicity, the theoretical rates from \cite{VinkSander2021} are very consistent with those from \cite{Vink2001} and thus have no impact on the wind momentum. When switching to theoretical mass-loss rates from \citet{Bjorklund2021}, however, there is a significant decrease in wind momentum on the order of $0.8\,$dex at any given time within the first 2.5 Myr. 

In the trend of wind momentum with metallicity, we can see that empirical terminal wind speed calibrations from \cite{Hawcroft2023}, which corresponds to a change from $Z^{0.13}$ to $Z^{0.22}$, has a relatively small impact, causing a downward shift in the wind momentum by $0.05$, $0.1$, and $0.15\,$dex in the LMC, SMC and IZw18 models respectively. This corresponds to a $0.08$, $0.12$, $0.18$ reduction in wind power. 

\subsection{UV slopes}

\begin{figure}[t!]
    \includegraphics[width=\columnwidth]{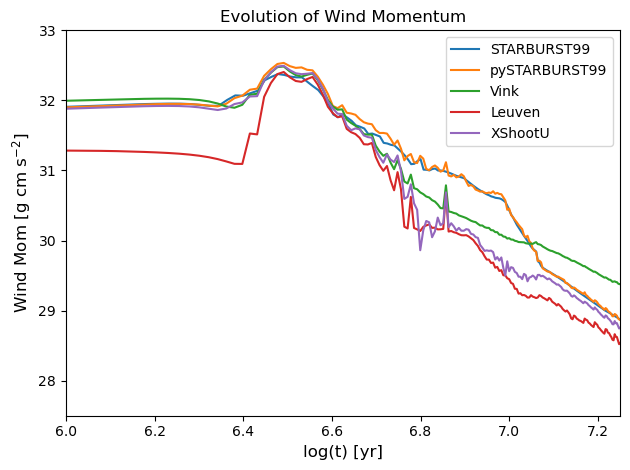}
    \caption{Wind momentum fluxes over time for a typical single stellar population with upper mass limit of 120$M_{\odot}$, no rotation comparing \starburst\ with \nstarburst\ and a few other mass loss and terminal wind speed recipes on the main sequence.}
    \label{fig: windmom_xshootu}
\end{figure}

The UV slope power index $\beta$ (where $F_{\lambda} \propto \lambda^{\beta}$) is commonly used as a gauge of the reddening of the stars due to dust attenuation, making it a powerful diagnostic tool especially at high redshift. 

We measure the $\beta$-slope with a linear fit in log space to the UV continuum in six windows across the range 1250\AA\, to 1750\AA\, as defined in \cite{Calzetti1994}\footnote{These are 1268-1284, 1309-1316, 1342-1371, 1435-1496, 1562-1583, 1677-1740\AA}. We find the $\beta$-slope trend with age is fairly consistent between all metallicities with significant deviations from this general trend arising when VMS are included as a result of their more extreme UV continua. This effect is more prominent in predictions with rotation. For all metallicities, $\beta$ lies within a range of -2.5 to -2.3 in the first $\sim$2.5-3 Myr, with higher metallicities having steeper slopes and the $\beta$ value being essentially constant for any given metallicity. We then find a non-monotonic increase in steepness of the slope until $\sim$10 Myr, after which the slope steadily decreases. 

\begin{figure}[t!]
    \includegraphics[width=\columnwidth]{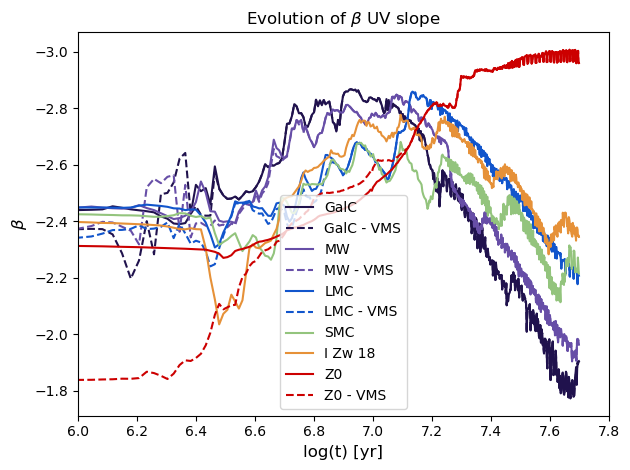}
    \includegraphics[width=\columnwidth]{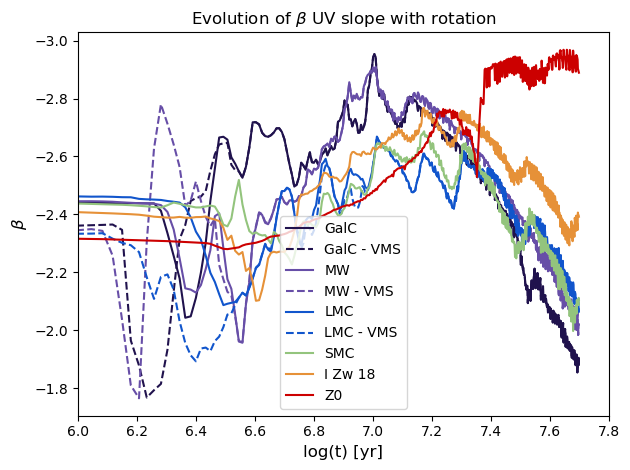}
    \caption{UV slopes computed from synthetic FUV SEDs, within the range 1250-1750\AA, including the contribution from the nebular continuum. }
    \label{fig: uv_slopes}
\end{figure}

\section{Conclusions and future work} \label{sec: conclusions}

We have updated the \starburst\, population synthesis code through three major avenues to improve its capacity to reproduce or predict the properties of star-forming galaxies and maximise ease of use for the community. We first translated the software from \fortran\, to python, and keeping in tradition with \starburst, make the code publicly available on the \starburst\, website/github. Currently not all modules of \starburst\, are available in python but work to add these capabilities are ongoing. The published version will be subject to regular updates. The other two major updates are to expand the metallicity range available and to increase the upper mass limit for both rotating and non-rotating populations. We are now able to offer an updated metallicity grid ($Z=0.02, 0.014, 0.006, 0.002, 0.0004$ and $0.0$) and an extended upper mass limit to $300 M_{\odot}$ for Z = 0.006 and 0.0 as well as models up to $500 M_{\odot}$ at $Z=0.014$ and $0.02$. These updates have been made possible using the latest suite of \genec\, stellar evolutionary models covering all the aforementioned metallicities and mass ranges both with and without the effects of rotation. We also generated a new library of synthetic spectra with the \fastwind\, code to complement this extended grid coverage in stellar evolutionary predictions. 

All the modules presented in this work are available for use in \nstarburst\, and the new stellar models have been implemented in the original \starburst\, \fortran\, code. We aim to add other \starburst\, functions to \nstarburst\ and make them available as they are tested and verified. The main module still to come is predicting high resolution synthetic spectra for the full unresolved stellar population which requires a larger library of more computationally expensive stellar atmosphere models which are currently in progress. Additional further work includes updates to the empirical spectral library which can be significantly expanded at low metallicities thanks to the ULLYSES programme. We also plan to add the functionality to compute mixed age populations.

Within the scope of this work, we have been able to make predictions for a number of properties for integrated stellar populations and how they are impacted by the added new metallicities and the change of upper mass limits, along with rotation. These are: Low resolution spectra or SEDs from 90-10,000\,\AA, ionising fluxes of hydrogen and helium as well as bolometric luminosity and hydrogen line equivalent widths, wind power and momentum (which is further updated with input from the latest literature of hot star winds) and UV $\beta$-slopes. 

We find that we are able to reproduce the results of the original \starburst\, models, including \genec\, models produced up to 2014 and the \wmbasic\, model atmosphere grid from \cite{Leitherer2010}, and are therefore confident in making predictions with new stellar atmosphere (\fastwind\, v10) and evolutionary (\genec\, up to 2023) models with \nstarburst. 

We find a steady shift in the peak SED flux to shorter wavelengths with decreasing metallicity at constant luminosity. As an example, for a 2 Myr old instantaneous burst, flux levels below 900 \AA \,increase by roughly a factor of two from solar to zero metallicity and the fluxes above 900 \AA \,decrease to compensate, although the shift above 900 \AA\, is much lower and becomes less significant with increasing wavelength.
Therefore the ionising flux of hydrogen (and helium) also increases with decreasing metallicity. Metal free populations have significantly higher ionising flux than other predictions at late times. Predictions with rotation maintain high ionising fluxes for slightly longer than non-rotating models due to the increase in stellar lifetime. The addition of VMS increases the ionising flux of any population, but only at early times (less than 3 Myr). The wind momentum decreases with decreasing metallicity but is boosted at early times if VMS are included. There is a more complex trend of the $\beta$-slope with age but generally higher metallicity populations have steeper UV slopes and the steepest slopes are observed in the first $\sim$10 Myr.

We would also like to highlight the limitations of population synthesis with only single star models. It is well established that almost all of the most massive stars are in binary systems \cite{Sana2013}, with over 50\% of these stars expected to undergo an interaction throughout their evolution \cite{Sana2012, Kummer2023}.
The impact of binarity is more pertinent for many of the global properties of stellar populations that aren't extensively discussed in this work, e.g. rates of supernova and compact object coalescence, but binary interaction is also important for the properties presented here particularly at later times (population ages $>10$Myr). For example, \cite{Eldridge2017} predict at least an order of magnitude increase in ionising flux output at 50Myr with the inclusion of binary systems in the BPASS code, with \cite{Gotberg2019} finding similar increases when adding a population of stripped stars to \starburst\ models. Since we find the inclusion of rotation provides the strongest boost to ionising flux, similarly to \cite{Byler2017}, at earlier times (from roughly 3Myr to 10Myr, with an almost 1 dex increase for a 6Myr population) it is clear that parameter studies with single star populations can still offer valuable insights. However, efforts to combine multiple important effects for massive star populations, including metallicity, the upper mass limit, rotation and binarity are incredibly important and will be essential to gain a complete picture of stellar populations throughout the universe. 
Binary interaction should not have a significant impact on the wind luminosity of a stellar population (excluding SN feedback), although non-conservative mass transfer could result in a small boost to wind yields during the stellar lifetime. \cite{Gotberg2019} predict that the H$\alpha$ EW is not strongly impacted by binary interaction in constant star formation scenarios, but that binary stripping has a significant impact on nebular H$\alpha$ if star forming has stopped for more than 10Myr. A similar impact is shown for a wider range of nebular diagnostics in e.g. \cite{Xiao2018, Lecroq2024}. \cite{Gotberg2019} also note that $\beta$ is not impacted by binary interaction below 100Myr, meaning single star predictions provide sufficient and robust for measurements of the UV slope for young populations.

Within this paper we do not provide much comparison to observations, in the interest of releasing the models and code as soon as possible. A few ready applications of these updated models for low metallicity and/or including VMS are to help to better reproduce the global properties of extreme populations at high redshift, for example increased ionising flux \cite{LLerena2024, Munoz2024} and extreme UV slopes \cite{Kumari2024, Dottorini2024, Cullen2025, Fujimoto2025}. However, an extensive assessment of the ability of the updated models to reproduce the observed properties of stellar populations is beyond the scope of this work, but is planned for an upcoming publication. We also plan to discuss the rest-frame UV spectra of star-forming galaxies in a future work which will update the theoretical and empirical high-resolution stellar libraries used in \nstarburst. Further work is also planned to interpret nebular diagnostics using photoionisation models produced from \nstarburst\ SEDs (Arangur\'e et al. in prep.)

\begin{acknowledgements}
Support for this work has been provided by NASA through grant No. AR-16623 from the Space Telescope Science Institute, which is operated by AURA, Inc., under NASA contract NAS5-26555. AACS is funded by the Deutsche Forschungsgemeinschaft (DFG, German Research Foundation) in the form of an Emmy Noether Research Group -- Project-ID 445674056 (SA4064/1-1, PI Sander). AACS acknowledges further support by the Federal Ministry of Education and Research (BMBF) and the Baden-Württemberg Ministry of Science as part of the Excellence Strategy of the German Federal and State Governments. A. W. acknowledges UNAM's DGAPA for the support in carrying out her sabbatical stay at UCSD, USA, through program PASPA. O. A. acknowledges support from CONAHCyT Beca Nacional para estudios de Posgrado, CVU 1142736. A.W. and O.A. acknowledge support from UNAM's DGAPA through program PAPIIT IN106922. GM and SE have received funding from the European Research Council (ERC) under the European Union's Horizon 2020 research and innovation programme (grant agreement No 833925, project STAREX). Based on observations obtained with the NASA/ESA Hubble Space Telescope, retrieved from the Mikulski Archive for Space Telescopes (MAST) at the Space Telescope Science Institute (STScI). STScI is operated by the Association of Universities for Research in Astronomy, Inc. under NASA contract NAS 5-26555. This research has made use of the SIMBAD database, operated at CDS, Strasbourg, France. 
\end{acknowledgements}

\bibliography{SB99}{}

\begin{thebibliography}{}
\expandafter\ifx\csname natexlab\endcsname\relax\def\natexlab#1{#1}\fi
\providecommand{\url}[1]{\href{#1}{#1}}
\providecommand{\dodoi}[1]{doi:~\href{http://doi.org/#1}{\nolinkurl{#1}}}
\providecommand{\doeprint}[1]{\href{http://ascl.net/#1}{\nolinkurl{http://ascl.net/#1}}}
\providecommand{\doarXiv}[1]{\href{https://arxiv.org/abs/#1}{\nolinkurl{https://arxiv.org/abs/#1}}}

\bibitem[{{Agrawal} {et~al.}(2021){Agrawal}, {Hurley}, {Stevenson}, {Sz{\'e}csi}, \& {Flynn}}]{Agrawal2021}
{Agrawal}, P., {Hurley}, J., {Stevenson}, S., {Sz{\'e}csi}, D., \& {Flynn}, C. 2021, in MOBSTER-1 virtual conference: Stellar Variability as a Probe of Magnetic Fields in Massive Stars, 22, \dodoi{10.5281/zenodo.5525465}

\bibitem[{{Asplund} {et~al.}(2009){Asplund}, {Grevesse}, {Sauval}, \& {Scott}}]{Asplund2009}
{Asplund}, M., {Grevesse}, N., {Sauval}, A.~J., \& {Scott}, P. 2009, \araa, 47, 481, \dodoi{10.1146/annurev.astro.46.060407.145222}

\bibitem[{{Backs} {et~al.}(2024){Backs}, {Brands}, {de Koter}, {Kaper}, {Vink}, {Puls}, {Sundqvist}, {Tramper}, {Sana}, {Bernini-Peron}, {Bestenlehner}, {Crowther}, {Hawcroft}, {Ignace}, {Kuiper}, {van Loon}, {Mahy}, {Marcolino}, {Najarro}, {Oskinova}, {Pauli}, {Ramachandran}, {Sander}, \& {Verhamme}}]{Backs2024}
{Backs}, F., {Brands}, S.~A., {de Koter}, A., {et~al.} 2024, arXiv e-prints, arXiv:2411.06884, \dodoi{10.48550/arXiv.2411.06884}

\bibitem[{{Berg} {et~al.}(2022){Berg}, {James}, {King}, {McDonald}, {Chen}, {Chisholm}, {Heckman}, {Martin}, {Stark}, {Aloisi}, {Amor{\'\i}n}, {Arellano-C{\'o}rdova}, {Bayliss}, {Bordoloi}, {Brinchmann}, {Charlot}, {Chevallard}, {Clark}, {Erb}, {Feltre}, {Gronke}, {Hayes}, {Henry}, {Hernandez}, {Jaskot}, {Jones}, {Kewley}, {Kumari}, {Leitherer}, {Llerena}, {Maseda}, {Mingozzi}, {Nanayakkara}, {Ouchi}, {Plat}, {Pogge}, {Ravindranath}, {Rigby}, {Sanders}, {Scarlata}, {Senchyna}, {Skillman}, {Steidel}, {Strom}, {Sugahara}, {Wilkins}, {Wofford}, {Xu}, \& {Classy Team}}]{Berg2022}
{Berg}, D.~A., {James}, B.~L., {King}, T., {et~al.} 2022, \apjs, 261, 31, \dodoi{10.3847/1538-4365/ac6c03}

\bibitem[{{Berg} {et~al.}(2024){Berg}, {Skillman}, {Chisholm}, {Pogge}, {Gazagnes}, {Rogers}, {Erb}, {Arellano-C{\'o}rdova}, {Leitherer}, {Appel}, \& {Moustakas}}]{Berg2024}
{Berg}, D.~A., {Skillman}, E.~D., {Chisholm}, J., {et~al.} 2024, \apj, 971, 87, \dodoi{10.3847/1538-4357/ad5292}

\bibitem[{Bestenlehner {et~al.}(2014)Bestenlehner, Gr{\"{a}}fener, Vink, Najarro, {De Koter}, Sana, Evans, Crowther, H{\'{e}}nault-Brunet, Herrero, Langer, Schneider, Sim{\'{o}}n-D{\'{i}}az, Taylor, \& Walborn}]{Bestenlehner2014}
Bestenlehner, J.~M., Gr{\"{a}}fener, G., Vink, J.~S., {et~al.} 2014, {The VLT-FLAMES tarantula survey: XVII. Physical and wind properties of massive stars at the top of the main sequence},  EDP Sciences, \dodoi{10.1051/0004-6361/201423643}

\bibitem[{{Bestenlehner} {et~al.}(2020){Bestenlehner}, {Crowther}, {Caballero-Nieves}, {Schneider}, {Sim{\'o}n-D{\'\i}az}, {Brands}, {de Koter}, {Gr{\"a}fener}, {Herrero}, {Langer}, {Lennon}, {Ma{\'\i}z Apell{\'a}niz}, {Puls}, \& {Vink}}]{Bestenlehner2020}
{Bestenlehner}, J.~M., {Crowther}, P.~A., {Caballero-Nieves}, S.~M., {et~al.} 2020, \mnras, 499, 1918, \dodoi{10.1093/mnras/staa2801}

\bibitem[{{Bj{\"o}rklund} {et~al.}(2021){Bj{\"o}rklund}, {Sundqvist}, {Puls}, \& {Najarro}}]{Bjorklund2021}
{Bj{\"o}rklund}, R., {Sundqvist}, J.~O., {Puls}, J., \& {Najarro}, F. 2021, \aap, 648, A36, \dodoi{10.1051/0004-6361/202038384}

\bibitem[{{Boquien} {et~al.}(2019){Boquien}, {Burgarella}, {Roehlly}, {Buat}, {Ciesla}, {Corre}, {Inoue}, \& {Salas}}]{Boquien2019}
{Boquien}, M., {Burgarella}, D., {Roehlly}, Y., {et~al.} 2019, \aap, 622, A103, \dodoi{10.1051/0004-6361/201834156}

\bibitem[{{Bouret} {et~al.}(2015){Bouret}, {Lanz}, {Hillier}, {Martins}, {Marcolino}, \& {Depagne}}]{Bouret2015}
{Bouret}, J.~C., {Lanz}, T., {Hillier}, D.~J., {et~al.} 2015, \mnras, 449, 1545, \dodoi{10.1093/mnras/stv379}

\bibitem[{{Brands} {et~al.}(2022){Brands}, {de Koter}, {Bestenlehner}, {Crowther}, {Sundqvist}, {Puls}, {Caballero-Nieves}, {Abdul-Masih}, {Driessen}, {Garc{\'\i}a}, {Geen}, {Gr{\"a}fener}, {Hawcroft}, {Kaper}, {Keszthelyi}, {Langer}, {Sana}, {Schneider}, {Shenar}, \& {Vink}}]{Brands2022}
{Brands}, S.~A., {de Koter}, A., {Bestenlehner}, J.~M., {et~al.} 2022, \aap, 663, A36, \dodoi{10.1051/0004-6361/202142742}

\bibitem[{{Broekgaarden} {et~al.}(2021){Broekgaarden}, {Berger}, {Neijssel}, {Vigna-G{\'o}mez}, {Chattopadhyay}, {Stevenson}, {Chruslinska}, {Justham}, {de Mink}, \& {Mandel}}]{Broekgaarden2021}
{Broekgaarden}, F.~S., {Berger}, E., {Neijssel}, C.~J., {et~al.} 2021, \mnras, 508, 5028, \dodoi{10.1093/mnras/stab2716}

\bibitem[{{Brott} {et~al.}(2011){Brott}, {de Mink}, {Cantiello}, {Langer}, {de Koter}, {Evans}, {Hunter}, {Trundle}, \& {Vink}}]{Brott2011}
{Brott}, I., {de Mink}, S.~E., {Cantiello}, M., {et~al.} 2011, \aap, 530, A115, \dodoi{10.1051/0004-6361/201016113}

\bibitem[{{Bruzual} \& {Charlot}(2003)}]{Bruzual2003}
{Bruzual}, G., \& {Charlot}, S. 2003, \mnras, 344, 1000, \dodoi{10.1046/j.1365-8711.2003.06897.x}

\bibitem[{{Byler} {et~al.}(2017){Byler}, {Dalcanton}, {Conroy}, \& {Johnson}}]{Byler2017}
{Byler}, N., {Dalcanton}, J.~J., {Conroy}, C., \& {Johnson}, B.~D. 2017, \apj, 840, 44, \dodoi{10.3847/1538-4357/aa6c66}

\bibitem[{{Byrne} {et~al.}(2025){Byrne}, {Eldridge}, \& {Stanway}}]{Byrne2025}
{Byrne}, C.~M., {Eldridge}, J.~J., \& {Stanway}, E.~R. 2025, \mnras, 537, 2433, \dodoi{10.1093/mnras/staf178}

\bibitem[{{Calzetti} {et~al.}(1994){Calzetti}, {Kinney}, \& {Storchi-Bergmann}}]{Calzetti1994}
{Calzetti}, D., {Kinney}, A.~L., \& {Storchi-Bergmann}, T. 1994, \apj, 429, 582, \dodoi{10.1086/174346}

\bibitem[{{Cameron} {et~al.}(2023){Cameron}, {Katz}, {Rey}, \& {Saxena}}]{Cameron2023}
{Cameron}, A.~J., {Katz}, H., {Rey}, M.~P., \& {Saxena}, A. 2023, \mnras, 523, 3516, \dodoi{10.1093/mnras/stad1579}

\bibitem[{Carneiro {et~al.}(2016)Carneiro, Puls, Sundqvist, \& Hoffmann}]{Carneiro2016}
Carneiro, L.~P., Puls, J., Sundqvist, J.~O., \& Hoffmann, T.~L. 2016, Astronomy and Astrophysics, 590, \dodoi{10.1051/0004-6361/201527718}

\bibitem[{{Chen} {et~al.}(2015){Chen}, {Bressan}, {Girardi}, {Marigo}, {Kong}, \& {Lanza}}]{Chen2015}
{Chen}, Y., {Bressan}, A., {Girardi}, L., {et~al.} 2015, \mnras, 452, 1068, \dodoi{10.1093/mnras/stv1281}

\bibitem[{{Chisholm} {et~al.}(2018){Chisholm}, {Gazagnes}, {Schaerer}, {Verhamme}, {Rigby}, {Bayliss}, {Sharon}, {Gladders}, \& {Dahle}}]{Chisholm2018}
{Chisholm}, J., {Gazagnes}, S., {Schaerer}, D., {et~al.} 2018, \aap, 616, A30, \dodoi{10.1051/0004-6361/201832758}

\bibitem[{{Choi} {et~al.}(2016){Choi}, {Dotter}, {Conroy}, {Cantiello}, {Paxton}, \& {Johnson}}]{Choi2016}
{Choi}, J., {Dotter}, A., {Conroy}, C., {et~al.} 2016, \apj, 823, 102, \dodoi{10.3847/0004-637X/823/2/102}

\bibitem[{{Conroy} \& {Gunn}(2010)}]{conroy2010}
{Conroy}, C., \& {Gunn}, J.~E. 2010, \apj, 712, 833, \dodoi{10.1088/0004-637X/712/2/833}

\bibitem[{{Conroy} {et~al.}(2009){Conroy}, {Gunn}, \& {White}}]{Conroy2009}
{Conroy}, C., {Gunn}, J.~E., \& {White}, M. 2009, \apj, 699, 486, \dodoi{10.1088/0004-637X/699/1/486}

\bibitem[{{Crowther} \& {Castro}(2024)}]{Crowther2024}
{Crowther}, P.~A., \& {Castro}, N. 2024, \mnras, 527, 9023, \dodoi{10.1093/mnras/stad3698}

\bibitem[{{Crowther} {et~al.}(2010){Crowther}, {Schnurr}, {Hirschi}, {Yusof}, {Parker}, {Goodwin}, \& {Kassim}}]{Crowther2010}
{Crowther}, P.~A., {Schnurr}, O., {Hirschi}, R., {et~al.} 2010, \mnras, 408, 731, \dodoi{10.1111/j.1365-2966.2010.17167.x}

\bibitem[{{Crowther} {et~al.}(2016){Crowther}, {Caballero-Nieves}, {Bostroem}, {Ma{\'\i}z Apell{\'a}niz}, {Schneider}, {Walborn}, {Angus}, {Brott}, {Bonanos}, {de Koter}, {de Mink}, {Evans}, {Gr{\"a}fener}, {Herrero}, {Howarth}, {Langer}, {Lennon}, {Puls}, {Sana}, \& {Vink}}]{Crowther2016}
{Crowther}, P.~A., {Caballero-Nieves}, S.~M., {Bostroem}, K.~A., {et~al.} 2016, \mnras, 458, 624, \dodoi{10.1093/mnras/stw273}

\bibitem[{{Cullen} {et~al.}(2021){Cullen}, {Shapley}, {McLure}, {Dunlop}, {Sanders}, {Topping}, {Reddy}, {Amor{\'\i}n}, {Begley}, {Bolzonella}, {Calabr{\`o}}, {Carnall}, {Castellano}, {Cimatti}, {Cirasuolo}, {Cresci}, {Fontana}, {Fontanot}, {Garilli}, {Guaita}, {Hamadouche}, {Hathi}, {Mannucci}, {McLeod}, {Pentericci}, {Saxena}, {Talia}, \& {Zamorani}}]{Cullen2021}
{Cullen}, F., {Shapley}, A.~E., {McLure}, R.~J., {et~al.} 2021, \mnras, 505, 903, \dodoi{10.1093/mnras/stab1340}

\bibitem[{{Cullen} {et~al.}(2025){Cullen}, {Carnall}, {Scholte}, {McLeod}, {McLure}, {Arellano-C{\'o}rdova}, {Stanton}, {Donnan}, {Dunlop}, {Shapley}, {Barrufet}, {Begley}, {Bondestam}, {Cirasuolo}, {Leung}, {Pollock}, \& {Stevenson}}]{Cullen2025}
{Cullen}, F., {Carnall}, A.~C., {Scholte}, D., {et~al.} 2025, arXiv e-prints, arXiv:2501.11099, \dodoi{10.48550/arXiv.2501.11099}

\bibitem[{{de Jager} {et~al.}(1988){de Jager}, {Nieuwenhuijzen}, \& {van der Hucht}}]{deJager1988}
{de Jager}, C., {Nieuwenhuijzen}, H., \& {van der Hucht}, K.~A. 1988, \aaps, 72, 259

\bibitem[{{Dotter}(2016)}]{Dotter2016}
{Dotter}, A. 2016, \apjs, 222, 8, \dodoi{10.3847/0067-0049/222/1/8}

\bibitem[{{Dottorini} {et~al.}(2024){Dottorini}, {Calabr{\`o}}, {Pentericci}, {Mascia}, {Llerena}, {Napolitano}, {Santini}, {Roberts-Borsani}, {Castellano}, {Amor{\'\i}n}, {Dickinson}, {Fontana}, {Hathi}, {Hirschmann}, {Koekemoer}, {Lucas}, {Merlin}, {Morales}, {Pacucci}, {Wilkins}, {Arrabal Haro}, {Bagley}, {Finkelstein}, {Kartaltepe}, {Papovich}, \& {Pirzkal}}]{Dottorini2024}
{Dottorini}, D., {Calabr{\`o}}, A., {Pentericci}, L., {et~al.} 2024, arXiv e-prints, arXiv:2412.01623, \dodoi{10.48550/arXiv.2412.01623}

\bibitem[{{Eggenberger} {et~al.}(2021){Eggenberger}, {Ekstr{\"o}m}, {Georgy}, {Martinet}, {Pezzotti}, {Nandal}, {Meynet}, {Buldgen}, {Salmon}, {Haemmerl{\'e}}, {Maeder}, {Hirschi}, {Yusof}, {Groh}, {Farrell}, {Murphy}, \& {Choplin}}]{Eggenberger2021}
{Eggenberger}, P., {Ekstr{\"o}m}, S., {Georgy}, C., {et~al.} 2021, \aap, 652, A137, \dodoi{10.1051/0004-6361/202141222}

\bibitem[{{Ekstr{\"o}m} {et~al.}(2012){Ekstr{\"o}m}, {Georgy}, {Eggenberger}, {Meynet}, {Mowlavi}, {Wyttenbach}, {Granada}, {Decressin}, {Hirschi}, {Frischknecht}, {Charbonnel}, \& {Maeder}}]{Ekstrom2012}
{Ekstr{\"o}m}, S., {Georgy}, C., {Eggenberger}, P., {et~al.} 2012, \aap, 537, A146, \dodoi{10.1051/0004-6361/201117751}

\bibitem[{{Eldridge} \& {Stanway}(2022)}]{Eldridge2022}
{Eldridge}, J.~J., \& {Stanway}, E.~R. 2022, arXiv e-prints, arXiv:2202.01413.
\newblock \doarXiv{2202.01413}

\bibitem[{{Eldridge} {et~al.}(2017){Eldridge}, {Stanway}, {Xiao}, {McClelland}, {Taylor}, {Ng}, {Greis}, \& {Bray}}]{Eldridge2017}
{Eldridge}, J.~J., {Stanway}, E.~R., {Xiao}, L., {et~al.} 2017, \pasa, 34, e058, \dodoi{10.1017/pasa.2017.51}

\bibitem[{{Eldridge} \& {Tout}(2004)}]{Eldridge2004}
{Eldridge}, J.~J., \& {Tout}, C.~A. 2004, \mnras, 353, 87, \dodoi{10.1111/j.1365-2966.2004.08041.x}

\bibitem[{{Figer} {et~al.}(2002){Figer}, {Najarro}, {Gilmore}, {Morris}, {Kim}, {Serabyn}, {McLean}, {Gilbert}, {Graham}, {Larkin}, {Levenson}, \& {Teplitz}}]{Figer2002}
{Figer}, D.~F., {Najarro}, F., {Gilmore}, D., {et~al.} 2002, \apj, 581, 258, \dodoi{10.1086/344154}

\bibitem[{{Flury} {et~al.}(2022){Flury}, {Jaskot}, {Ferguson}, {Worseck}, {Makan}, {Chisholm}, {Saldana-Lopez}, {Schaerer}, {McCandliss}, {Wang}, {Ford}, {Heckman}, {Ji}, {Giavalisco}, {Amorin}, {Atek}, {Blaizot}, {Borthakur}, {Carr}, {Castellano}, {Cristiani}, {De Barros}, {Dickinson}, {Finkelstein}, {Fleming}, {Fontanot}, {Garel}, {Grazian}, {Hayes}, {Henry}, {Mauerhofer}, {Micheva}, {Oey}, {Ostlin}, {Papovich}, {Pentericci}, {Ravindranath}, {Rosdahl}, {Rutkowski}, {Santini}, {Scarlata}, {Teplitz}, {Thuan}, {Trebitsch}, {Vanzella}, {Verhamme}, \& {Xu}}]{Flury2022}
{Flury}, S.~R., {Jaskot}, A.~E., {Ferguson}, H.~C., {et~al.} 2022, \apjs, 260, 1, \dodoi{10.3847/1538-4365/ac5331}

\bibitem[{{Fragos} {et~al.}(2023){Fragos}, {Andrews}, {Bavera}, {Berry}, {Coughlin}, {Dotter}, {Giri}, {Kalogera}, {Katsaggelos}, {Kovlakas}, {Lalvani}, {Misra}, {Srivastava}, {Qin}, {Rocha}, {Rom{\'a}n-Garza}, {Serra}, {Stahle}, {Sun}, {Teng}, {Trajcevski}, {Tran}, {Xing}, {Zapartas}, \& {Zevin}}]{Fragos2023}
{Fragos}, T., {Andrews}, J.~J., {Bavera}, S.~S., {et~al.} 2023, \apjs, 264, 45, \dodoi{10.3847/1538-4365/ac90c1}

\bibitem[{{Fujimoto} {et~al.}(2025){Fujimoto}, {Naidu}, {Chisholm}, {Atek}, {Endsley}, {Kokorev}, {Furtak}, {Pan}, {Liu}, {Bromm}, {Venditti}, {Visbal}, {Sarmento}, {Weibel}, {Oesch}, {Brammer}, {Schaerer}, {Adamo}, {Berg}, {Bezanson}, {Chemerynska}, {Claeyssens}, {Dessauges-Zavadsky}, {Frebel}, {Korber}, {Labbe}, {Marques-Chaves}, {Matthee}, {McQuinn}, {Mu{\~n}oz}, {Natarajan}, {Saldana-Lopez}, {Suess}, {Volonteri}, \& {Zitrin}}]{Fujimoto2025}
{Fujimoto}, S., {Naidu}, R.~P., {Chisholm}, J., {et~al.} 2025, arXiv e-prints, arXiv:2501.11678, \dodoi{10.48550/arXiv.2501.11678}

\bibitem[{{Garcia} {et~al.}(2014){Garcia}, {Herrero}, {Najarro}, {Lennon}, \& {Alejandro Urbaneja}}]{Garcia2014}
{Garcia}, M., {Herrero}, A., {Najarro}, F., {Lennon}, D.~J., \& {Alejandro Urbaneja}, M. 2014, \apj, 788, 64, \dodoi{10.1088/0004-637X/788/1/64}

\bibitem[{{Geen} {et~al.}(2023){Geen}, {Agrawal}, {Crowther}, {Keller}, {de Koter}, {Keszthelyi}, {van de Voort}, {Ali}, {Backs}, {Bonne}, {Brugaletta}, {Derkink}, {Ekstr{\"o}m}, {Fichtner}, {Grassitelli}, {G{\"o}tberg}, {Higgins}, {Laplace}, {You Liow}, {Lorenzo}, {McLeod}, {Meynet}, {Newsome}, {Oliva}, {Ramachandran}, {Rey}, {Rieder}, {Romano-D{\'\i}az}, {Sabhahit}, {Sander}, {Sarwar}, {Stinshoff}, {Stoop}, {Sz{\'e}csi}, {Trebitsch}, {Vink}, \& {Winch}}]{Geen2023}
{Geen}, S., {Agrawal}, P., {Crowther}, P.~A., {et~al.} 2023, \pasp, 135, 021001, \dodoi{10.1088/1538-3873/acb6b5}

\bibitem[{{Georgy} {et~al.}(2013){Georgy}, {Ekstr{\"o}m}, {Eggenberger}, {Meynet}, {Haemmerl{\'e}}, {Maeder}, {Granada}, {Groh}, {Hirschi}, {Mowlavi}, {Yusof}, {Charbonnel}, {Decressin}, \& {Barblan}}]{Georgy2013}
{Georgy}, C., {Ekstr{\"o}m}, S., {Eggenberger}, P., {et~al.} 2013, \aap, 558, A103, \dodoi{10.1051/0004-6361/201322178}

\bibitem[{{G{\"o}tberg} {et~al.}(2019){G{\"o}tberg}, {de Mink}, {Groh}, {Leitherer}, \& {Norman}}]{Gotberg2019}
{G{\"o}tberg}, Y., {de Mink}, S.~E., {Groh}, J.~H., {Leitherer}, C., \& {Norman}, C. 2019, \aap, 629, A134, \dodoi{10.1051/0004-6361/201834525}

\bibitem[{{G{\"o}tberg} {et~al.}(2020){G{\"o}tberg}, {de Mink}, {McQuinn}, {Zapartas}, {Groh}, \& {Norman}}]{Gotberg2020}
{G{\"o}tberg}, Y., {de Mink}, S.~E., {McQuinn}, M., {et~al.} 2020, \aap, 634, A134, \dodoi{10.1051/0004-6361/201936669}

\bibitem[{{Gr{\"a}fener} \& {Hamann}(2008)}]{Grafener2008}
{Gr{\"a}fener}, G., \& {Hamann}, W.~R. 2008, \aap, 482, 945, \dodoi{10.1051/0004-6361:20066176}

\bibitem[{Gr{\"{a}}fener {et~al.}(2002)Gr{\"{a}}fener, Koesterke, \& Hamann}]{Grafener2002}
Gr{\"{a}}fener, G., Koesterke, L., \& Hamann, W.~R. 2002, Astronomy and Astrophysics, 387, 244, \dodoi{10.1051/0004-6361:20020269}

\bibitem[{{Grasha} {et~al.}(2021){Grasha}, {Roy}, {Sutherland}, \& {Kewley}}]{Grasha2021}
{Grasha}, K., {Roy}, A., {Sutherland}, R.~S., \& {Kewley}, L.~J. 2021, \apj, 908, 241, \dodoi{10.3847/1538-4357/abd6bf}

\bibitem[{{Groh} {et~al.}(2019){Groh}, {Ekstr{\"o}m}, {Georgy}, {Meynet}, {Choplin}, {Eggenberger}, {Hirschi}, {Maeder}, {Murphy}, {Boian}, \& {Farrell}}]{Groh2019}
{Groh}, J.~H., {Ekstr{\"o}m}, S., {Georgy}, C., {et~al.} 2019, \aap, 627, A24, \dodoi{10.1051/0004-6361/201833720}

\bibitem[{{Hainich} {et~al.}(2019){Hainich}, {Ramachandran}, {Shenar}, {Sander}, {Todt}, {Gruner}, {Oskinova}, \& {Hamann}}]{Hainich2019}
{Hainich}, R., {Ramachandran}, V., {Shenar}, T., {et~al.} 2019, \aap, 621, A85, \dodoi{10.1051/0004-6361/201833787}

\bibitem[{{Hawcroft} {et~al.}(2024{\natexlab{a}}){Hawcroft}, {Sana}, {Mahy}, {Sundqvist}, {de Koter}, {Crowther}, {Bestenlehner}, {Brands}, {David-Uraz}, {Decin}, {Erba}, {Garcia}, {Hamann}, {Herrero}, {Ignace}, {Kee}, {Kub{\'a}tov{\'a}}, {Lefever}, {Moffat}, {Najarro}, {Oskinova}, {Pauli}, {Prinja}, {Puls}, {Sander}, {Shenar}, {St-Louis}, {ud-Doula}, \& {Vink}}]{Hawcroft2023}
{Hawcroft}, C., {Sana}, H., {Mahy}, L., {et~al.} 2024{\natexlab{a}}, \aap, 688, A105, \dodoi{10.1051/0004-6361/202245588}

\bibitem[{{Hawcroft} {et~al.}(2024{\natexlab{b}}){Hawcroft}, {Mahy}, {Sana}, {Sundqvist}, {Abdul-Masih}, {Brands}, {Decin}, {de Koter}, \& {Puls}}]{Hawcroft2024}
{Hawcroft}, C., {Mahy}, L., {Sana}, H., {et~al.} 2024{\natexlab{b}}, \aap, 690, A126, \dodoi{10.1051/0004-6361/202348478}

\bibitem[{Hillier \& Miller(1998)}]{Hillier1998}
Hillier, D.~J., \& Miller, D.~L. 1998, The Astrophysical Journal, 496, 407, \dodoi{10.1086/305350}

\bibitem[{{Izotov} {et~al.}(2016){Izotov}, {Schaerer}, {Thuan}, {Worseck}, {Guseva}, {Orlitov{\'a}}, \& {Verhamme}}]{Izotov2016}
{Izotov}, Y.~I., {Schaerer}, D., {Thuan}, T.~X., {et~al.} 2016, \mnras, 461, 3683, \dodoi{10.1093/mnras/stw1205}

\bibitem[{{Kalari} {et~al.}(2022){Kalari}, {Horch}, {Salinas}, {Vink}, {Andersen}, {Bestenlehner}, \& {Rubio}}]{Kalari2022}
{Kalari}, V.~M., {Horch}, E.~P., {Salinas}, R., {et~al.} 2022, \apj, 935, 162, \dodoi{10.3847/1538-4357/ac8424}

\bibitem[{{Kimm} {et~al.}(2022){Kimm}, {Bieri}, {Geen}, {Rosdahl}, {Blaizot}, {Michel-Dansac}, \& {Garel}}]{Kimm2022}
{Kimm}, T., {Bieri}, R., {Geen}, S., {et~al.} 2022, \apjs, 259, 21, \dodoi{10.3847/1538-4365/ac426d}

\bibitem[{{K{\"o}hler} {et~al.}(2015){K{\"o}hler}, {Langer}, {de Koter}, {de Mink}, {Crowther}, {Evans}, {Gr{\"a}fener}, {Sana}, {Sanyal}, {Schneider}, \& {Vink}}]{Kohler2015}
{K{\"o}hler}, K., {Langer}, N., {de Koter}, A., {et~al.} 2015, \aap, 573, A71, \dodoi{10.1051/0004-6361/201424356}

\bibitem[{{Kotulla} {et~al.}(2009){Kotulla}, {Fritze}, {Weilbacher}, \& {Anders}}]{Kotulla2009}
{Kotulla}, R., {Fritze}, U., {Weilbacher}, P., \& {Anders}, P. 2009, \mnras, 396, 462, \dodoi{10.1111/j.1365-2966.2009.14717.x}

\bibitem[{{Kroupa}(2002)}]{Kroupa2002}
{Kroupa}, P. 2002, Science, 295, 82, \dodoi{10.1126/science.1067524}

\bibitem[{{Krti{\v{c}}ka} \& {Kub{\'a}t}(2017)}]{Krticka2017}
{Krti{\v{c}}ka}, J., \& {Kub{\'a}t}, J. 2017, \aap, 606, A31, \dodoi{10.1051/0004-6361/201730723}

\bibitem[{{Krumholz} {et~al.}(2015){Krumholz}, {Fumagalli}, {da Silva}, {Rendahl}, \& {Parra}}]{Krumholz2015}
{Krumholz}, M.~R., {Fumagalli}, M., {da Silva}, R.~L., {Rendahl}, T., \& {Parra}, J. 2015, \mnras, 452, 1447, \dodoi{10.1093/mnras/stv1374}

\bibitem[{{Kudritzki} \& {Puls}(2000)}]{Kudritzki2000}
{Kudritzki}, R.-P., \& {Puls}, J. 2000, \araa, 38, 613, \dodoi{10.1146/annurev.astro.38.1.613}

\bibitem[{{Kumari} {et~al.}(2024){Kumari}, {Smit}, {Witstok}, {Sirianni}, {Maiolino}, {Bunker}, {Bhatawdekar}, {Boyett}, {Cameron}, {Carniani}, {Charlot}, {Curti}, {Curtis-Lake}, {D'Eugenio}, {Eisenstein}, {Hainline}, {Ji}, {Jones}, {Robertson}, {Saxena}, {Scholtz}, {Simmonds}, {Williams}, \& {Willmer}}]{Kumari2024}
{Kumari}, N., {Smit}, R., {Witstok}, J., {et~al.} 2024, arXiv e-prints, arXiv:2406.11997, \dodoi{10.48550/arXiv.2406.11997}

\bibitem[{{Kummer} {et~al.}(2023){Kummer}, {Toonen}, \& {de Koter}}]{Kummer2023}
{Kummer}, F., {Toonen}, S., \& {de Koter}, A. 2023, \aap, 678, A60, \dodoi{10.1051/0004-6361/202347179}

\bibitem[{{Langer}(2012)}]{Langer2012}
{Langer}, N. 2012, \araa, 50, 107, \dodoi{10.1146/annurev-astro-081811-125534}

\bibitem[{{Le Borgne} {et~al.}(2004){Le Borgne}, {Rocca-Volmerange}, {Prugniel}, {Lan{\c{c}}on}, {Fioc}, \& {Soubiran}}]{LeBorgne2004}
{Le Borgne}, D., {Rocca-Volmerange}, B., {Prugniel}, P., {et~al.} 2004, \aap, 425, 881, \dodoi{10.1051/0004-6361:200400044}

\bibitem[{{Lecroq} {et~al.}(2024){Lecroq}, {Charlot}, {Bressan}, {Bruzual}, {Costa}, {Iorio}, {Spera}, {Mapelli}, {Chen}, {Chevallard}, \& {Dall'Amico}}]{Lecroq2024}
{Lecroq}, M., {Charlot}, S., {Bressan}, A., {et~al.} 2024, \mnras, 527, 9480, \dodoi{10.1093/mnras/stad3838}

\bibitem[{{Leitherer}(2020)}]{Leitherer2020}
{Leitherer}, C. 2020, Galaxies, 8, 13, \dodoi{10.3390/galaxies8010013}

\bibitem[{{Leitherer} {et~al.}(2014){Leitherer}, {Ekstr{\"o}m}, {Meynet}, {Schaerer}, {Agienko}, \& {Levesque}}]{Leitherer2014}
{Leitherer}, C., {Ekstr{\"o}m}, S., {Meynet}, G., {et~al.} 2014, \apjs, 212, 14, \dodoi{10.1088/0067-0049/212/1/14}

\bibitem[{{Leitherer} {et~al.}(2010){Leitherer}, {Ortiz Ot{\'a}lvaro}, {Bresolin}, {Kudritzki}, {Lo Faro}, {Pauldrach}, {Pettini}, \& {Rix}}]{Leitherer2010}
{Leitherer}, C., {Ortiz Ot{\'a}lvaro}, P.~A., {Bresolin}, F., {et~al.} 2010, \apjs, 189, 309, \dodoi{10.1088/0067-0049/189/2/309}

\bibitem[{{Leitherer} {et~al.}(1992){Leitherer}, {Robert}, \& {Drissen}}]{Leitherer1992}
{Leitherer}, C., {Robert}, C., \& {Drissen}, L. 1992, \apj, 401, 596, \dodoi{10.1086/172089}

\bibitem[{{Leitherer} {et~al.}(1999){Leitherer}, {Schaerer}, {Goldader}, {Delgado}, {Robert}, {Kune}, {de Mello}, {Devost}, \& {Heckman}}]{Leitherer1999}
{Leitherer}, C., {Schaerer}, D., {Goldader}, J.~D., {et~al.} 1999, \apjs, 123, 3, \dodoi{10.1086/313233}

\bibitem[{{Lejeune} {et~al.}(1997){Lejeune}, {Cuisinier}, \& {Buser}}]{lejeune1997}
{Lejeune}, T., {Cuisinier}, F., \& {Buser}, R. 1997, \aaps, 125, 229, \dodoi{10.1051/aas:1997373}

\bibitem[{{Levesque} {et~al.}(2012){Levesque}, {Leitherer}, {Ekstrom}, {Meynet}, \& {Schaerer}}]{Levesque2012}
{Levesque}, E.~M., {Leitherer}, C., {Ekstrom}, S., {Meynet}, G., \& {Schaerer}, D. 2012, \apj, 751, 67, \dodoi{10.1088/0004-637X/751/1/67}

\bibitem[{{Llerena} {et~al.}(2024){Llerena}, {Pentericci}, {Napolitano}, {Mascia}, {Amor{\'\i}n}, {Calabr{\`o}}, {Castellano}, {Cleri}, {Giavalisco}, {Grogin}, {Hathi}, {Hirschmann}, {Koekemoer}, {Nanayakkara}, {Pacucci}, {Shen}, {Wilkins}, {Yoon}, {Yung}, {Bhatawdekar}, {Lucas}, {Wang}, {Arrabal Haro}, {Bagley}, {Finkelstein}, {Kartaltepe}, {Merlin}, {Papovich}, \& {Pirzkal}}]{LLerena2024}
{Llerena}, M., {Pentericci}, L., {Napolitano}, L., {et~al.} 2024, arXiv e-prints, arXiv:2412.01358, \dodoi{10.48550/arXiv.2412.01358}

\bibitem[{{Lohr} {et~al.}(2018){Lohr}, {Clark}, {Najarro}, {Patrick}, {Crowther}, \& {Evans}}]{Lohr2018}
{Lohr}, M.~E., {Clark}, J.~S., {Najarro}, F., {et~al.} 2018, \aap, 617, A66, \dodoi{10.1051/0004-6361/201832670}

\bibitem[{{Maeder} \& {Meynet}(1994)}]{Maeder1994}
{Maeder}, A., \& {Meynet}, G. 1994, \aap, 287, 803

\bibitem[{{Maeder} \& {Meynet}(2000)}]{Maeder2000}
---. 2000, \aap, 361, 159, \dodoi{10.48550/arXiv.astro-ph/0006405}

\bibitem[{{Marchant} \& {Bodensteiner}(2024)}]{Marchant2024}
{Marchant}, P., \& {Bodensteiner}, J. 2024, \araa, 62, 21, \dodoi{10.1146/annurev-astro-052722-105936}

\bibitem[{{Marchant} {et~al.}(2017){Marchant}, {Langer}, {Podsiadlowski}, {Tauris}, {de Mink}, {Mandel}, \& {Moriya}}]{Marchant2017}
{Marchant}, P., {Langer}, N., {Podsiadlowski}, P., {et~al.} 2017, \aap, 604, A55, \dodoi{10.1051/0004-6361/201630188}

\bibitem[{{Marcolino} {et~al.}(2022){Marcolino}, {Bouret}, {Rocha-Pinto}, {Bernini-Peron}, \& {Vink}}]{Marcolino2022}
{Marcolino}, W.~L.~F., {Bouret}, J.~C., {Rocha-Pinto}, H.~J., {Bernini-Peron}, M., \& {Vink}, J.~S. 2022, \mnras, 511, 5104, \dodoi{10.1093/mnras/stac452}

\bibitem[{{Marques-Chaves} {et~al.}(2022){Marques-Chaves}, {Schaerer}, {Amor{\'\i}n}, {Atek}, {Borthakur}, {Chisholm}, {Fern{\'a}ndez}, {Flury}, {Giavalisco}, {Grazian}, {Hayes}, {Heckman}, {Henry}, {Izotov}, {Jaskot}, {Ji}, {McCandliss}, {Oey}, {{\"O}stlin}, {Ravindranath}, {Rutkowski}, {Saldana-Lopez}, {Teplitz}, {Thuan}, {Verhamme}, {Wang}, {Worseck}, \& {Xu}}]{Marques-Chaves2022}
{Marques-Chaves}, R., {Schaerer}, D., {Amor{\'\i}n}, R.~O., {et~al.} 2022, \aap, 663, L1, \dodoi{10.1051/0004-6361/202243598}

\bibitem[{{Martinet} {et~al.}(2023){Martinet}, {Meynet}, {Ekstr{\"o}m}, {Georgy}, \& {Hirschi}}]{Martinet2023}
{Martinet}, S., {Meynet}, G., {Ekstr{\"o}m}, S., {Georgy}, C., \& {Hirschi}, R. 2023, \aap, 679, A137, \dodoi{10.1051/0004-6361/202347514}

\bibitem[{{Martins} {et~al.}(2008){Martins}, {Hillier}, {Paumard}, {Eisenhauer}, {Ott}, \& {Genzel}}]{Martins2008}
{Martins}, F., {Hillier}, D.~J., {Paumard}, T., {et~al.} 2008, \aap, 478, 219, \dodoi{10.1051/0004-6361:20078469}

\bibitem[{{Martins} \& {Palacios}(2022)}]{Martins2022}
{Martins}, F., \& {Palacios}, A. 2022, arXiv e-prints, arXiv:2202.13703.
\newblock \doarXiv{2202.13703}

\bibitem[{{Martins} {et~al.}(2025){Martins}, {Palacios}, {Schaerer}, \& {Marques-Chaves}}]{Martins2025}
{Martins}, F., {Palacios}, A., {Schaerer}, D., \& {Marques-Chaves}, R. 2025, arXiv e-prints, arXiv:2505.02993, \dodoi{10.48550/arXiv.2505.02993}

\bibitem[{{Martins} {et~al.}(2023){Martins}, {Schaerer}, {Marques-Chaves}, \& {Upadhyaya}}]{Martins2023}
{Martins}, F., {Schaerer}, D., {Marques-Chaves}, R., \& {Upadhyaya}, A. 2023, \aap, 678, A159, \dodoi{10.1051/0004-6361/202346732}

\bibitem[{{Me{\v{s}}tri{\'c}} {et~al.}(2023){Me{\v{s}}tri{\'c}}, {Vanzella}, {Upadhyaya}, {Martins}, {Marques-Chaves}, {Schaerer}, {Guibert}, {Zanella}, {Grillo}, {Rosati}, {Calura}, {Caminha}, {Bolamperti}, {Meneghetti}, {Bergamini}, {Mercurio}, {Nonino}, \& {Pascale}}]{Mestric2023}
{Me{\v{s}}tri{\'c}}, U., {Vanzella}, E., {Upadhyaya}, A., {et~al.} 2023, \aap, 673, A50, \dodoi{10.1051/0004-6361/202345895}

\bibitem[{{Mill{\'a}n-Irigoyen} {et~al.}(2021){Mill{\'a}n-Irigoyen}, {Moll{\'a}}, {Cervi{\~n}o}, {Ascasibar}, {Garc{\'\i}a-Vargas}, \& {Coelho}}]{Millan-Irigoyen2021}
{Mill{\'a}n-Irigoyen}, I., {Moll{\'a}}, M., {Cervi{\~n}o}, M., {et~al.} 2021, \mnras, 506, 4781, \dodoi{10.1093/mnras/stab1969}

\bibitem[{{Mokiem} {et~al.}(2004){Mokiem}, {Mart{\'\i}n-Hern{\'a}ndez}, {Lenorzer}, {de Koter}, \& {Tielens}}]{Mokiem2004}
{Mokiem}, M.~R., {Mart{\'\i}n-Hern{\'a}ndez}, N.~L., {Lenorzer}, A., {de Koter}, A., \& {Tielens}, A.~G.~G.~M. 2004, \aap, 419, 319, \dodoi{10.1051/0004-6361:20040074}

\bibitem[{{Mokiem} {et~al.}(2007){Mokiem}, {de Koter}, {Vink}, {Puls}, {Evans}, {Smartt}, {Crowther}, {Herrero}, {Langer}, {Lennon}, {Najarro}, \& {Villamariz}}]{Mokiem2007}
{Mokiem}, M.~R., {de Koter}, A., {Vink}, J.~S., {et~al.} 2007, \aap, 473, 603, \dodoi{10.1051/0004-6361:20077545}

\bibitem[{{Mowlavi} {et~al.}(1998){Mowlavi}, {Schaerer}, {Meynet}, {Bernasconi}, {Charbonnel}, \& {Maeder}}]{Mowlavi1998}
{Mowlavi}, N., {Schaerer}, D., {Meynet}, G., {et~al.} 1998, \aaps, 128, 471, \dodoi{10.1051/aas:1998388}

\bibitem[{{Mu{\~n}oz} {et~al.}(2024){Mu{\~n}oz}, {Mirocha}, {Chisholm}, {Furlanetto}, \& {Mason}}]{Munoz2024}
{Mu{\~n}oz}, J.~B., {Mirocha}, J., {Chisholm}, J., {Furlanetto}, S.~R., \& {Mason}, C. 2024, \mnras, 535, L37, \dodoi{10.1093/mnrasl/slae086}

\bibitem[{{Murphy} {et~al.}(2021){Murphy}, {Groh}, {Ekstr{\"o}m}, {Meynet}, {Pezzotti}, {Georgy}, {Choplin}, {Eggenberger}, {Farrell}, {Haemmerl{\'e}}, {Hirschi}, {Maeder}, \& {Martinet}}]{Murphy2021}
{Murphy}, L.~J., {Groh}, J.~H., {Ekstr{\"o}m}, S., {et~al.} 2021, \mnras, 501, 2745, \dodoi{10.1093/mnras/staa3803}

\bibitem[{{Najarro} {et~al.}(2004){Najarro}, {Figer}, {Hillier}, \& {Kudritzki}}]{Najarro2004}
{Najarro}, F., {Figer}, D.~F., {Hillier}, D.~J., \& {Kudritzki}, R.~P. 2004, \apjl, 611, L105, \dodoi{10.1086/423955}

\bibitem[{{Nandal} {et~al.}(2024){Nandal}, {Meynet}, {Ekstr{\"o}m}, {Moyano}, {Eggenberger}, {Choplin}, {Georgy}, {Farrell}, \& {Maeder}}]{Nandal2024}
{Nandal}, D., {Meynet}, G., {Ekstr{\"o}m}, S., {et~al.} 2024, \aap, 684, A169, \dodoi{10.1051/0004-6361/202346979}

\bibitem[{{Nugis} \& {Lamers}(2000)}]{NugisLamers2000}
{Nugis}, T., \& {Lamers}, H.~J.~G.~L.~M. 2000, \aap, 360, 227

\bibitem[{{Pahl} {et~al.}(2020){Pahl}, {Shapley}, {Faisst}, {Capak}, {Du}, {Reddy}, {Laursen}, \& {Topping}}]{Pahl2020}
{Pahl}, A.~J., {Shapley}, A., {Faisst}, A.~L., {et~al.} 2020, \mnras, 493, 3194, \dodoi{10.1093/mnras/staa355}

\bibitem[{Pauldrach {et~al.}(2001)Pauldrach, Hoffmann, \& Lennon}]{Pauldrach2001}
Pauldrach, A.~W., Hoffmann, T.~L., \& Lennon, M. 2001, Astronomy and Astrophysics, 375, 161, \dodoi{10.1051/0004-6361:20010805}

\bibitem[{{Pauli} {et~al.}(2022){Pauli}, {Langer}, {Aguilera-Dena}, {Wang}, \& {Marchant}}]{Pauli2022b}
{Pauli}, D., {Langer}, N., {Aguilera-Dena}, D.~R., {Wang}, C., \& {Marchant}, P. 2022, \aap, 667, A58, \dodoi{10.1051/0004-6361/202243965}

\bibitem[{{Paxton} {et~al.}(2013){Paxton}, {Cantiello}, {Arras}, {Bildsten}, {Brown}, {Dotter}, {Mankovich}, {Montgomery}, {Stello}, {Timmes}, \& {Townsend}}]{Paxton2013}
{Paxton}, B., {Cantiello}, M., {Arras}, P., {et~al.} 2013, \apjs, 208, 4, \dodoi{10.1088/0067-0049/208/1/4}

\bibitem[{{Pietrinferni} {et~al.}(2021){Pietrinferni}, {Hidalgo}, {Cassisi}, {Salaris}, {Savino}, {Mucciarelli}, {Verma}, {Silva Aguirre}, {Aparicio}, \& {Ferguson}}]{Pietrinferni2021}
{Pietrinferni}, A., {Hidalgo}, S., {Cassisi}, S., {et~al.} 2021, \apj, 908, 102, \dodoi{10.3847/1538-4357/abd4d5}

\bibitem[{{Plat} {et~al.}(2019){Plat}, {Charlot}, {Bruzual}, {Feltre}, {Vidal-Garc{\'\i}a}, {Morisset}, {Chevallard}, \& {Todt}}]{Plat2019}
{Plat}, A., {Charlot}, S., {Bruzual}, G., {et~al.} 2019, \mnras, 490, 978, \dodoi{10.1093/mnras/stz2616}

\bibitem[{{Puls} {et~al.}(2020){Puls}, {Najarro}, {Sundqvist}, \& {Sen}}]{Puls2020}
{Puls}, J., {Najarro}, F., {Sundqvist}, J.~O., \& {Sen}, K. 2020, \aap, 642, A172, \dodoi{10.1051/0004-6361/202038464}

\bibitem[{Puls {et~al.}(2005)Puls, Urbaneja, Venero, Repolust, Springmann, Jokuthy, \& Mokiem}]{Puls2005}
Puls, J., Urbaneja, M.~A., Venero, R., {et~al.} 2005, Astronomy and Astrophysics, 435, 669, \dodoi{10.1051/0004-6361:20042365}

\bibitem[{{Rivera-Thorsen} {et~al.}(2024){Rivera-Thorsen}, {Chisholm}, {Welch}, {Rigby}, {Hutchison}, {Florian}, {Sharon}, {Choe}, {Dahle}, {Bayliss}, {Khullar}, {Gladders}, {Hayes}, {Adamo}, {Owens}, \& {Kim}}]{Rivera-Thorsen2024}
{Rivera-Thorsen}, T.~E., {Chisholm}, J., {Welch}, B., {et~al.} 2024, arXiv e-prints, arXiv:2404.08884, \dodoi{10.48550/arXiv.2404.08884}

\bibitem[{{Roman-Duval} {et~al.}(2020){Roman-Duval}, {Proffitt}, {Taylor}, {Monroe}, {Fischer}, {Fischer}, {Fullerton}, {Aloisi}, {Britt}, {Busko}, {Carlberg}, {De Rosa}, {Jedrzejewski}, {Lockwood}, {Frazer}, {Hernandez}, {James}, {Oliveira}, {Plesha}, {Riedel}, {Riley}, {Sahnow}, {Sankrit}, {Shaw}, {Smith}, {Sohn}, {Som}, {Ubeda}, \& {Welty}}]{RomanDuval2020}
{Roman-Duval}, J., {Proffitt}, C.~R., {Taylor}, J.~M., {et~al.} 2020, Research Notes of the American Astronomical Society, 4, 205, \dodoi{10.3847/2515-5172/abca2f}

\bibitem[{{Roy} {et~al.}(2024){Roy}, {Heckman}, {Henry}, {Chisholm}, {Flury}, {Leitherer}, {Hayes}, {Jaskot}, {Ji}, {Schaerer}, {Wang}, {Borthakur}, {Xu}, \& {{\"O}stlin}}]{Roy2024}
{Roy}, N., {Heckman}, T., {Henry}, A., {et~al.} 2024, arXiv e-prints, arXiv:2410.13254, \dodoi{10.48550/arXiv.2410.13254}

\bibitem[{{Sabhahit} {et~al.}(2022){Sabhahit}, {Vink}, {Higgins}, \& {Sander}}]{Sabhahit2022}
{Sabhahit}, G.~N., {Vink}, J.~S., {Higgins}, E.~R., \& {Sander}, A. A.~C. 2022, \mnras, 514, 3736, \dodoi{10.1093/mnras/stac1410}

\bibitem[{Sana {et~al.}(2012)Sana, de~Mink, de~Koter, Langer, Evans, Gieles, Gosset, Izzard, Bouquin, \& Schneider}]{Sana2012}
Sana, H., de~Mink, S.~E., de~Koter, A., {et~al.} 2012, Science, 337, 444, \dodoi{10.1126/science.1223344}

\bibitem[{{Sana} {et~al.}(2013){Sana}, {de Koter}, {de Mink}, {Dunstall}, {Evans}, {H{\'e}nault-Brunet}, {Ma{\'\i}z Apell{\'a}niz}, {Ram{\'\i}rez-Agudelo}, {Taylor}, {Walborn}, {Clark}, {Crowther}, {Herrero}, {Gieles}, {Langer}, {Lennon}, \& {Vink}}]{Sana2013}
{Sana}, H., {de Koter}, A., {de Mink}, S.~E., {et~al.} 2013, \aap, 550, A107, \dodoi{10.1051/0004-6361/201219621}

\bibitem[{{S{\'a}nchez} {et~al.}(2022){S{\'a}nchez}, {Barrera-Ballesteros}, {Lacerda}, {Mej{\'\i}a-Narvaez}, {Camps-Fari{\~n}a}, {Bruzual}, {Espinosa-Ponce}, {Rodr{\'\i}guez-Puebla}, {Calette}, {Ibarra-Medel}, {Avila-Reese}, {Hernandez-Toledo}, {Bershady}, {Cano-Diaz}, \& {Munguia-Cordova}}]{Sanchez2022}
{S{\'a}nchez}, S.~F., {Barrera-Ballesteros}, J.~K., {Lacerda}, E., {et~al.} 2022, \apjs, 262, 36, \dodoi{10.3847/1538-4365/ac7b8f}

\bibitem[{{Sander} {et~al.}(2015){Sander}, {Shenar}, {Hainich}, {G{\'\i}menez-Garc{\'\i}a}, {Todt}, \& {Hamann}}]{Sander2015}
{Sander}, A., {Shenar}, T., {Hainich}, R., {et~al.} 2015, \aap, 577, A13, \dodoi{10.1051/0004-6361/201425356}

\bibitem[{{Sander} \& {Vink}(2020)}]{SanderVink2020}
{Sander}, A. A.~C., \& {Vink}, J.~S. 2020, \mnras, 499, 873, \dodoi{10.1093/mnras/staa2712}

\bibitem[{{Sander} {et~al.}(2024){Sander}, {Bouret}, {Bernini-Peron}, {Puls}, {Backs}, {Berlanas}, {Bestenlehner}, {Brands}, {Herrero}, {Martins}, {Maryeva}, {Pauli}, {Ramachandran}, {Crowther}, {G{\'o}mez-Gonz{\'a}lez}, {Gormaz-Matamala}, {Hamann}, {Hillier}, {Kuiper}, {Larkin}, {Lefever}, {Mehner}, {Najarro}, {Oskinova}, {Sch{\"o}sser}, {Shenar}, {Todt}, {ud-Doula}, \& {Vink}}]{Sander2024}
{Sander}, A.~A.~C., {Bouret}, J.~C., {Bernini-Peron}, M., {et~al.} 2024, \aap, 689, A30, \dodoi{10.1051/0004-6361/202449829}

\bibitem[{Santolaya-Rey {et~al.}(1997)Santolaya-Rey, Puls, \& Herrero}]{Santolaya-Rey1997a}
Santolaya-Rey, A., Puls, J., \& Herrero, A. 1997, $\backslash$Aap, 323, 488, \dodoi{10.1051/0004-6361/201832993}

\bibitem[{{Schaerer} {et~al.}(2024){Schaerer}, {Guibert}, {Marques-Chaves}, \& {Martins}}]{Schaerer2024}
{Schaerer}, D., {Guibert}, J., {Marques-Chaves}, R., \& {Martins}, F. 2024, arXiv e-prints, arXiv:2407.12122, \dodoi{10.48550/arXiv.2407.12122}

\bibitem[{{Schnurr} {et~al.}(2008){Schnurr}, {Casoli}, {Chen{\'e}}, {Moffat}, \& {St-Louis}}]{Schnurr2008}
{Schnurr}, O., {Casoli}, J., {Chen{\'e}}, A.~N., {Moffat}, A.~F.~J., \& {St-Louis}, N. 2008, \mnras, 389, L38, \dodoi{10.1111/j.1745-3933.2008.00517.x}

\bibitem[{{Sch{\"o}sser} {et~al.}(2025){Sch{\"o}sser}, {Ramachandran}, {Sander}, {Gallagher}, {Bernini-Peron}, {Gonz{\'a}lez-Tor{\`a}}, {Josiek}, {Lefever}, {Hamann}, \& {Oskinova}}]{Schosser2025}
{Sch{\"o}sser}, E.~C., {Ramachandran}, V., {Sander}, A.~A.~C., {et~al.} 2025, \aap, 696, L3, \dodoi{10.1051/0004-6361/202554027}

\bibitem[{{Senchyna} {et~al.}(2021){Senchyna}, {Stark}, {Charlot}, {Chevallard}, {Bruzual}, \& {Vidal-Garc{\'\i}a}}]{Senchyna2021}
{Senchyna}, P., {Stark}, D.~P., {Charlot}, S., {et~al.} 2021, \mnras, 503, 6112, \dodoi{10.1093/mnras/stab884}

\bibitem[{{Smith} {et~al.}(2016){Smith}, {Crowther}, {Calzetti}, \& {Sidoli}}]{Smith2016}
{Smith}, L.~J., {Crowther}, P.~A., {Calzetti}, D., \& {Sidoli}, F. 2016, \apj, 823, 38, \dodoi{10.3847/0004-637X/823/1/38}

\bibitem[{{Smith} {et~al.}(2023){Smith}, {Oey}, {Hernandez}, {Ryon}, {Leitherer}, {Charlot}, {Bruzual}, {Calzetti}, {Chu}, {Hayes}, {James}, {Jaskot}, \& {{\"O}stlin}}]{Smith2023}
{Smith}, L.~J., {Oey}, M.~S., {Hernandez}, S., {et~al.} 2023, \apj, 958, 194, \dodoi{10.3847/1538-4357/ad00b4}

\bibitem[{{Steidel} {et~al.}(2016){Steidel}, {Strom}, {Pettini}, {Rudie}, {Reddy}, \& {Trainor}}]{Steidel2016}
{Steidel}, C.~C., {Strom}, A.~L., {Pettini}, M., {et~al.} 2016, \apj, 826, 159, \dodoi{10.3847/0004-637X/826/2/159}

\bibitem[{{Stevenson} {et~al.}(2017){Stevenson}, {Vigna-G{\'o}mez}, {Mandel}, {Barrett}, {Neijssel}, {Perkins}, \& {de Mink}}]{Stevenson2017}
{Stevenson}, S., {Vigna-G{\'o}mez}, A., {Mandel}, I., {et~al.} 2017, Nature Communications, 8, 14906, \dodoi{10.1038/ncomms14906}

\bibitem[{{Strom} {et~al.}(2022){Strom}, {Rudie}, {Steidel}, \& {Trainor}}]{Strom2022}
{Strom}, A.~L., {Rudie}, G.~C., {Steidel}, C.~C., \& {Trainor}, R.~F. 2022, \apj, 925, 116, \dodoi{10.3847/1538-4357/ac38a3}

\bibitem[{{Sundqvist} \& {Puls}(2018)}]{Sundqvist2018}
{Sundqvist}, J.~O., \& {Puls}, J. 2018, \aap, 619, A59, \dodoi{10.1051/0004-6361/201832993}

\bibitem[{{Sylvester} {et~al.}(1998){Sylvester}, {Skinner}, \& {Barlow}}]{Sylvester1998}
{Sylvester}, R.~J., {Skinner}, C.~J., \& {Barlow}, M.~J. 1998, \mnras, 301, 1083, \dodoi{10.1046/j.1365-8711.1998.02078.x}

\bibitem[{{Sz{\'e}csi} {et~al.}(2022){Sz{\'e}csi}, {Agrawal}, {W{\"u}nsch}, \& {Langer}}]{Szecsi2022}
{Sz{\'e}csi}, D., {Agrawal}, P., {W{\"u}nsch}, R., \& {Langer}, N. 2022, \aap, 658, A125, \dodoi{10.1051/0004-6361/202141536}

\bibitem[{{Telford} {et~al.}(2024){Telford}, {Chisholm}, {Sander}, {Ramachandran}, {McQuinn}, \& {Berg}}]{Telford2024}
{Telford}, O.~G., {Chisholm}, J., {Sander}, A. A.~C., {et~al.} 2024, \apj, 974, 85, \dodoi{10.3847/1538-4357/ad697e}

\bibitem[{{Upadhyaya} {et~al.}(2024){Upadhyaya}, {Marques-Chaves}, {Schaerer}, {Martins}, {P{\'e}rez-Fournon}, {Palacios}, \& {Stanway}}]{Upadhyaya2024}
{Upadhyaya}, A., {Marques-Chaves}, R., {Schaerer}, D., {et~al.} 2024, \aap, 686, A185, \dodoi{10.1051/0004-6361/202449184}

\bibitem[{{van Loon} {et~al.}(1999){van Loon}, {Groenewegen}, {de Koter}, {Trams}, {Waters}, {Zijlstra}, {Whitelock}, \& {Loup}}]{vanLoon1999}
{van Loon}, J.~T., {Groenewegen}, M.~A.~T., {de Koter}, A., {et~al.} 1999, \aap, 351, 559, \dodoi{10.48550/arXiv.astro-ph/9909416}

\bibitem[{{Vazdekis} {et~al.}(2016){Vazdekis}, {Koleva}, {Ricciardelli}, {R{\"o}ck}, \& {Falc{\'o}n-Barroso}}]{Vazdekis2016}
{Vazdekis}, A., {Koleva}, M., {Ricciardelli}, E., {R{\"o}ck}, B., \& {Falc{\'o}n-Barroso}, J. 2016, \mnras, 463, 3409, \dodoi{10.1093/mnras/stw2231}

\bibitem[{{Vink} {et~al.}(2001){Vink}, {de Koter}, \& {Lamers}}]{Vink2001}
{Vink}, J.~S., {de Koter}, A., \& {Lamers}, H.~J.~G.~L.~M. 2001, \aap, 369, 574, \dodoi{10.1051/0004-6361:20010127}

\bibitem[{{Vink} {et~al.}(2011){Vink}, {Muijres}, {Anthonisse}, {de Koter}, {Gr{\"a}fener}, \& {Langer}}]{Vink2011}
{Vink}, J.~S., {Muijres}, L.~E., {Anthonisse}, B., {et~al.} 2011, \aap, 531, A132, \dodoi{10.1051/0004-6361/201116614}

\bibitem[{{Vink} \& {Sander}(2021)}]{VinkSander2021}
{Vink}, J.~S., \& {Sander}, A. A.~C. 2021, \mnras, 504, 2051, \dodoi{10.1093/mnras/stab902}

\bibitem[{{Vink} {et~al.}(2023){Vink}, {Mehner}, {Crowther}, {Fullerton}, {Garcia}, {Martins}, {Morrell}, {Oskinova}, {St-Louis}, {ud-Doula}, {Sander}, {Sana}, {Bouret}, {Kub{\'a}tov{\'a}}, {Marchant}, {Martins}, {Wofford}, {van Loon}, {Grace Telford}, {G{\"o}tberg}, {Bowman}, {Erba}, {Kalari}, {Abdul-Masih}, {Alkousa}, {Backs}, {Barbosa}, {Berlanas}, {Bernini-Peron}, {Bestenlehner}, {Blomme}, {Bodensteiner}, {Brands}, {Evans}, {David-Uraz}, {Driessen}, {Dsilva}, {Geen}, {G{\'o}mez-Gonz{\'a}lez}, {Grassitelli}, {Hamann}, {Hawcroft}, {Herrero}, {Higgins}, {John Hillier}, {Ignace}, {Istrate}, {Kaper}, {Kee}, {Kehrig}, {Keszthelyi}, {Klencki}, {de Koter}, {Kuiper}, {Laplace}, {Larkin}, {Lefever}, {Leitherer}, {Lennon}, {Mahy}, {Ma{\'\i}z Apell{\'a}niz}, {Maravelias}, {Marcolino}, {McLeod}, {de Mink}, {Najarro}, {Oey}, {Parsons}, {Pauli}, {Pedersen}, {Prinja}, {Ramachandran}, {Ram{\'\i}rez-Tannus}, {Sabhahit}, {Schootemeijer}, {Reyero Serantes}, {Shenar}, {Stringfellow}, {Sudnik}, {Tramper}, \& {Wang}}]{Vink2023}
{Vink}, J.~S., {Mehner}, A., {Crowther}, P.~A., {et~al.} 2023, \aap, 675, A154, \dodoi{10.1051/0004-6361/202245650}

\bibitem[{{Welch} {et~al.}(2025){Welch}, {Rivera-Thorsen}, {Rigby}, {Hutchison}, {Olivier}, {Berg}, {Sharon}, {Dahle}, {Owens}, {Bayliss}, {Khullar}, {Chisholm}, {Hayes}, \& {Kim}}]{Welch2025}
{Welch}, B., {Rivera-Thorsen}, T.~E., {Rigby}, J.~R., {et~al.} 2025, \apj, 980, 33, \dodoi{10.3847/1538-4357/ada76c}

\bibitem[{{Wofford} {et~al.}(2014){Wofford}, {Leitherer}, {Chandar}, \& {Bouret}}]{Wofford2014}
{Wofford}, A., {Leitherer}, C., {Chandar}, R., \& {Bouret}, J.-C. 2014, \apj, 781, 122, \dodoi{10.1088/0004-637X/781/2/122}

\bibitem[{{Wofford} {et~al.}(2023){Wofford}, {Sixtos}, {Charlot}, {Bruzual}, {Cullen}, {Stanton}, {Hern{\'a}ndez}, {Smith}, \& {Hayes}}]{Wofford2023}
{Wofford}, A., {Sixtos}, A., {Charlot}, S., {et~al.} 2023, \mnras, 523, 3949, \dodoi{10.1093/mnras/stad1622}

\bibitem[{{Xiao} {et~al.}(2018){Xiao}, {Stanway}, \& {Eldridge}}]{Xiao2018}
{Xiao}, L., {Stanway}, E.~R., \& {Eldridge}, J.~J. 2018, \mnras, 477, 904, \dodoi{10.1093/mnras/sty646}

\bibitem[{{Yusof} {et~al.}(2022){Yusof}, {Hirschi}, {Eggenberger}, {Ekstr{\"o}m}, {Georgy}, {Sibony}, {Crowther}, {Meynet}, {Kassim}, {Harun}, {Maeder}, {Groh}, {Farrell}, \& {Murphy}}]{Yusof2022}
{Yusof}, N., {Hirschi}, R., {Eggenberger}, P., {et~al.} 2022, \mnras, 511, 2814, \dodoi{10.1093/mnras/stac230}

\end{thebibliography}
\bibliographystyle{aasjournal}

\begin{appendix}

There are a number of additional \starburst\ modules which have not been presented in this work, some of which have already been implemented in the \nstarburst\ code and can reproduce \starburst\ outputs but further updates are planned for these modules, while others have not yet been implemented. The high resolution UV spectral synthesis, spectral type, supernova rate and colour predictions are all in the former state. The optical spectral synthesis, empirical spectral synthesis, line equivalent width and chemical yield routines are in the latter case. In these appendices we present an example prediction of colours using \nstarburst. 
The UV spectral synthesis will be presented in an upcoming work with the incorporation of a new grid of high resolution theoretical spectra. The spectral type module is available in the \nstarburst\ code but will be updated with new spectral type calibrations. The implementation of the remaining modules is still in progress but regular updates will be provided on the \nstarburst\ webpage.



\section{Example values}

Here we present tables of example output values for the predictions shown in Figs \ref{fig: ionflux_metallicity} - \ref{fig: uv_slopes}.

\begin{deluxetable}{cccccccccccccccccccc}
\tabletypesize{\scriptsize}
\label{tbl:HIionflux}
\tablecolumns{7}
\tablecaption{\ion{H}{1} ionising fluxes in units of photons s$^{-1}$.}
\tablehead{
\colhead{t} & \colhead{GalC} & \colhead{MW} & \colhead{LMC} & \colhead{SMC} & \colhead{IZw18} & \colhead{Z0} & \colhead{GalC} & \colhead{MW} & \colhead{LMC} & \colhead{SMC} & \colhead{IZw18} & \colhead{Z0} & \colhead{GalC} & \colhead{MW} & \colhead{LMC} & \colhead{Z0} & \colhead{GalC} & \colhead{MW} & \colhead{LMC}\\
\colhead{[Myr]} & \colhead{v00} & \colhead{v00} & \colhead{v00} & \colhead{v00} & \colhead{v00} & \colhead{v00} & \colhead{v40} & \colhead{v40} & \colhead{v40} & \colhead{v40} & \colhead{v40} & \colhead{v40} & \colhead{M300} & \colhead{M300} & \colhead{M300} & \colhead{M300} & \colhead{M300v40} & \colhead{M300v40} & \colhead{M300v40}
}
\startdata
1.0	&	52.68	&	52.71	&	52.74	&	52.77	&	52.78	&	52.82	&	52.69	&	52.70	&	52.73	&	52.75	&	52.76	&	52.80	&	52.96	&	52.99	&	53.04	&	53.08	&	52.97	&	53.02	&	53.05	\\
1.6	&	52.67	&	52.69	&	52.76	&	52.80	&	52.82	&	52.85	&	52.71	&	52.73	&	52.76	&	52.79	&	52.80	&	52.84	&	52.90	&	52.99	&	53.02	&	53.10	&	52.97	&	53.02	&	53.07	\\	
2.5	&	52.43	&	52.52	&	52.67	&	52.80	&	52.87	&	52.91	&	52.73	&	52.77	&	52.84	&	52.85	&	52.84	&	52.90	&	52.63	&	52.74	&	52.88	&	53.17	&	52.76	&	52.90	&	53.10	\\	
4.0	&	52.05	&	52.15	&	52.30	&	52.36	&	52.46	&	52.72	&	52.56	&	52.69	&	52.82	&	52.63	&	52.58	&	52.78	&	52.03	&	52.14	&	52.32	&	52.73	&	52.52	&	52.67	&	52.81	\\	
6.3	&	50.97	&	51.07	&	51.37	&	51.64	&	51.84	&	52.38	&	51.94	&	52.01	&	51.89	&	51.94	&	52.05	&	52.47	&	50.95	&	51.06	&	51.45	&	52.37	&	51.88	&	52.00	&	52.05	\\	
10.0	&	49.81	&	49.98	&	50.26	&	50.55	&	50.90	&	52.11	&	50.90	&	50.66	&	50.67	&	51.00	&	51.53	&	52.23	&	49.79	&	49.96	&	50.24	&	52.08	&	50.89	&	50.64	&	50.74	\\	
15.8	&	48.86	&	49.06	&	49.29	&	49.53	&	49.80	&	51.70	&	49.39	&	49.41	&	49.65	&	49.97	&	50.23	&	51.81	&	48.85	&	49.04	&	49.27	&	51.69	&	49.30	&	49.40	&	49.66	\\	
25.1	&	48.07	&	48.25	&	48.57	&	48.80	&	49.05	&	50.92	&	48.49	&	48.59	&	48.84	&	49.14	&	49.38	&	51.11	&	48.05	&	48.23	&	48.55	&	50.90	&	48.43	&	48.58	&	48.84	\\	
\enddata
\end{deluxetable}

\begin{deluxetable}{cccccccccccccccccccc}
\tabletypesize{\scriptsize}
\label{tbl:HeIionflux}
\tablecolumns{7}
\tablecaption{\ion{He}{1} ionising fluxes in units of photons s$^{-1}$.}
\tablehead{
\colhead{t} & \colhead{GalC} & \colhead{MW} & \colhead{LMC} & \colhead{SMC} & \colhead{IZw18} & \colhead{Z0} & \colhead{GalC} & \colhead{MW} & \colhead{LMC} & \colhead{SMC} & \colhead{IZw18} & \colhead{Z0} & \colhead{GalC} & \colhead{MW} & \colhead{LMC} & \colhead{Z0} & \colhead{GalC} & \colhead{MW} & \colhead{LMC}\\
\colhead{[Myr]} & \colhead{v00} & \colhead{v00} & \colhead{v00} & \colhead{v00} & \colhead{v00} & \colhead{v00} & \colhead{v40} & \colhead{v40} & \colhead{v40} & \colhead{v40} & \colhead{v40} & \colhead{v40} & \colhead{M300} & \colhead{M300} & \colhead{M300} & \colhead{M300} & \colhead{M300v40} & \colhead{M300v40} & \colhead{M300v40}
}
\startdata
1.0	&	51.91	&	52.01	&	52.11	&	52.18	&	52.25	&	52.35	&	51.95	&	52.00	&	52.08	&	52.15	&	52.23	&	52.34	&	52.22	&	52.29	&	52.46	&	52.84	&	52.29	&	52.41	&	52.49	\\
1.6	&	51.81	&	51.89	&	52.08	&	52.21	&	52.29	&	52.38	&	51.97	&	52.03	&	52.11	&	52.19	&	52.27	&	52.37	&	52.02	&	52.21	&	52.31	&	52.86	&	52.22	&	52.29	&	52.50	\\
2.5	&	51.22	&	51.51	&	51.83	&	52.09	&	52.31	&	52.44	&	51.93	&	52.00	&	52.20	&	52.20	&	52.24	&	52.43	&	51.80	&	52.05	&	52.31	&	52.89	&	51.91	&	52.10	&	52.47	\\
4.0	&	51.14	&	51.17	&	51.48	&	51.48	&	51.78	&	52.24	&	51.73	&	51.87	&	52.16	&	51.81	&	51.85	&	52.29	&	51.12	&	51.15	&	51.48	&	52.36	&	51.46	&	51.86	&	52.16	\\
6.3	&	48.47	&	48.49	&	49.17	&	50.03	&	50.80	&	51.91	&	50.76	&	50.98	&	50.37	&	50.64	&	51.15	&	51.99	&	48.45	&	48.48	&	49.65	&	51.98	&	50.11	&	50.96	&	50.88	\\
10.0	&	46.01	&	46.44	&	47.16	&	47.83	&	48.63	&	51.63	&	48.67	&	47.30	&	47.90	&	48.63	&	50.21	&	51.74	&	45.99	&	46.42	&	47.14	&	51.64	&	48.73	&	47.29	&	48.61	\\
15.8	&	44.11	&	44.62	&	45.39	&	46.02	&	46.69	&	51.07	&	45.15	&	45.19	&	45.94	&	46.78	&	47.39	&	51.24	&	44.09	&	44.60	&	45.37	&	51.07	&	44.92	&	45.18	&	45.98	\\
25.1	&	41.70	&	42.49	&	43.69	&	44.47	&	45.26	&	49.83	&	42.77	&	43.35	&	44.20	&	45.14	&	45.86	&	50.12	&	41.68	&	42.47	&	43.67	&	49.74	&	42.69	&	43.33	&	44.20	\\
\enddata
\end{deluxetable}

\begin{deluxetable}{cccccccccccccccccccc}
\tabletypesize{\scriptsize}
\label{tbl:HeIIionflux}
\tablecolumns{7}
\tablecaption{\ion{He}{2} ionising fluxes in units of photons s$^{-1}$.}
\tablehead{
\colhead{t} & \colhead{GalC} & \colhead{MW} & \colhead{LMC} & \colhead{SMC} & \colhead{IZw18} & \colhead{Z0} & \colhead{GalC} & \colhead{MW} & \colhead{LMC} & \colhead{SMC} & \colhead{IZw18} & \colhead{Z0} & \colhead{GalC} & \colhead{MW} & \colhead{LMC} & \colhead{Z0} & \colhead{GalC} & \colhead{MW} & \colhead{LMC}\\
\colhead{[Myr]} & \colhead{v00} & \colhead{v00} & \colhead{v00} & \colhead{v00} & \colhead{v00} & \colhead{v00} & \colhead{v40} & \colhead{v40} & \colhead{v40} & \colhead{v40} & \colhead{v40} & \colhead{v40} & \colhead{M300} & \colhead{M300} & \colhead{M300} & \colhead{M300} & \colhead{M300v40} & \colhead{M300v40} & \colhead{M300v40}
}
\startdata
1.0	&	47.25	&	47.60	&	47.84	&	48.02	&	48.48	&	48.80	&	47.50	&	47.60	&	47.80	&	47.98	&	48.45	&	48.79	&	47.71	&	47.92	&	48.57	&	51.41	&	48.03	&	48.52	&	48.75	\\
1.6	&	46.74	&	46.89	&	47.65	&	48.04	&	48.52	&	48.83	&	47.55	&	47.63	&	47.81	&	48.02	&	48.49	&	48.82	&	47.03	&	47.71	&	47.82	&	51.46	&	47.80	&	48.10	&	48.66	\\
2.5	&	44.22	&	45.93	&	46.80	&	47.47	&	48.37	&	48.89	&	47.35	&	47.47	&	47.79	&	47.80	&	48.10	&	48.88	&	44.20	&	45.92	&	50.48	&	51.23	&	47.33	&	47.80	&	49.83	\\
4.0	&	42.80	&	43.09	&	49.93	&	46.35	&	46.93	&	48.65	&	46.18	&	46.48	&	50.13	&	46.85	&	46.95	&	48.70	&	42.79	&	43.08	&	49.91	&	49.23	&	46.19	&	46.66	&	50.12	\\
6.3	&	37.74	&	38.57	&	40.54	&	42.84	&	44.37	&	48.29	&	41.55	&	41.92	&	43.01	&	44.50	&	44.88	&	48.36	&	37.72	&	38.56	&	41.13	&	48.47	&	41.41	&	41.90	&	44.28	\\
10.0	&	-	&	-	&	-	&	37.82	&	39.46	&	47.89	&	35.63	&	34.24	&	37.59	&	39.80	&	43.51	&	48.02	&	-	&	-	&	-	&	47.78	&	35.72	&	34.22	&	40.03	\\
15.8	&	-	&	-	&	-	&	-	&	-	&	45.98	&	-	&	-	&	-	&	36.08	&	37.67	&	46.92	&	-	&	-	&	-	&	45.95	&	-	&	-	&	-	\\
25.1	&	-	&	-	&	-	&	-	&	-	&	41.91	&	-	&	-	&	-	&	-	&	-	&	42.40	&	-	&	-	&	-	&	41.79	&	-	&	-	&	-	\\
\enddata
\end{deluxetable}

\begin{deluxetable}{cccccccccccccccccccc}
\tabletypesize{\scriptsize}
\label{tbl:Lbol}
\tablecolumns{7}
\tablecaption{Bolometric luminosities in units of log$L_{\rm{Bol}}$ erg s$_{-1}$.}
\tablehead{
\colhead{t} & \colhead{GalC} & \colhead{MW} & \colhead{LMC} & \colhead{SMC} & \colhead{IZw18} & \colhead{Z0} & \colhead{GalC} & \colhead{MW} & \colhead{LMC} & \colhead{SMC} & \colhead{IZw18} & \colhead{Z0} & \colhead{GalC} & \colhead{MW} & \colhead{LMC} & \colhead{Z0} & \colhead{GalC} & \colhead{MW} & \colhead{LMC}\\
\colhead{[Myr]} & \colhead{v00} & \colhead{v00} & \colhead{v00} & \colhead{v00} & \colhead{v00} & \colhead{v00} & \colhead{v40} & \colhead{v40} & \colhead{v40} & \colhead{v40} & \colhead{v40} & \colhead{v40} & \colhead{M300} & \colhead{M300} & \colhead{M300} & \colhead{M300} & \colhead{M300v40} & \colhead{M300v40} & \colhead{M300v40}
}
\startdata
1.0	&	42.58	&	42.58	&	42.59	&	42.60	&	42.60	&	42.61	&	42.57	&	42.58	&	42.58	&	42.59	&	42.59	&	42.60	&	42.79	&	42.81	&	42.82	&	42.82	&	42.79	&	42.81	&	42.83	\\
1.6	&	42.59	&	42.61	&	42.62	&	42.63	&	42.64	&	42.64	&	42.60	&	42.61	&	42.62	&	42.62	&	42.63	&	42.63	&	42.80	&	42.82	&	42.86	&	42.85	&	42.80	&	42.83	&	42.88	\\
2.5	&	42.63	&	42.65	&	42.67	&	42.69	&	42.70	&	42.70	&	42.63	&	42.66	&	42.69	&	42.69	&	42.69	&	42.69	&	42.76	&	42.80	&	42.86	&	42.91	&	42.78	&	42.83	&	42.91	\\
4.0	&	42.41	&	42.44	&	42.49	&	42.51	&	42.50	&	42.53	&	42.58	&	42.65	&	42.69	&	42.69	&	42.66	&	42.59	&	42.39	&	42.42	&	42.47	&	42.51	&	42.56	&	42.63	&	42.68	\\
6.3	&	42.08	&	42.10	&	42.14	&	42.17	&	42.17	&	42.19	&	42.30	&	42.33	&	42.32	&	42.32	&	42.31	&	42.28	&	42.06	&	42.08	&	42.13	&	42.18	&	42.28	&	42.31	&	42.31	\\
10.0	&	41.74	&	41.77	&	41.82	&	41.84	&	41.85	&	41.96	&	41.93	&	41.94	&	41.97	&	41.98	&	42.04	&	42.08	&	41.73	&	41.75	&	41.80	&	41.94	&	41.91	&	41.92	&	41.95	\\
15.8	&	41.43	&	41.45	&	41.52	&	41.57	&	41.57	&	41.69	&	41.64	&	41.65	&	41.68	&	41.67	&	41.70	&	41.82	&	41.41	&	41.44	&	41.50	&	41.68	&	41.60	&	41.63	&	41.64	\\
25.1	&	41.23	&	41.20	&	41.29	&	41.36	&	41.36	&	41.23	&	41.36	&	41.39	&	41.45	&	41.44	&	41.48	&	41.33	&	41.22	&	41.18	&	41.28	&	41.23	&	41.34	&	41.37	&	41.40	\\
\enddata
\end{deluxetable}

\begin{deluxetable}{cccccccccccccccccccc}
\tabletypesize{\scriptsize}
\label{tbl:Ha_ew}
\tablecolumns{7}
\tablecaption{H$\alpha$ equivalent widths in units of \AA.}
\tablehead{
\colhead{t} & \colhead{GalC} & \colhead{MW} & \colhead{LMC} & \colhead{SMC} & \colhead{IZw18} & \colhead{Z0} & \colhead{GalC} & \colhead{MW} & \colhead{LMC} & \colhead{SMC} & \colhead{IZw18} & \colhead{Z0} & \colhead{GalC} & \colhead{MW} & \colhead{LMC} & \colhead{Z0} & \colhead{GalC} & \colhead{MW} & \colhead{LMC}\\
\colhead{[Myr]} & \colhead{v00} & \colhead{v00} & \colhead{v00} & \colhead{v00} & \colhead{v00} & \colhead{v00} & \colhead{v40} & \colhead{v40} & \colhead{v40} & \colhead{v40} & \colhead{v40} & \colhead{v40} & \colhead{M300} & \colhead{M300} & \colhead{M300} & \colhead{M300} & \colhead{M300v40} & \colhead{M300v40} & \colhead{M300v40}
}
\startdata
1.0	&	3.45	&	3.47	&	3.50	&	3.51	&	3.53	&	3.56	&	3.45	&	3.46	&	3.49	&	3.51	&	3.53	&	3.56	&	3.49	&	3.51	&	3.54	&	3.65	&	3.51	&	3.53	&	3.55	\\
1.6	&	3.41	&	3.43	&	3.48	&	3.52	&	3.53	&	3.56	&	3.45	&	3.46	&	3.49	&	3.51	&	3.53	&	3.56	&	3.38	&	3.46	&	3.49	&	3.65	&	3.45	&	3.43	&	3.51	\\
2.5	&	3.08	&	3.17	&	3.32	&	3.45	&	3.52	&	3.56	&	3.37	&	3.38	&	3.48	&	3.50	&	3.49	&	3.56	&	3.09	&	3.19	&	3.28	&	3.64	&	3.17	&	3.22	&	3.46	\\
4.0	&	2.83	&	2.83	&	3.04	&	2.97	&	2.99	&	3.54	&	3.11	&	3.27	&	3.42	&	3.13	&	3.03	&	3.53	&	2.83	&	2.84	&	3.07	&	3.58	&	3.11	&	3.27	&	3.43	\\
6.3	&	1.72	&	1.81	&	2.08	&	2.54	&	2.99	&	3.54	&	2.78	&	2.87	&	2.59	&	2.67	&	2.96	&	3.54	&	1.72	&	1.81	&	2.17	&	3.57	&	2.78	&	2.87	&	2.75	\\
10.0	&	0.94	&	0.99	&	1.33	&	1.85	&	2.35	&	3.52	&	1.89	&	1.56	&	1.63	&	2.06	&	2.77	&	3.51	&	0.94	&	0.99	&	1.33	&	3.52	&	1.89	&	1.56	&	1.69	\\
15.8	&	0.18	&	0.41	&	0.39	&	0.90	&	1.44	&	3.41	&	0.74	&	0.56	&	0.92	&	1.43	&	1.83	&	3.37	&	0.18	&	0.41	&	0.39	&	3.41	&	0.74	&	0.56	&	0.92	\\
25.1	&	-0.46	&	-0.19	&	-0.05	&	0.22	&	0.76	&	3.06	&	-0.21	&	-0.02	&	0.17	&	0.71	&	1.14	&	3.17	&	-0.46	&	-0.19	&	-0.05	&	3.06	&	-0.21	&	-0.02	&	0.17	\\
\enddata
\end{deluxetable}

\begin{deluxetable}{cccccccccccccccccccc}
\tabletypesize{\scriptsize}
\label{tbl:windpower}
\tablecolumns{7}
\tablecaption{Wind powers in units of log$L_{\rm{mech}}$ erg s$_{-1}$.}
\tablehead{
\colhead{t} & \colhead{GalC} & \colhead{MW} & \colhead{LMC} & \colhead{SMC} & \colhead{IZw18} & \colhead{Z0} & \colhead{GalC} & \colhead{MW} & \colhead{LMC} & \colhead{SMC} & \colhead{IZw18} & \colhead{Z0} & \colhead{GalC} & \colhead{MW} & \colhead{LMC} & \colhead{Z0} & \colhead{GalC} & \colhead{MW} & \colhead{LMC}\\
\colhead{[Myr]} & \colhead{v00} & \colhead{v00} & \colhead{v00} & \colhead{v00} & \colhead{v00} & \colhead{v00} & \colhead{v40} & \colhead{v40} & \colhead{v40} & \colhead{v40} & \colhead{v40} & \colhead{v40} & \colhead{M300} & \colhead{M300} & \colhead{M300} & \colhead{M300} & \colhead{M300v40} & \colhead{M300v40} & \colhead{M300v40}
}
\startdata
1.0	&	40.30	&	40.13	&	39.73	&	39.20	&	38.40	&	-	&	40.32	&	40.14	&	39.69	&	39.08	&	38.16	&	-	&	40.73	&	40.58	&	40.22	&	33.73	&	40.75	&	40.57	&	40.16	\\
1.6	&	40.29	&	40.13	&	39.78	&	39.28	&	38.49	&	-	&	40.33	&	40.17	&	39.75	&	39.16	&	38.28	&	-	&	40.65	&	40.50	&	40.16	&	33.76	&	40.70	&	40.53	&	40.19	\\
2.5	&	40.28	&	40.11	&	39.76	&	39.29	&	38.59	&	-	&	40.37	&	40.22	&	39.81	&	39.25	&	38.47	&	35.07	&	40.98	&	40.83	&	40.57	&	33.99	&	40.65	&	40.86	&	40.73	\\
4.0	&	40.30	&	40.16	&	39.84	&	39.06	&	38.42	&	-	&	40.64	&	40.63	&	40.36	&	38.98	&	38.30	&	35.27	&	40.29	&	40.15	&	39.83	&	34.16	&	40.62	&	40.61	&	40.35	\\
6.3	&	38.77	&	38.84	&	38.95	&	38.30	&	37.33	&	-	&	39.89	&	39.95	&	39.00	&	38.57	&	37.46	&	-	&	38.75	&	38.82	&	38.93	&	33.72	&	39.87	&	39.94	&	38.98	\\
10.0	&	37.98	&	38.30	&	37.91	&	37.63	&	36.88	&	-	&	39.05	&	38.70	&	38.49	&	37.77	&	36.91	&	-	&	37.97	&	38.29	&	37.89	&	33.49	&	39.04	&	38.68	&	38.47	\\
15.8	&	37.27	&	36.85	&	36.61	&	36.90	&	36.36	&	-	&	37.59	&	37.45	&	37.48	&	37.08	&	36.48	&	-	&	37.25	&	36.83	&	36.59	&	33.47	&	37.57	&	37.43	&	37.46	\\
25.1	&	36.48	&	36.19	&	35.79	&	36.12	&	35.63	&	-	&	36.65	&	36.50	&	36.64	&	36.38	&	35.75	&	-	&	36.47	&	36.17	&	35.77	&	32.57	&	36.63	&	36.48	&	36.62	\\
\enddata
\end{deluxetable}

\begin{deluxetable}{cccccccccccccccccccc}
\tabletypesize{\scriptsize}
\label{tbl:uvslope}
\tablecolumns{7}
\tablecaption{$\beta$ UV slopes.}
\tablehead{
\colhead{t} & \colhead{GalC} & \colhead{MW} & \colhead{LMC} & \colhead{SMC} & \colhead{IZw18} & \colhead{Z0} & \colhead{GalC} & \colhead{MW} & \colhead{LMC} & \colhead{SMC} & \colhead{IZw18} & \colhead{Z0} & \colhead{GalC} & \colhead{MW} & \colhead{LMC} & \colhead{Z0} & \colhead{GalC} & \colhead{MW} & \colhead{LMC}\\
\colhead{[Myr]} & \colhead{v00} & \colhead{v00} & \colhead{v00} & \colhead{v00} & \colhead{v00} & \colhead{v00} & \colhead{v40} & \colhead{v40} & \colhead{v40} & \colhead{v40} & \colhead{v40} & \colhead{v40} & \colhead{M300} & \colhead{M300} & \colhead{M300} & \colhead{M300} & \colhead{M300v40} & \colhead{M300v40} & \colhead{M300v40}
}
\startdata
1.0	&	-2.44	&	-2.45	&	-2.45	&	-2.42	&	-2.40	&	-2.31	&	-2.44	&	-2.45	&	-2.45	&	-2.42	&	-2.40	&	-2.31	&	-2.37	&	-2.37	&	-2.34	&	-1.84	&	-2.36	&	-2.35	&	-2.33	\\
1.6	&	-2.44	&	-2.46	&	-2.45	&	-2.42	&	-2.39	&	-2.31	&	-2.44	&	-2.46	&	-2.45	&	-2.42	&	-2.39	&	-2.31	&	-2.26	&	-2.32	&	-2.39	&	-1.84	&	-1.88	&	-1.76	&	-2.27	\\
2.5	&	-2.53	&	-2.47	&	-2.44	&	-2.41	&	-2.37	&	-2.30	&	-2.53	&	-2.47	&	-2.44	&	-2.41	&	-2.37	&	-2.30	&	-2.45	&	-2.39	&	-2.33	&	-1.92	&	-2.46	&	-2.51	&	-1.89	\\
4.0	&	-2.51	&	-2.46	&	-2.45	&	-2.35	&	-2.32	&	-2.31	&	-2.51	&	-2.46	&	-2.45	&	-2.35	&	-2.32	&	-2.31	&	-2.51	&	-2.45	&	-2.44	&	-2.25	&	-2.72	&	-2.31	&	-2.18	\\
6.3	&	-2.78	&	-2.67	&	-2.51	&	-2.53	&	-2.58	&	-2.40	&	-2.78	&	-2.67	&	-2.51	&	-2.53	&	-2.58	&	-2.40	&	-2.78	&	-2.67	&	-2.46	&	-2.46	&	-2.58	&	-2.54	&	-2.33	\\
10.0	&	-2.84	&	-2.77	&	-2.64	&	-2.64	&	-2.74	&	-2.51	&	-2.84	&	-2.77	&	-2.64	&	-2.64	&	-2.74	&	-2.51	&	-2.84	&	-2.77	&	-2.64	&	-2.58	&	-2.88	&	-2.89	&	-2.52	\\
15.8	&	-2.60	&	-2.74	&	-2.82	&	-2.46	&	-2.67	&	-2.80	&	-2.60	&	-2.74	&	-2.82	&	-2.46	&	-2.67	&	-2.80	&	-2.60	&	-2.74	&	-2.82	&	-2.80	&	-2.73	&	-2.76	&	-2.58	\\
25.1	&	-2.28	&	-2.41	&	-2.62	&	-2.44	&	-2.56	&	-2.93	&	-2.28	&	-2.41	&	-2.62	&	-2.44	&	-2.56	&	-2.93	&	-2.28	&	-2.41	&	-2.62	&	-2.93	&	-2.50	&	-2.57	&	-2.54	\\
\enddata
\end{deluxetable}

\section{Colours} \label{sec: colours} 

\nstarburst\ can produce predictions for all colours as defined in \starburst, these include the F130M, F210M, U, B, V, R, I, J, H, K and L bands. We also intend to extend the colours available in \nstarburst. Here, as an example, we show a comparison of $M_{V}$ predicted from \nstarburst\ including all stars between $120M_{\odot}$ and $300M_{\odot}$. A significant increase in $M_{V}$ is predicted with the addition of VMS.

\begin{figure}[t!]
    \includegraphics[width=\columnwidth]{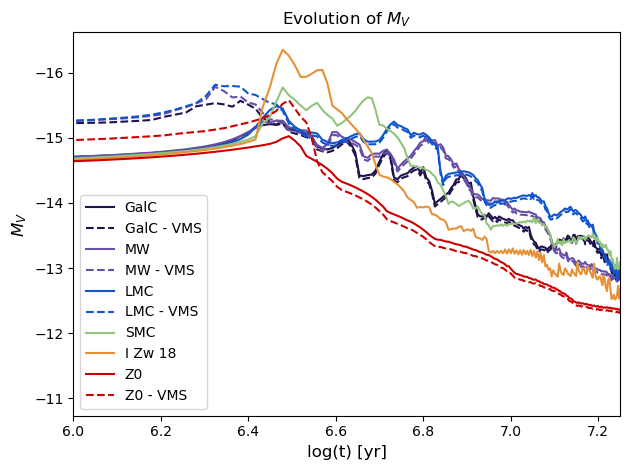} 
    \caption{As described in Fig. \ref{fig: ionflux_metallicity} but showing M$_{V}$.}
    \label{fig: colour_MV}
\end{figure}





\section{Additional comparisons} \label{sec: app-plots}

For completeness we include here a number of comparisons between \starburst\ and \nstarburst, such as the SEDs with similar inputs from both code versions, the SEDs produced using \wmbasic\ and \fastwind, and the SEDs including VMS at various metallicities.

\begin{figure}[t!]
    \includegraphics[scale=0.7]{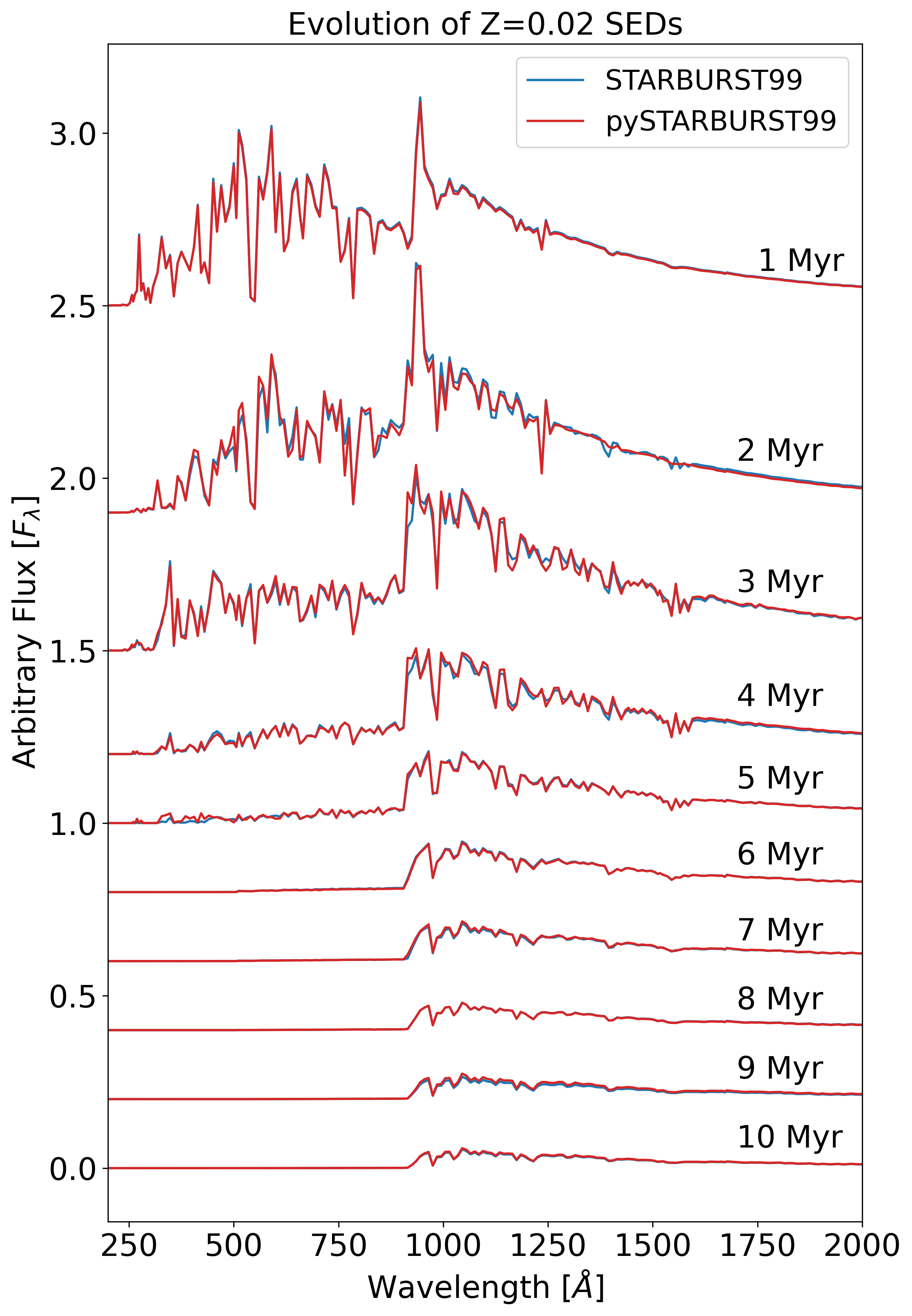}
    \caption{Synthetic FUV SEDs from \starburst\, compared to those produced with \nstarburst, in both cases utilising the \cite{Yusof2022} stellar evolutionary models at $Z=0.02$ and \wmbasic\, spectra at $Z=0.02$. The evolution with time is shown from 1 Myr to 10 Myr at intervals of 1 Myr.}
    \label{fig: py_fort_comp_mwc}
\end{figure}

\begin{figure}[t!]
    \includegraphics[scale=0.7]{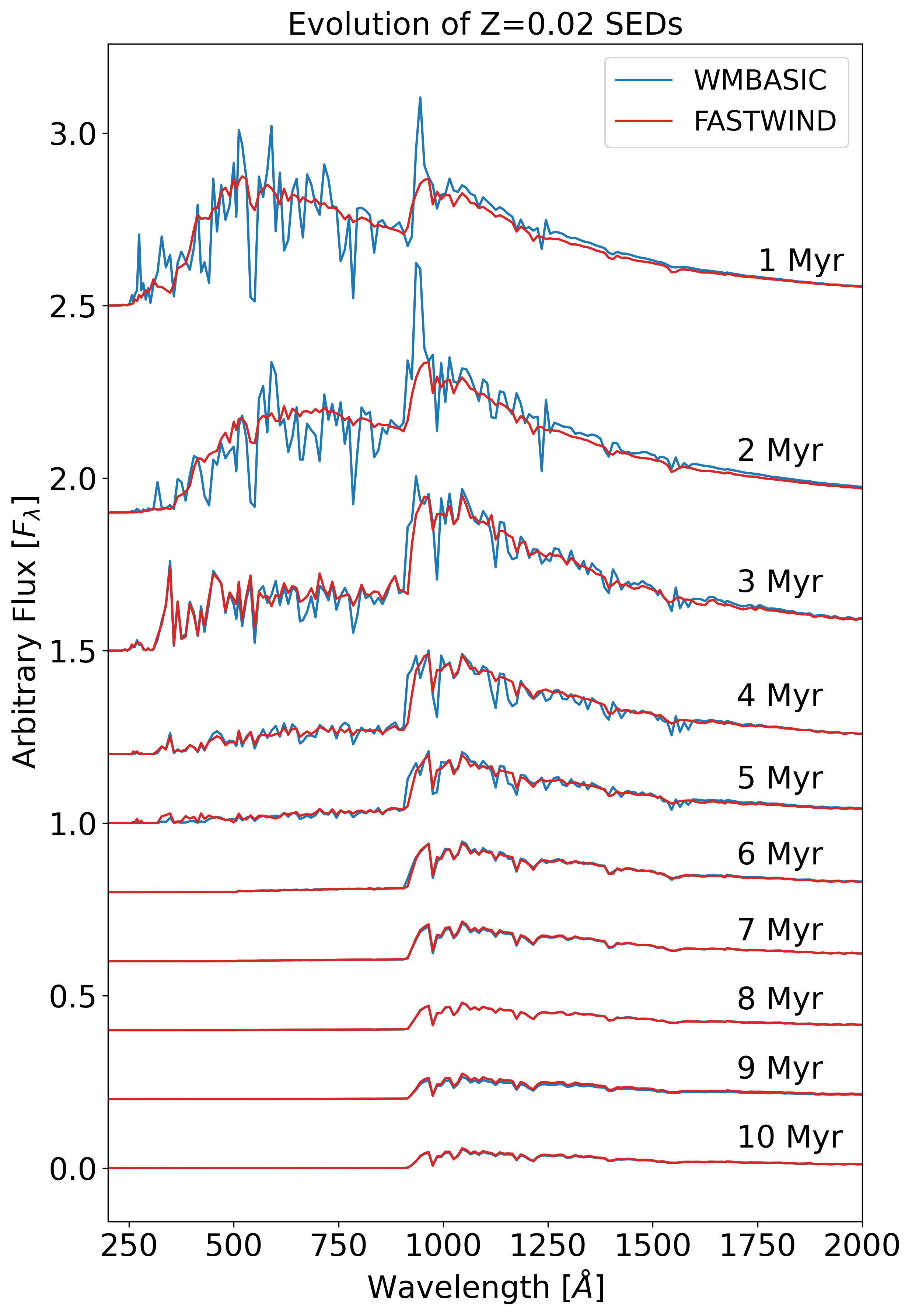}
    \caption{Synthetic FUV SEDs, utilising the \cite{Yusof2022} stellar evolutionary models at $Z=0.02$ for both, with \wmbasic\, spectra at $Z=0.02$ compared with \fastwind\, spectra at $Z=0.02$. The evolution with time is shown from 1 Myr to 10 Myr at intervals of 1 Myr.}
    \label{fig: wm_fw_comp_mwc}
\end{figure}

\begin{figure}[t!]
    \includegraphics[scale=0.7]{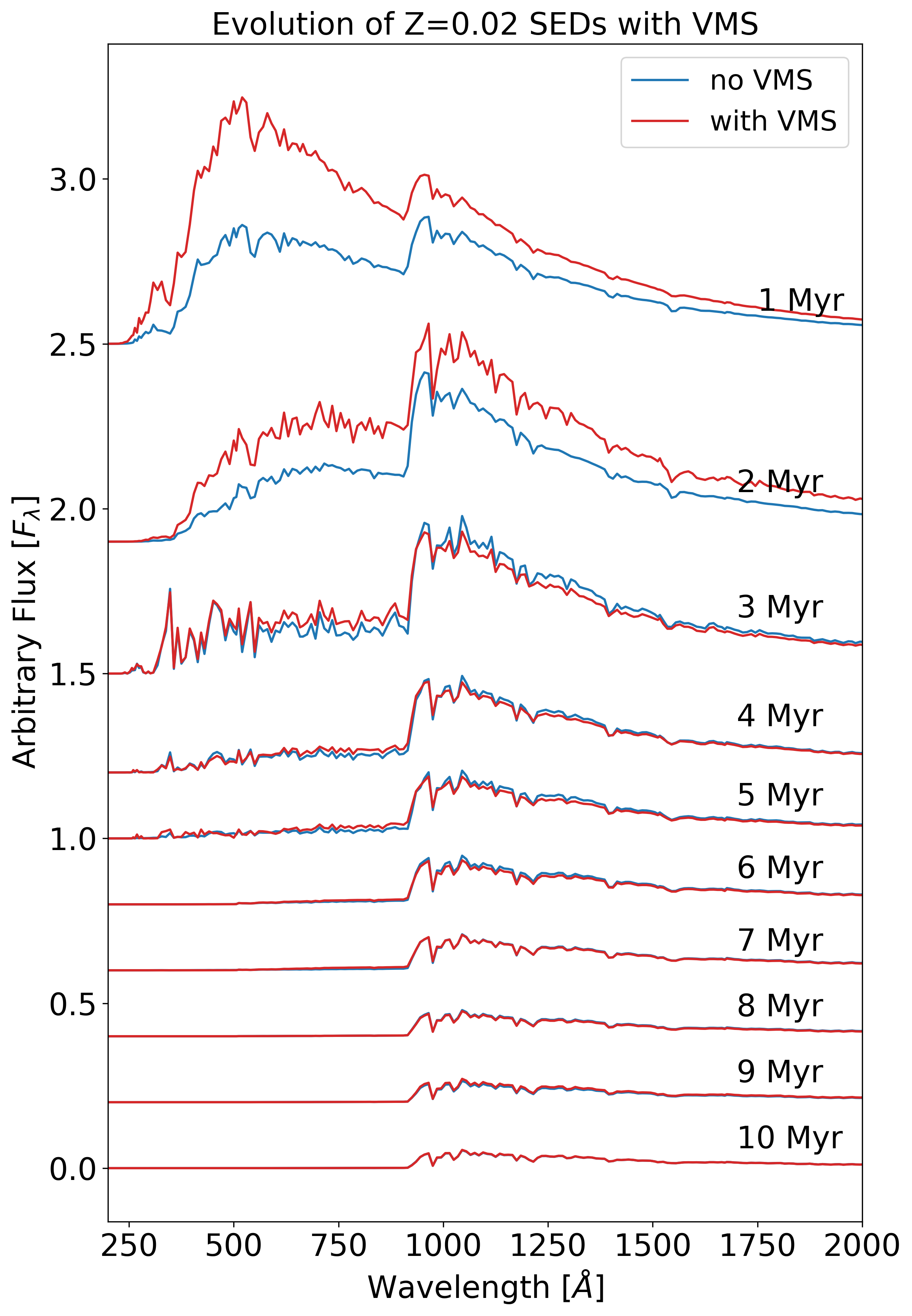}
    \caption{Synthetic FUV SEDs from \nstarburst, utilising the \cite{Yusof2022} stellar evolutionary models at Z=0.02 and \fastwind\, spectra at Z=0.02. This is compared with the addition of VMS evolutionary tracks up to $300M_{\odot}$ from \cite{Martinet2023} and \fastwind\, models tailored to match the extended parameter space coverage from VMS evolutionary tracks. The evolution with time is shown from 1 Myr to 10 Myr at intervals of 1 Myr.}
    \label{fig: sed_VMS_mwc}
\end{figure}

\begin{figure}[t!]
    \includegraphics[scale=0.7]{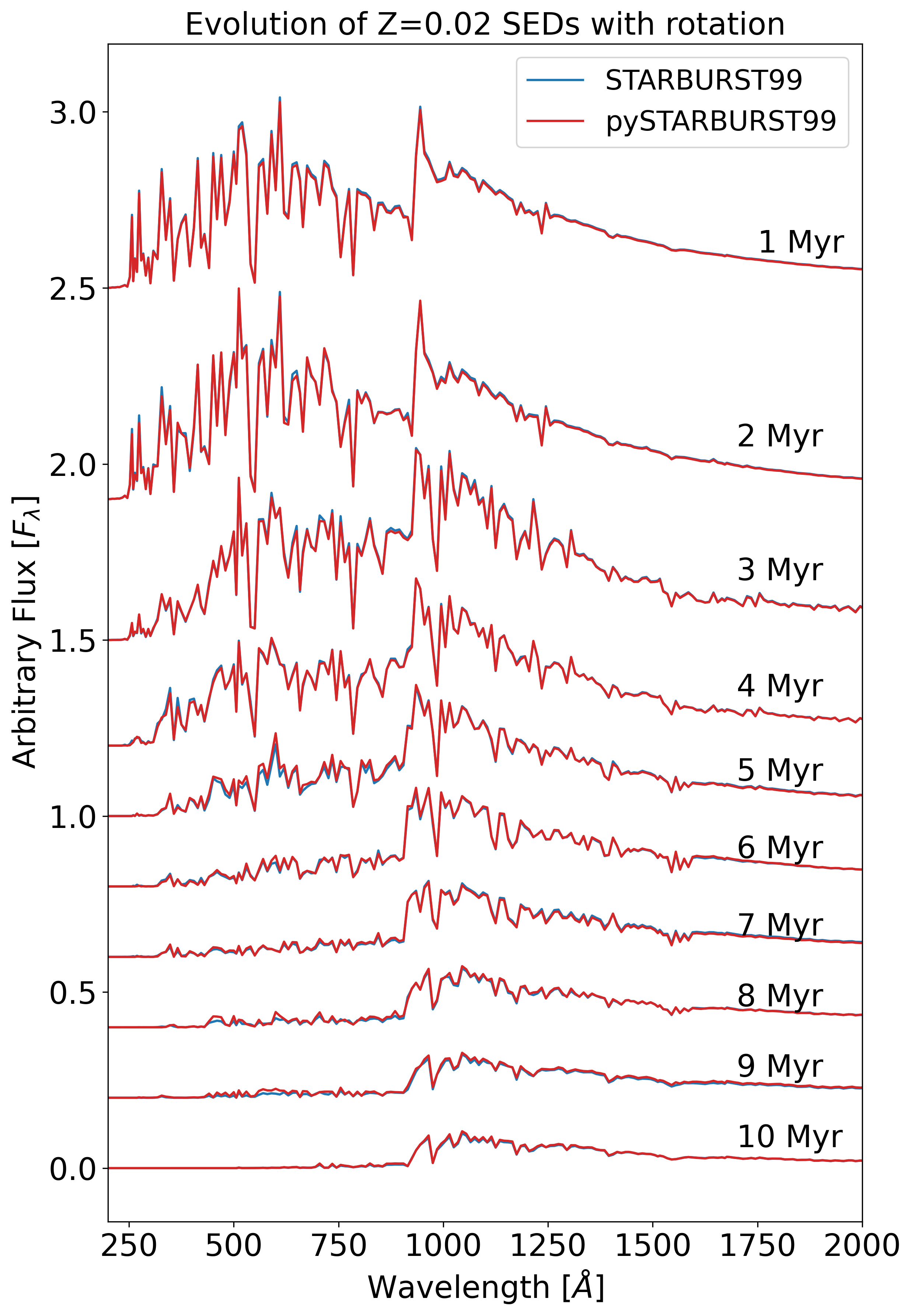}
    \caption{Synthetic FUV SEDs from \starburst\, compared to those produced with \nstarburst, in both cases utilising the \cite{Yusof2022} stellar evolutionary models including rotation at $Z=0.02$ and \wmbasic\, spectra at $Z=0.02$. The evolution with time is shown from 1 Myr to 10 Myr at intervals of 1 Myr.}
    \label{fig: py_fort_comp_rot_mwc}
\end{figure}

\begin{figure}[t!]
    \includegraphics[scale=0.7]{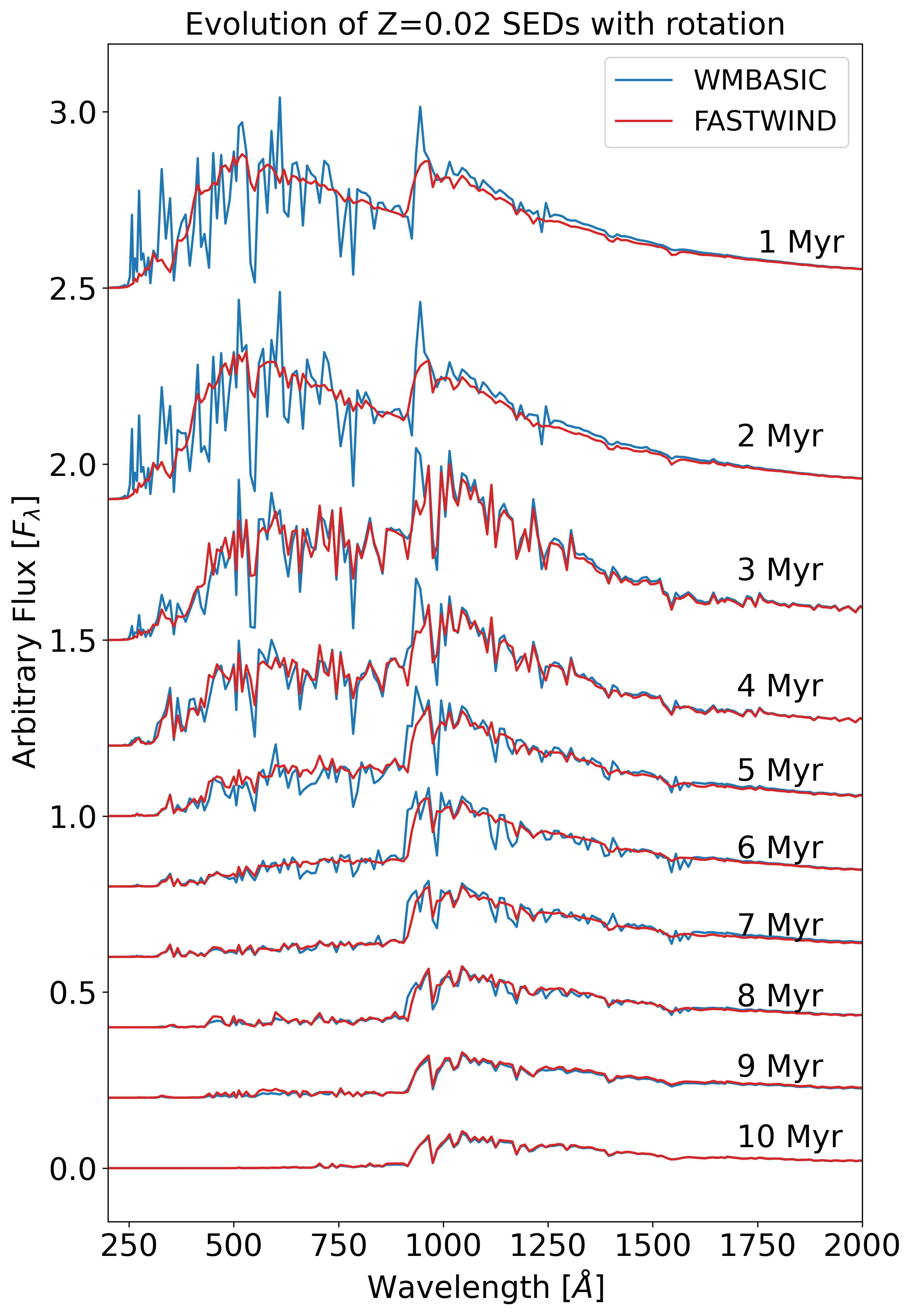}
    \caption{Synthetic FUV SEDs, utilising the \cite{Yusof2022} stellar evolutionary models at $Z=0.02$ with rotation for both, with \wmbasic\, spectra at $Z=0.02$ compared with \fastwind\, spectra at $Z=0.02$. The evolution with time is shown from 1 Myr to 10 Myr at intervals of 1 Myr.}
    \label{fig: wm_fw_comp_rot_mwc}
\end{figure}

\begin{figure}[t!]
    \includegraphics[scale=0.7]{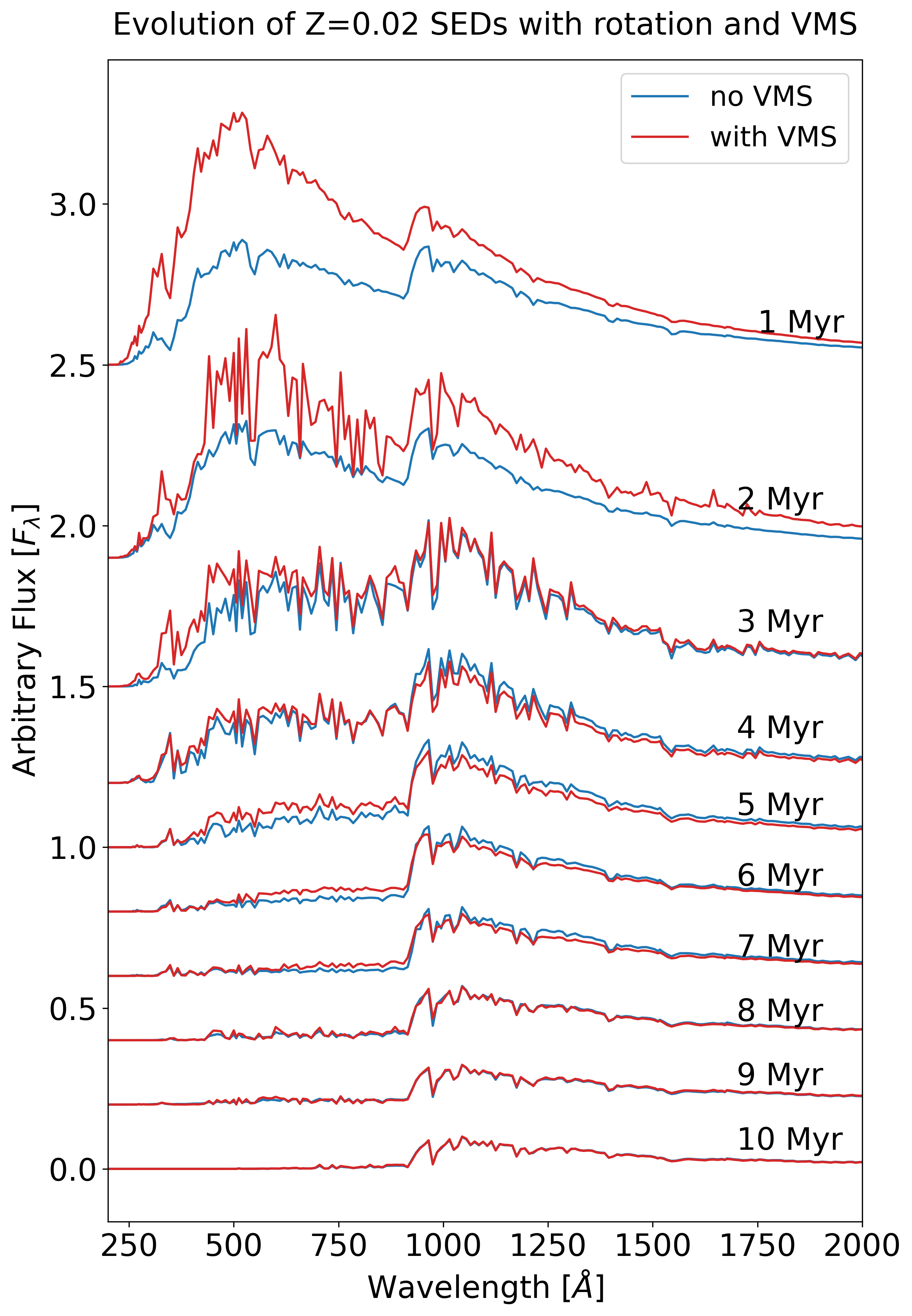}
    \caption{Synthetic FUV SEDs from \nstarburst, utilising the \cite{Yusof2022} stellar evolutionary models at Z=0.02 with rotation and \fastwind\, spectra at Z=0.02. This is compared with the addition of VMS evolutionary tracks up to $300M_{\odot}$ from \cite{Martinet2023} and \fastwind\, models tailored to match the extended parameter space coverage from VMS evolutionary tracks. The evolution with time is shown from 1 Myr to 10 Myr at intervals of 1 Myr.}
    \label{fig: sed_VMS_rot_mwc}
\end{figure}

\begin{figure}[t!]
    \includegraphics[scale=0.7]{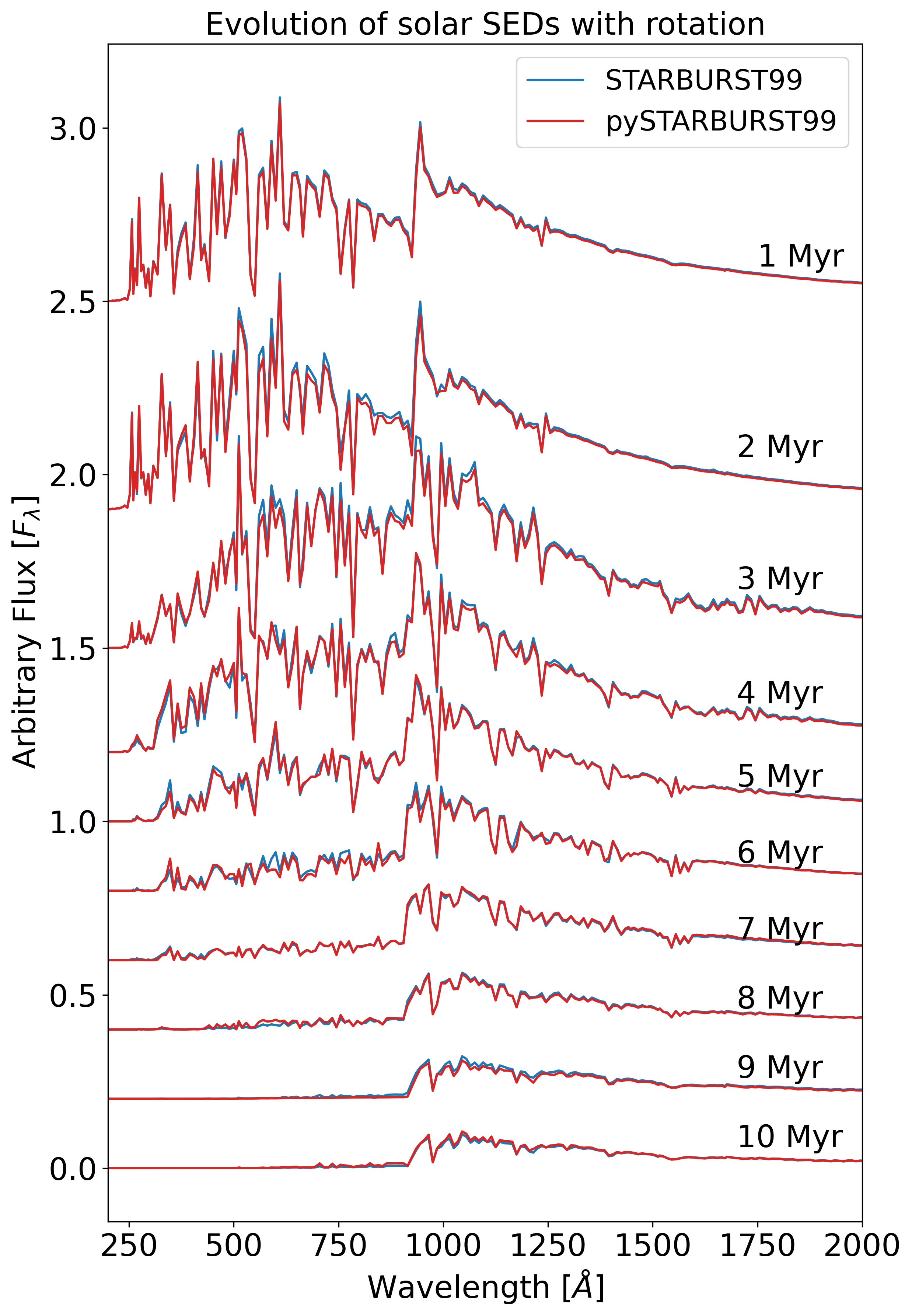}
    \caption{Synthetic FUV SEDs from \starburst\, compared to those produced with \nstarburst, in both cases utilising the \cite{Ekstrom2012} stellar evolutionary models including rotation at Z=0.014 and \wmbasic\, spectra at Z=0.2. The evolution with time is shown from 1 Myr to 10 Myr at intervals of 1 Myr.}
    \label{fig: py_fort_comp_rot}
\end{figure}

\begin{figure}[t!]
    \includegraphics[scale=0.7]{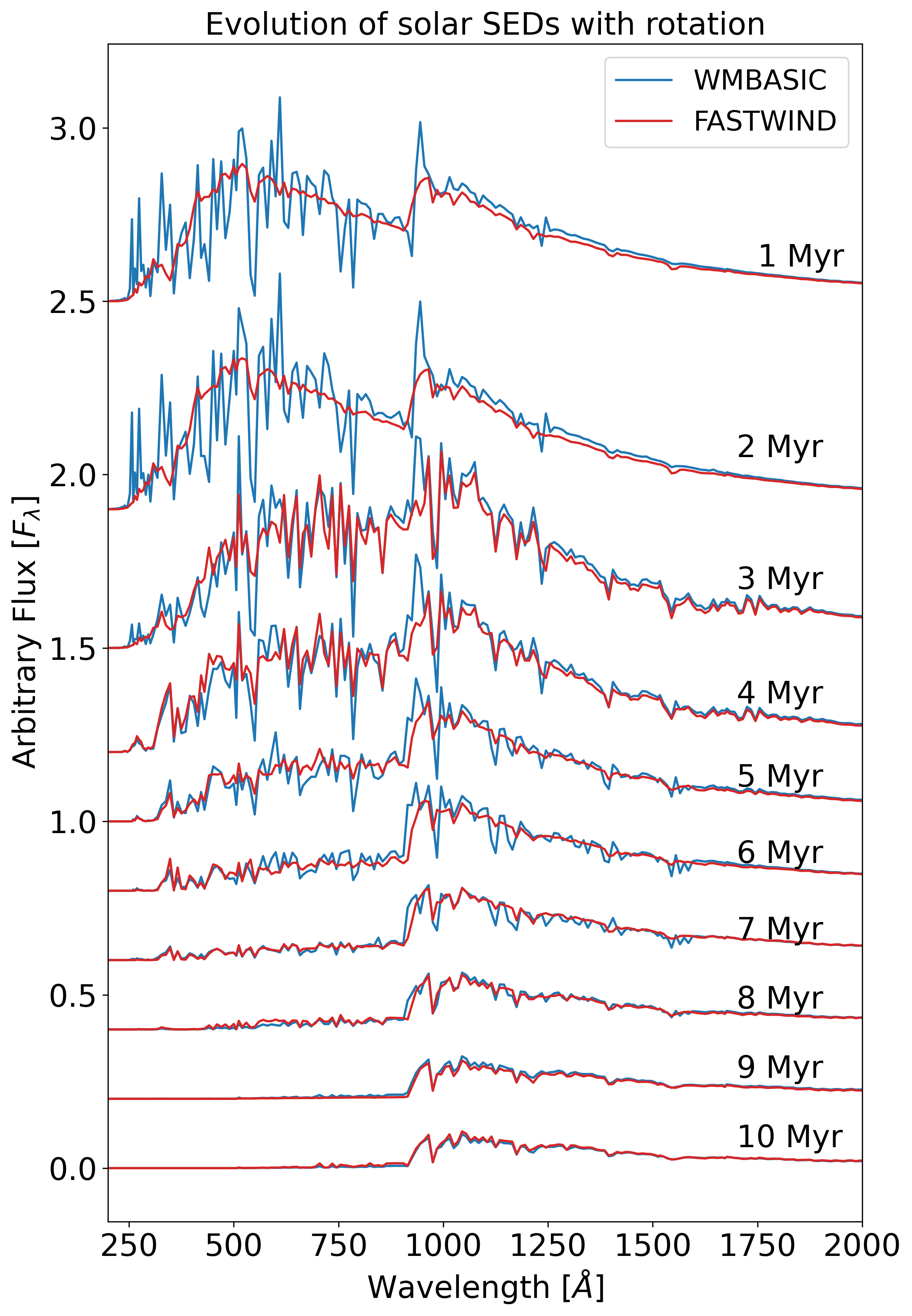}
    \caption{Synthetic FUV SEDs, utilising the \cite{Ekstrom2012} stellar evolutionary models at $Z=0.014$ with rotation for both, with \wmbasic\, spectra at $Z=0.02$ compared with \fastwind\, spectra at $Z=0.014$. The evolution with time is shown from 1 Myr to 10 Myr at intervals of 1 Myr.}
    \label{fig: wm_fw_comp_rot}
\end{figure}

\begin{figure}[t!]
    \includegraphics[scale=0.7]{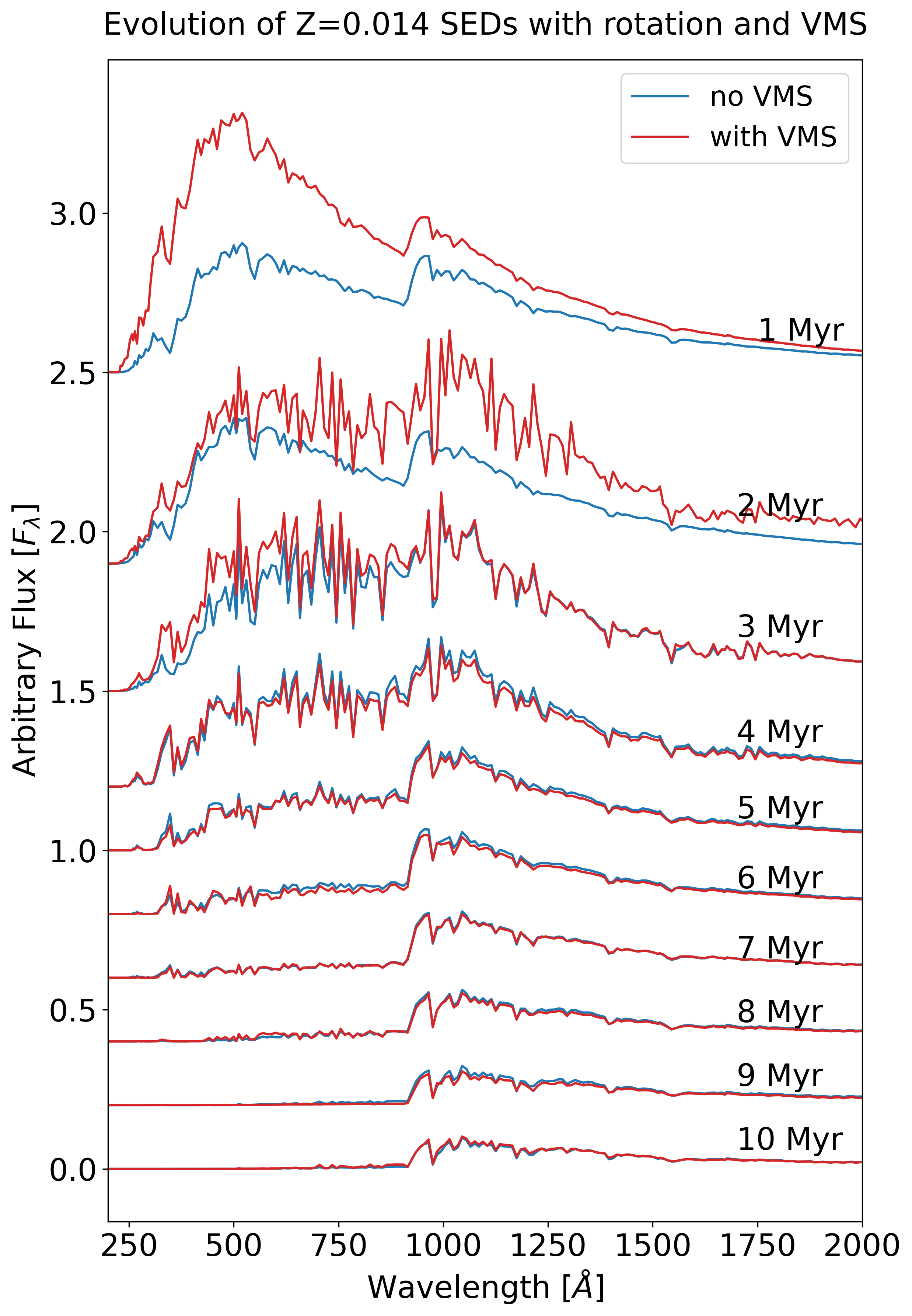}
    \caption{Synthetic FUV SEDs from \nstarburst, utilising the \cite{Ekstrom2012} stellar evolutionary models at Z=0.014 with rotation and \fastwind\, spectra at Z=0.014. This is compared with the addition of VMS evolutionary tracks up to $300M_{\odot}$ from \cite{Martinet2023} and \fastwind\, models tailored to match the extended parameter space coverage from VMS evolutionary tracks. The evolution with time is shown from 1 Myr to 10 Myr at intervals of 1 Myr.}
    \label{fig: sed_VMS_rot}
\end{figure}

\begin{figure}[t!]
    \includegraphics[scale=0.7]{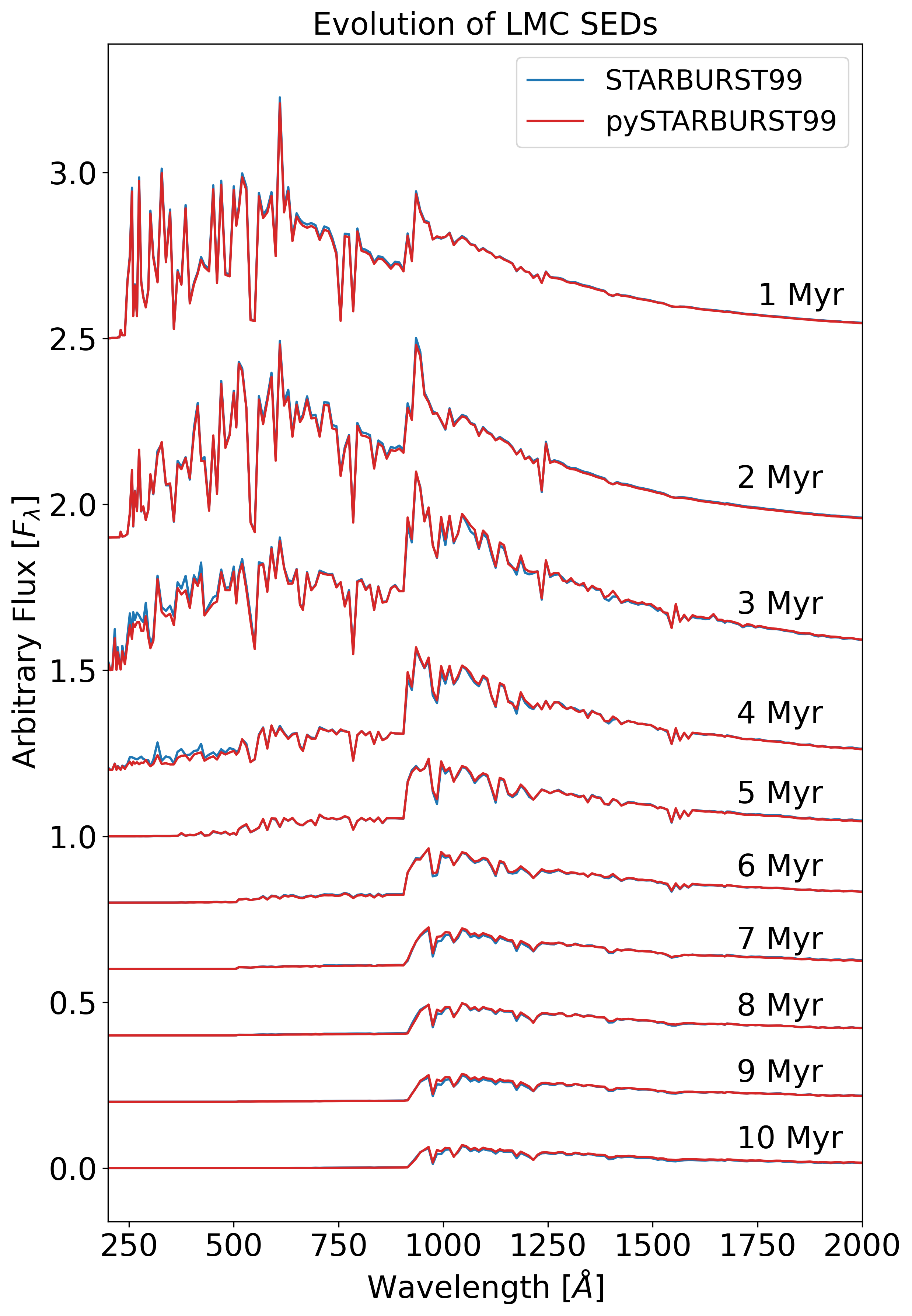}
    \caption{Synthetic FUV SEDs from \starburst\, compared to those produced with \nstarburst, in both cases utilising the \cite{Eggenberger2021} stellar evolutionary models at Z=0.006 and \wmbasic\, spectra at Z=0.008. The evolution with time is shown from 1 Myr to 10 Myr at intervals of 1 Myr.}
    \label{fig: py_fort_comp_LMC}
\end{figure}

\begin{figure}[t!]
    \includegraphics[scale=0.7]{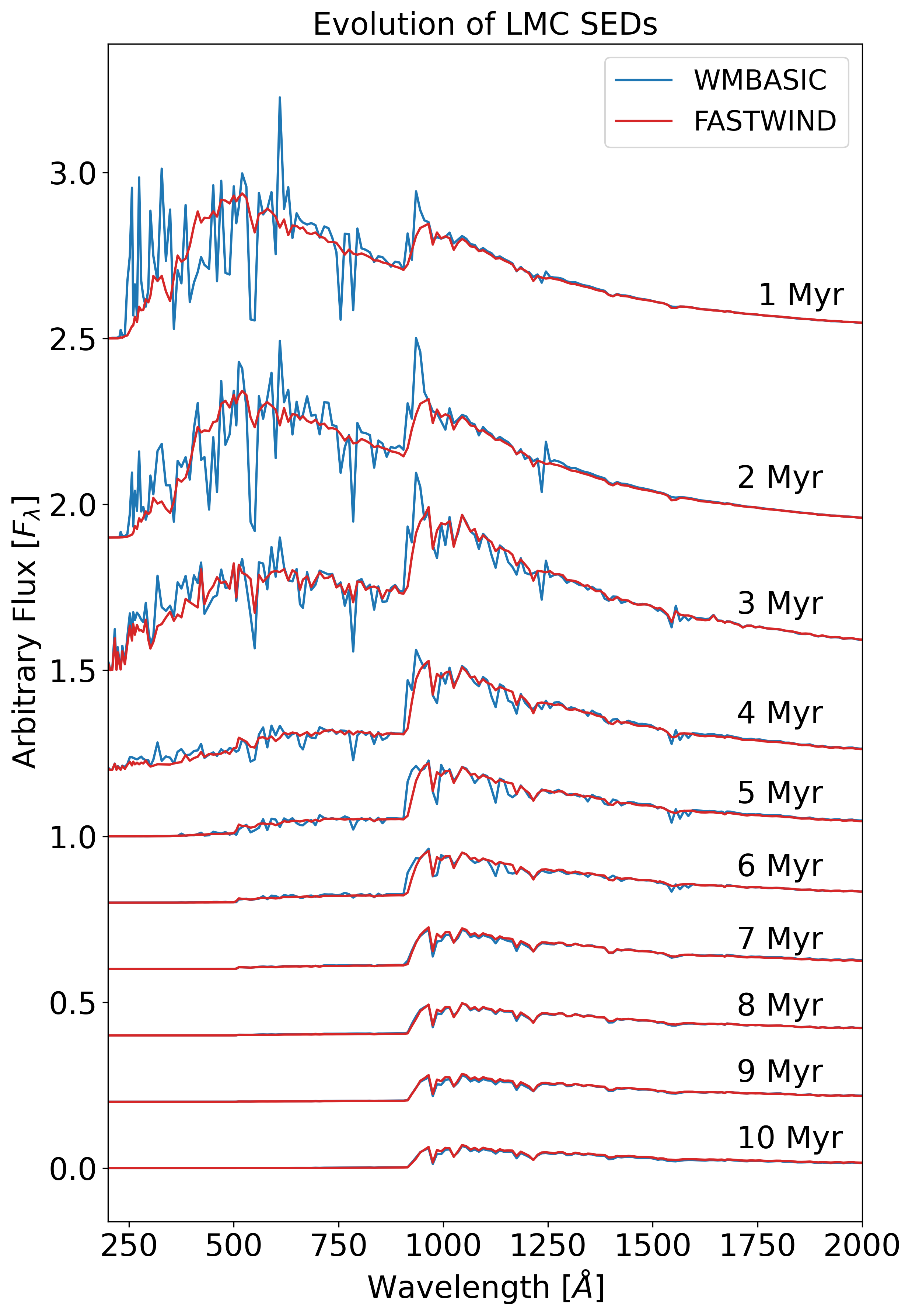}
    \caption{Synthetic FUV SEDs, utilising the \cite{Eggenberger2021} stellar evolutionary models at $Z=0.006$ for both, with \wmbasic\, spectra at $Z=0.008$ compared with \fastwind\, spectra at $Z=0.006$. The evolution with time is shown from 1 Myr to 10 Myr at intervals of 1 Myr.}
    \label{fig: wm_fw_comp_lmc}
\end{figure}

\begin{figure}[t!]
    \includegraphics[scale=0.7]{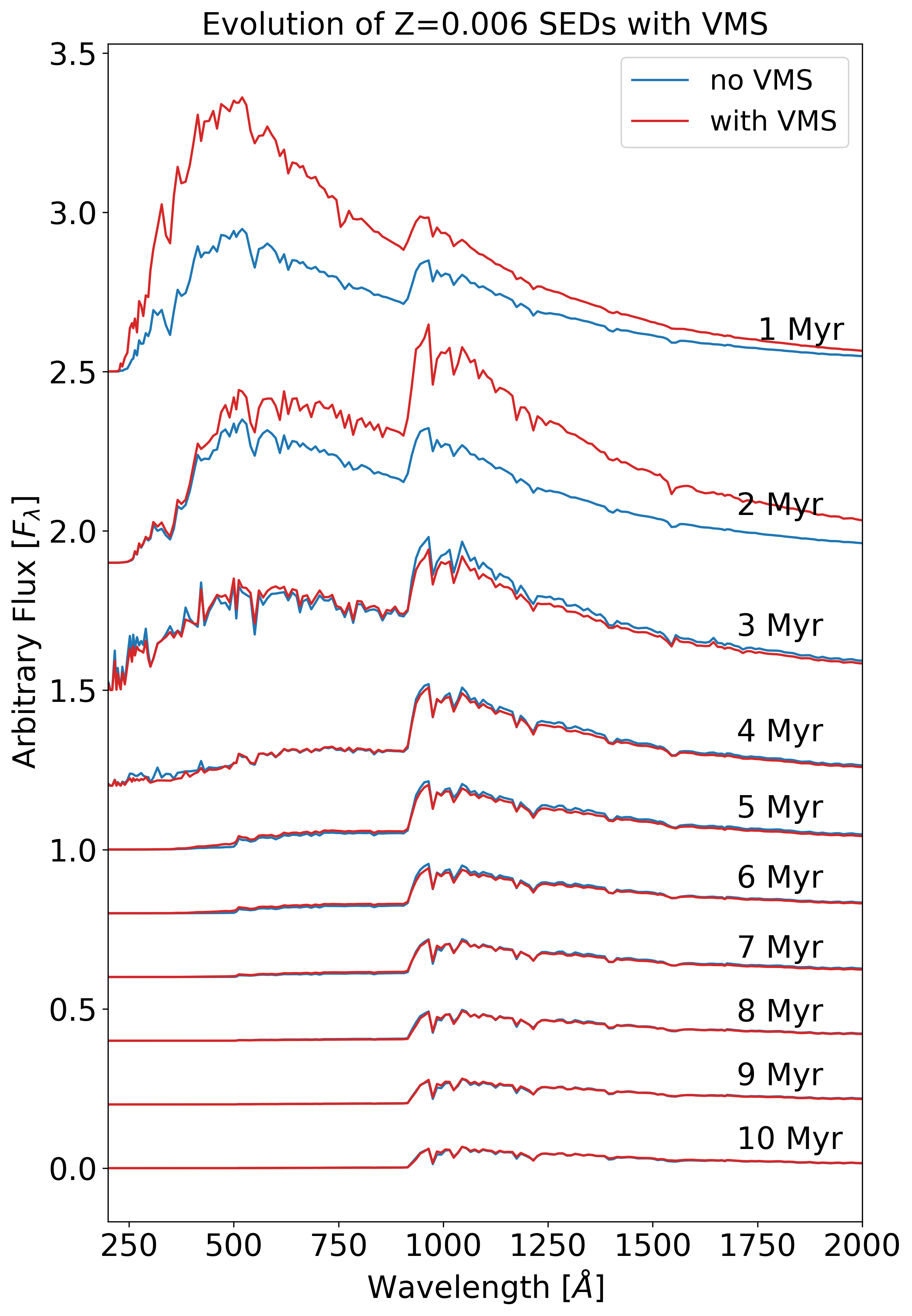}
    \caption{Synthetic FUV SEDs from \nstarburst, utilising the \cite{Eggenberger2021} stellar evolutionary models at $Z=0.006$ and \fastwind\, spectra at $Z=0.006$. This is compared with the addition of VMS evolutionary tracks up to $300M_{\odot}$ from \cite{Martinet2023} and \fastwind\, models tailored to match the extended parameter space coverage from VMS evolutionary tracks. The evolution with time is shown from 1 Myr to 10 Myr at intervals of 1 Myr.}
    \label{fig: sed_lmc_VMS_norot}
\end{figure}

\begin{figure}[t!]
    \includegraphics[scale=0.7]{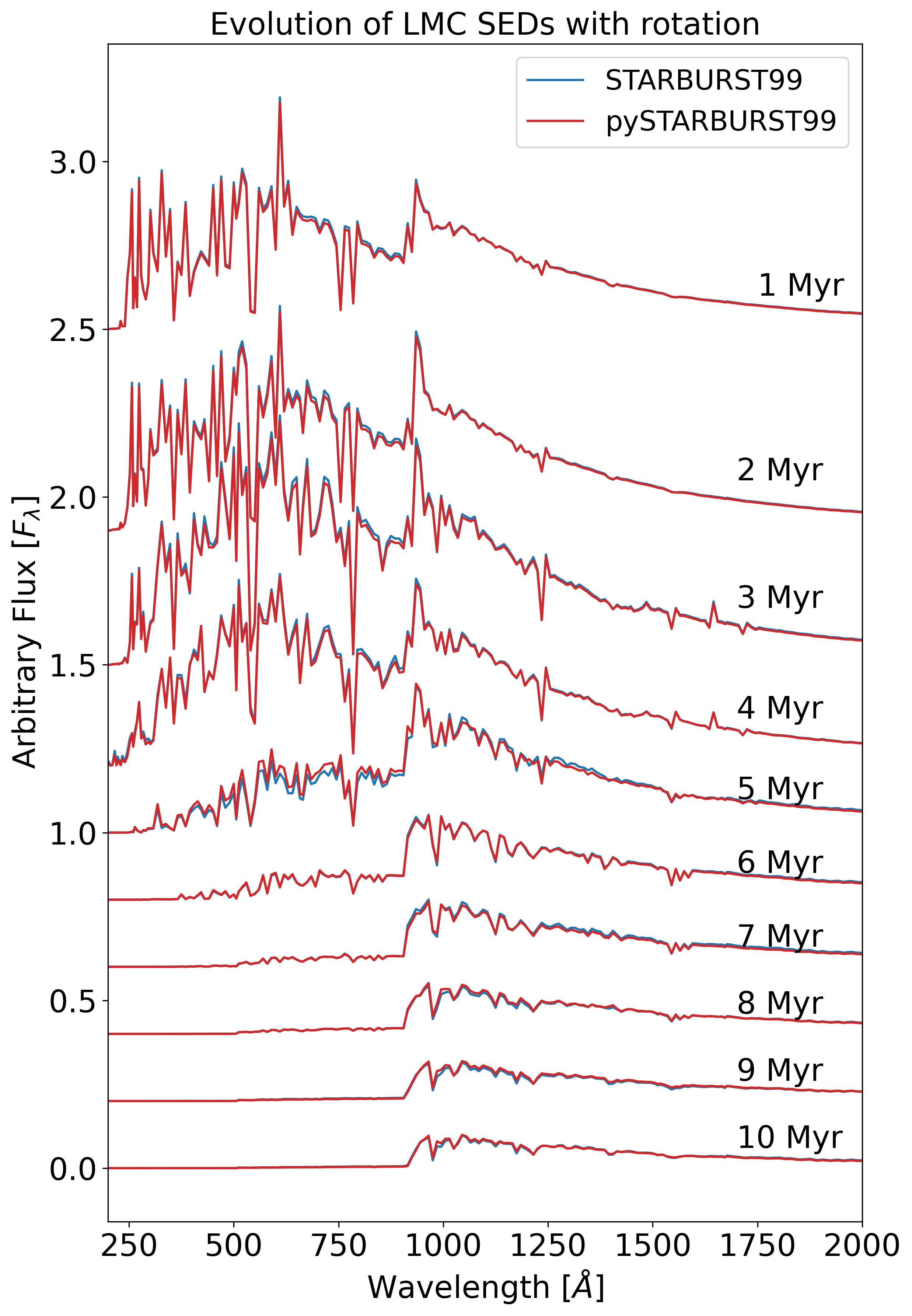}
    \caption{Synthetic FUV SEDs from \starburst\, compared to those produced with \nstarburst, in both cases utilising the \cite{Eggenberger2021} stellar evolutionary models including rotation at $Z=0.006$ and \wmbasic\, spectra at $Z=0.008$. The evolution with time is shown from 1 Myr to 10 Myr at intervals of 1 Myr.}
    \label{fig: py_fort_comp_rot_lmc}
\end{figure}

\begin{figure}[t!]
    \includegraphics[scale=0.7]{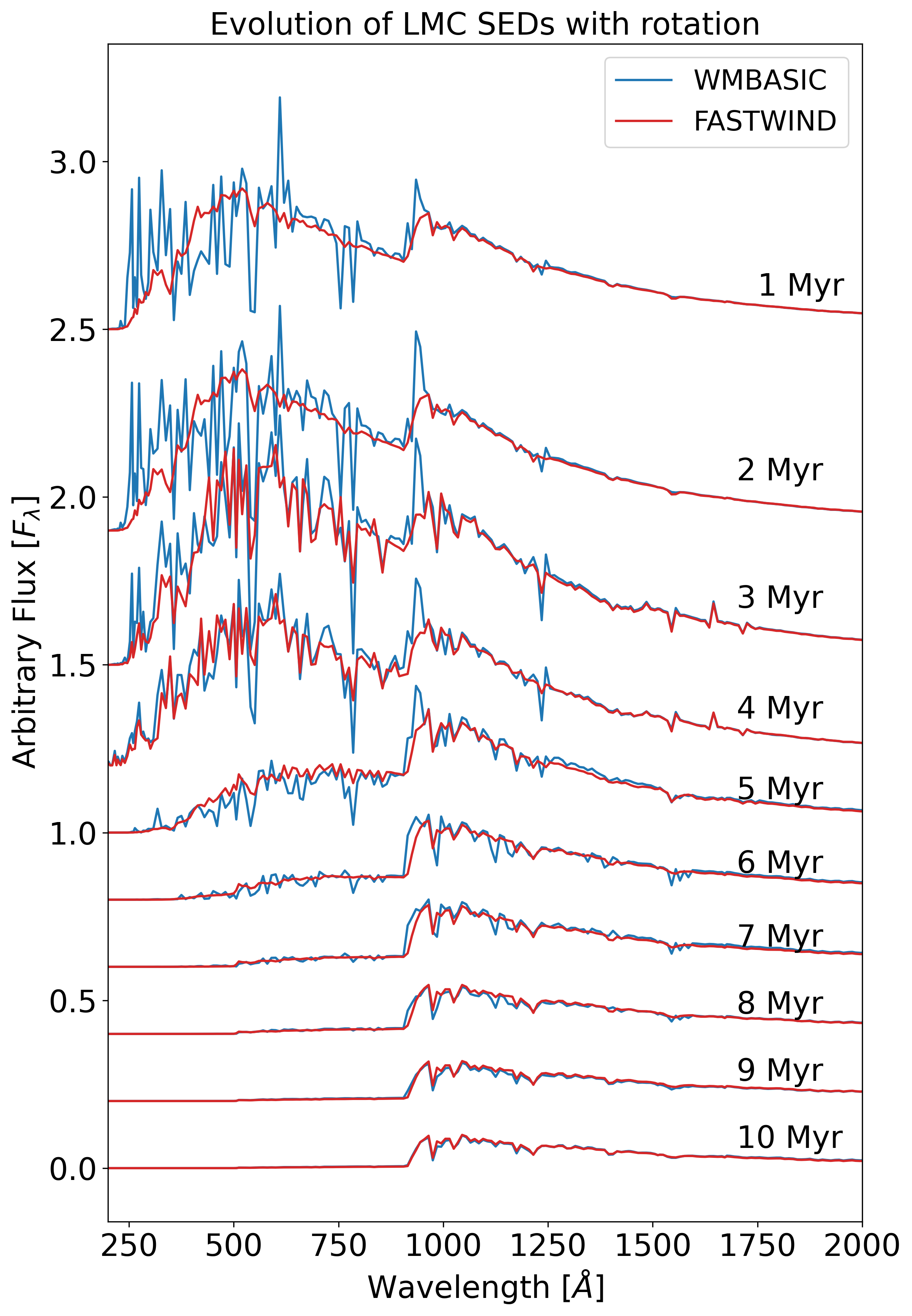}
    \caption{Synthetic FUV SEDs, utilising the \cite{Eggenberger2021} stellar evolutionary models at $Z=0.006$ with rotation for both, with \wmbasic\, spectra at $Z=0.008$ compared with \fastwind\, spectra at $Z=0.006$. The evolution with time is shown from 1 Myr to 10 Myr at intervals of 1 Myr.}
    \label{fig: wm_fw_comp_lmc_rot}
\end{figure}

\begin{figure}[t!]
    \includegraphics[scale=0.7]{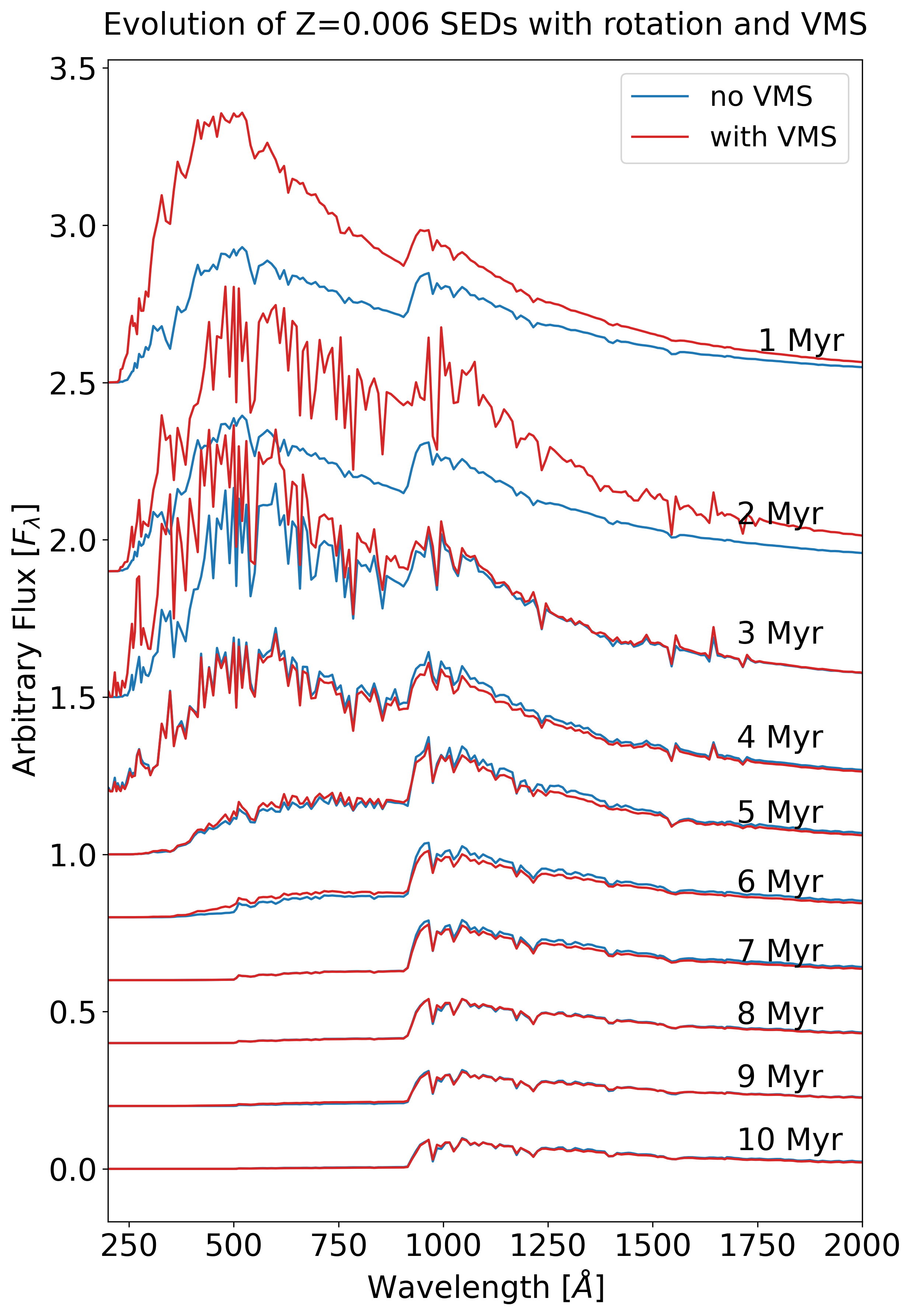}
    \caption{Synthetic FUV SEDs from \nstarburst, utilising the \cite{Eggenberger2021} stellar evolutionary models at $Z=0.006$ with rotation and \fastwind\, spectra at $Z=0.006$. This is compared with the addition of VMS evolutionary tracks up to $300M_{\odot}$ from \cite{Martinet2023} and \fastwind\, models tailored to match the extended parameter space coverage from VMS evolutionary tracks. The evolution with time is shown from 1 Myr to 10 Myr at intervals of 1 Myr.}
    \label{fig: sed_lmc_VMS_rot}
\end{figure}

\begin{figure}[t!]
    \includegraphics[scale=0.7]{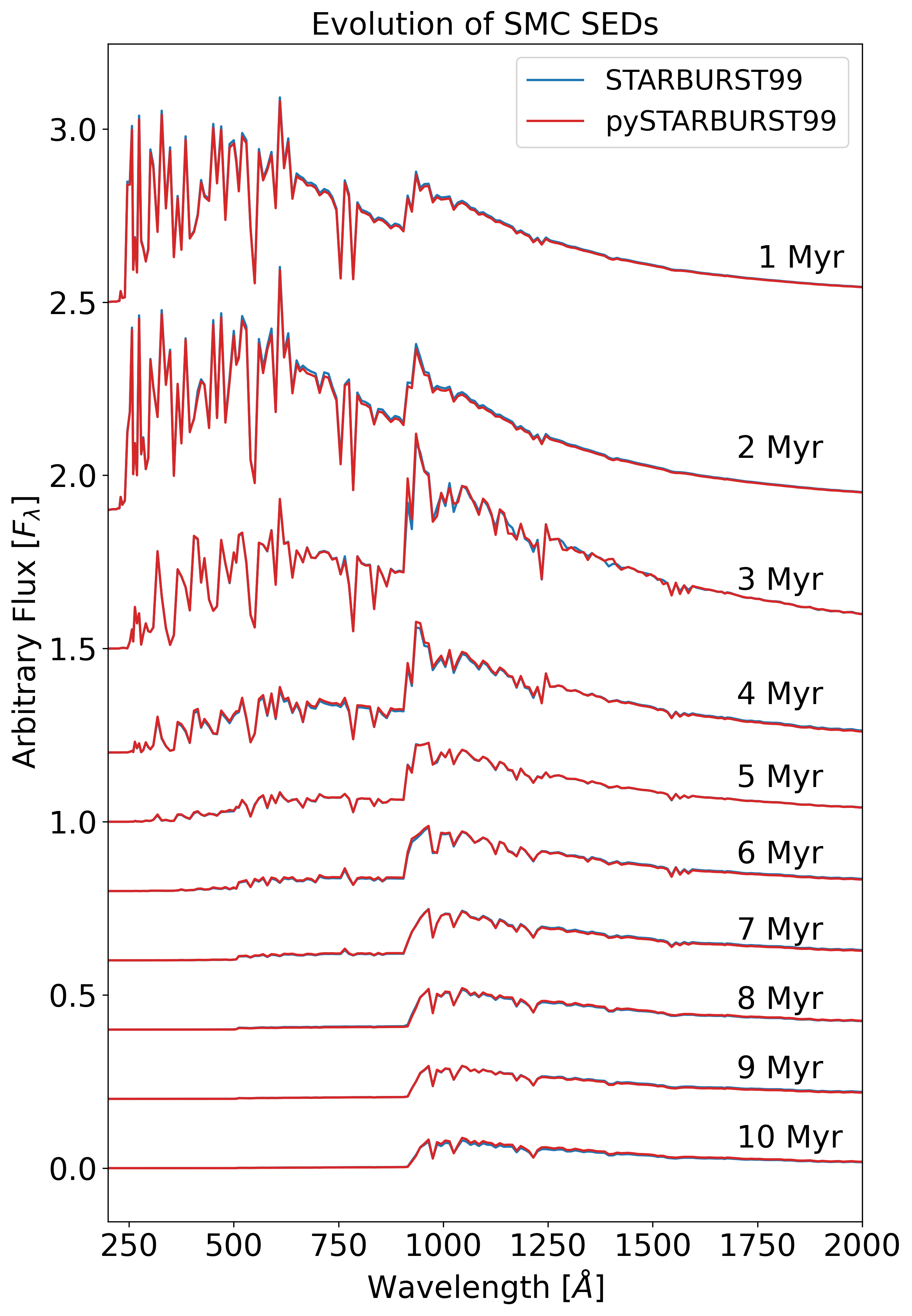}
    \caption{Synthetic FUV SEDs from \starburst\, compared to those produced with \nstarburst, in both cases utilising the \cite{Georgy2013} stellar evolutionary models at $Z=0.002$ and \wmbasic\, spectra at $Z=0.004$. The evolution with time is shown from 1 Myr to 10 Myr at intervals of 1 Myr.}
    \label{fig: py_fort_comp_SMC}
\end{figure}

\begin{figure}[t!]
    \includegraphics[scale=0.7]{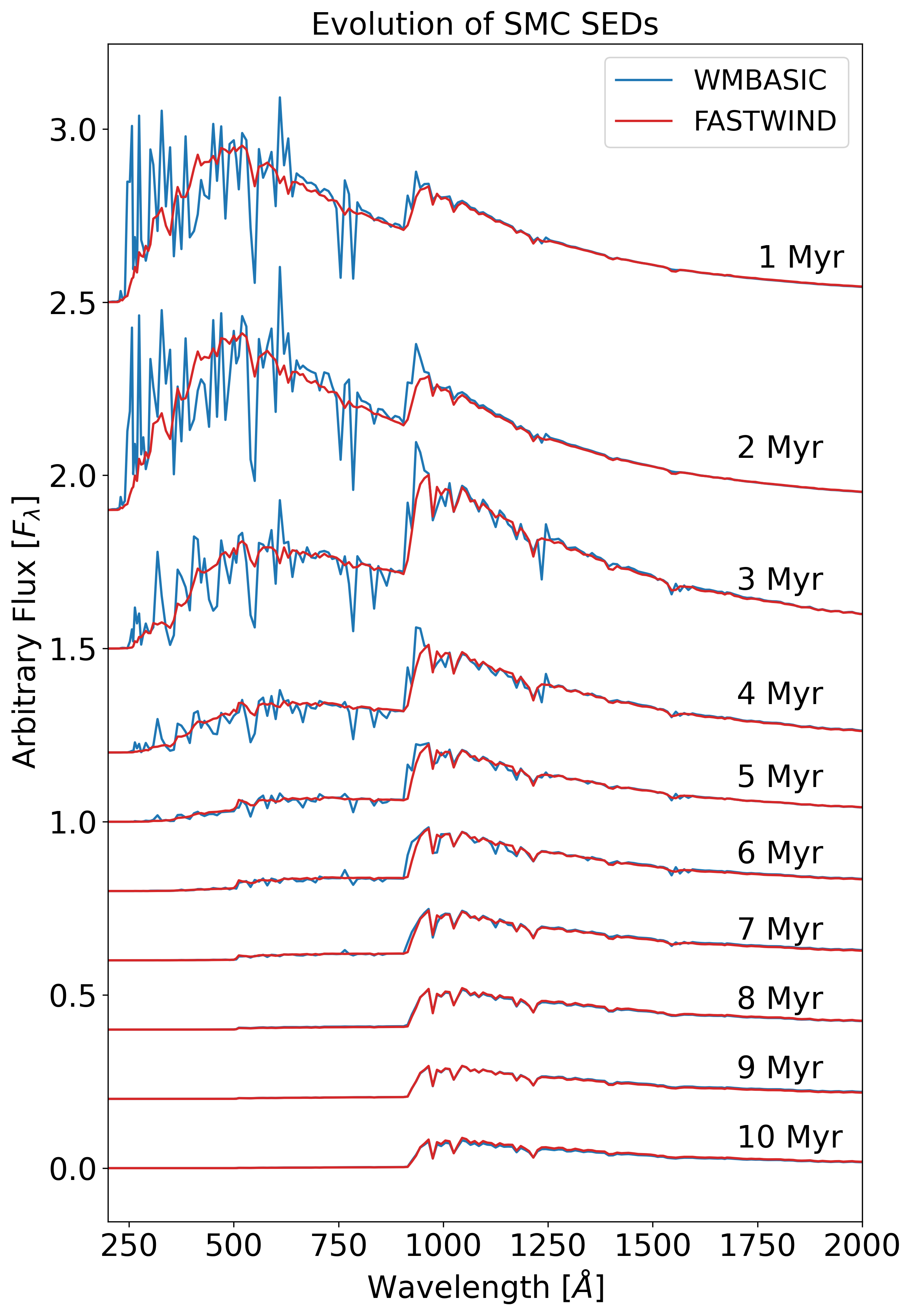}
    \caption{Synthetic FUV SEDs, utilising the \cite{Georgy2013} stellar evolutionary models at $Z=0.002$ for both, with \wmbasic\, spectra at $Z=0.004$ compared with \fastwind\, spectra at $Z=0.002$. The evolution with time is shown from 1 Myr to 10 Myr at intervals of 1 Myr.}
    \label{fig: wm_fw_comp_smc}
\end{figure}

\begin{figure}[t!]
    \includegraphics[scale=0.7]{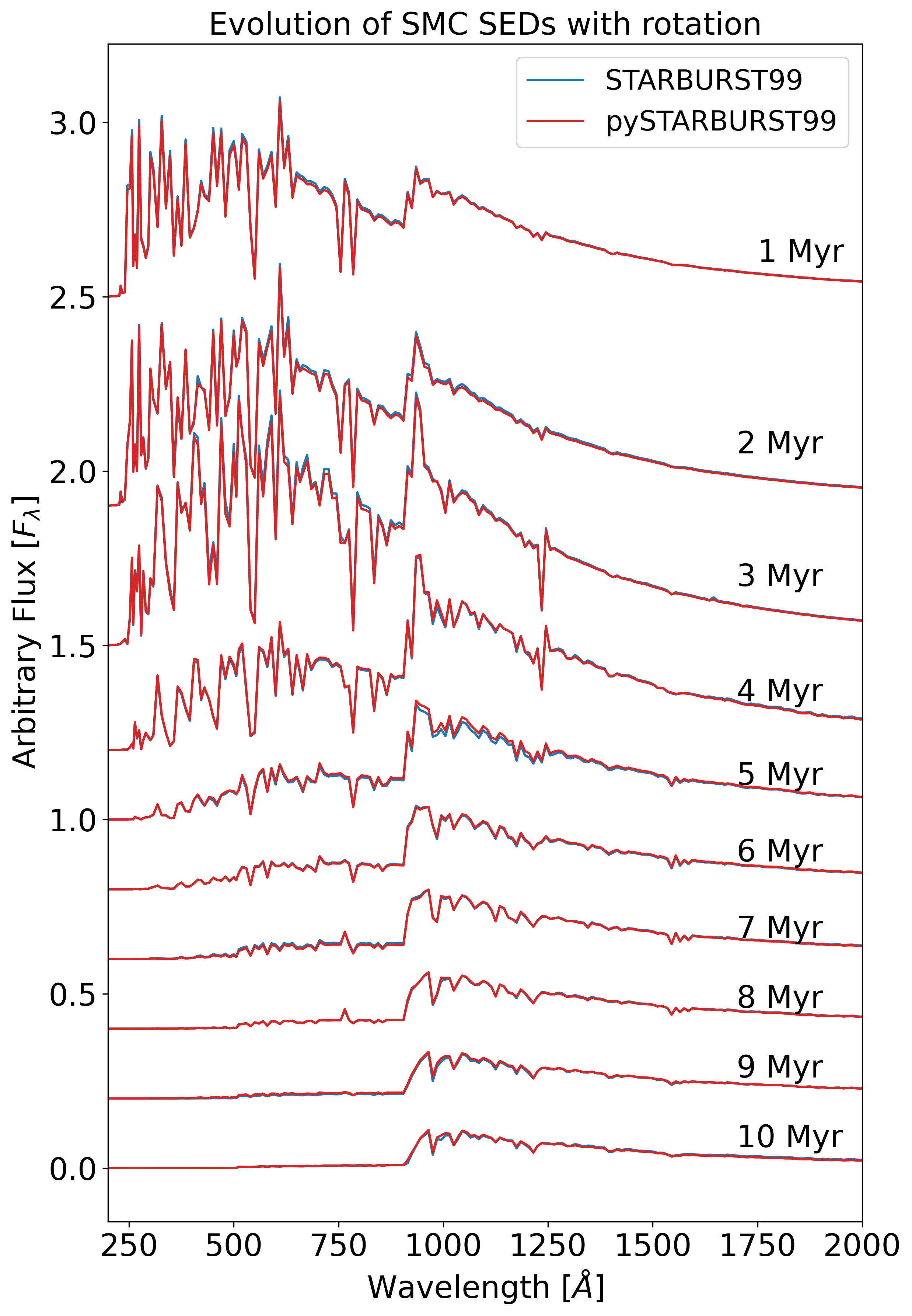}
    \caption{Synthetic FUV SEDs from \starburst\, compared to those produced with \nstarburst, in both cases utilising the \cite{Georgy2013} stellar evolutionary models at $Z=0.002$ with rotation and \wmbasic\, spectra at $Z=0.004$. The evolution with time is shown from 1 Myr to 10 Myr at intervals of 1 Myr.}
    \label{fig: py_fort_comp_SMC_rot}
\end{figure}

\begin{figure}[t!]
    \includegraphics[scale=0.7]{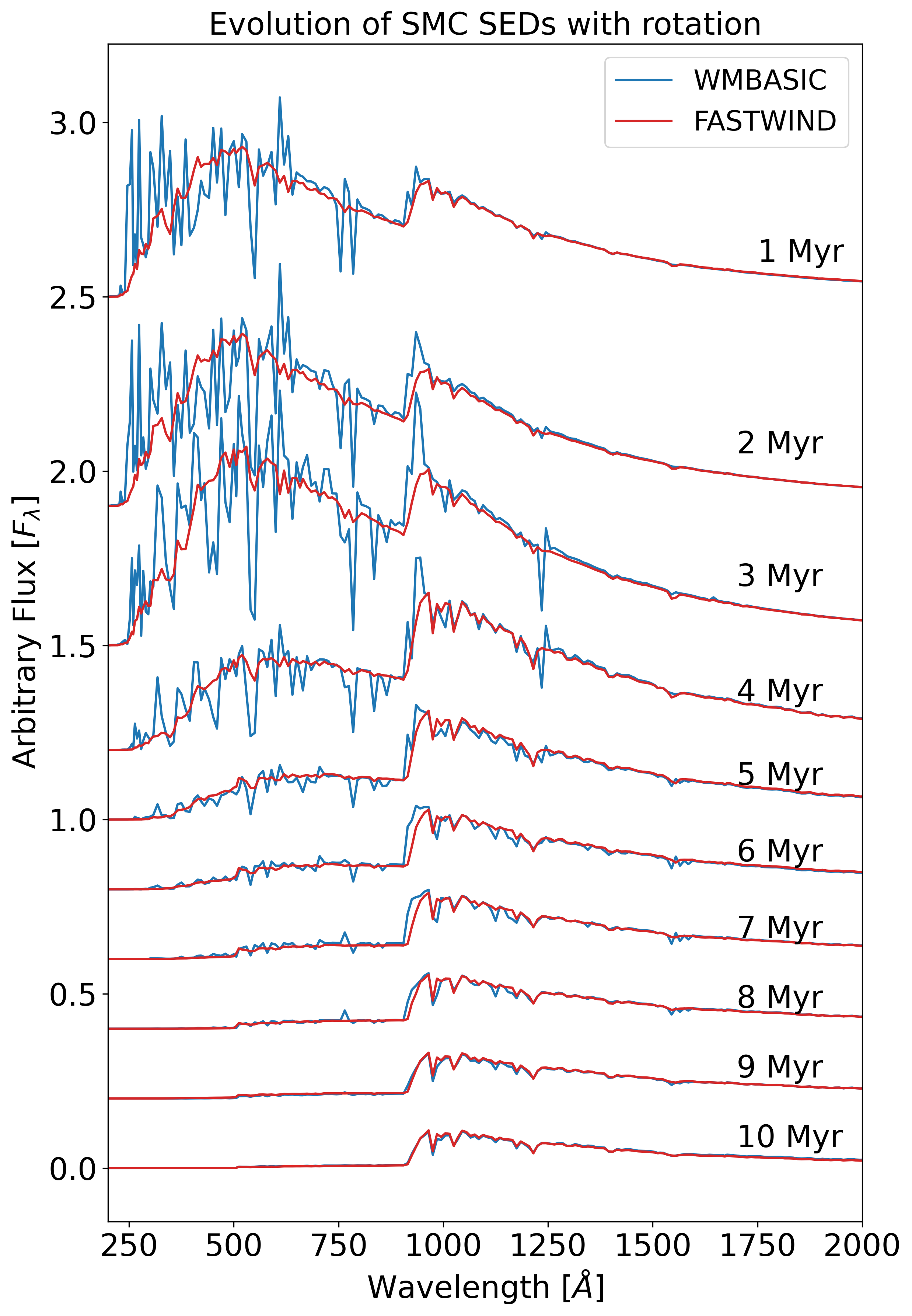}
    \caption{Synthetic FUV SEDs, utilising the \cite{Georgy2013} stellar evolutionary models at $Z=0.002$ with rotation for both, with \wmbasic\, spectra at $Z=0.004$ compared with \fastwind\, spectra at $Z=0.002$. The evolution with time is shown from 1 Myr to 10 Myr at intervals of 1 Myr.}
    \label{fig: wm_fw_comp_smc_rot}
\end{figure}

\begin{figure}[t!]
    \includegraphics[scale=0.7]{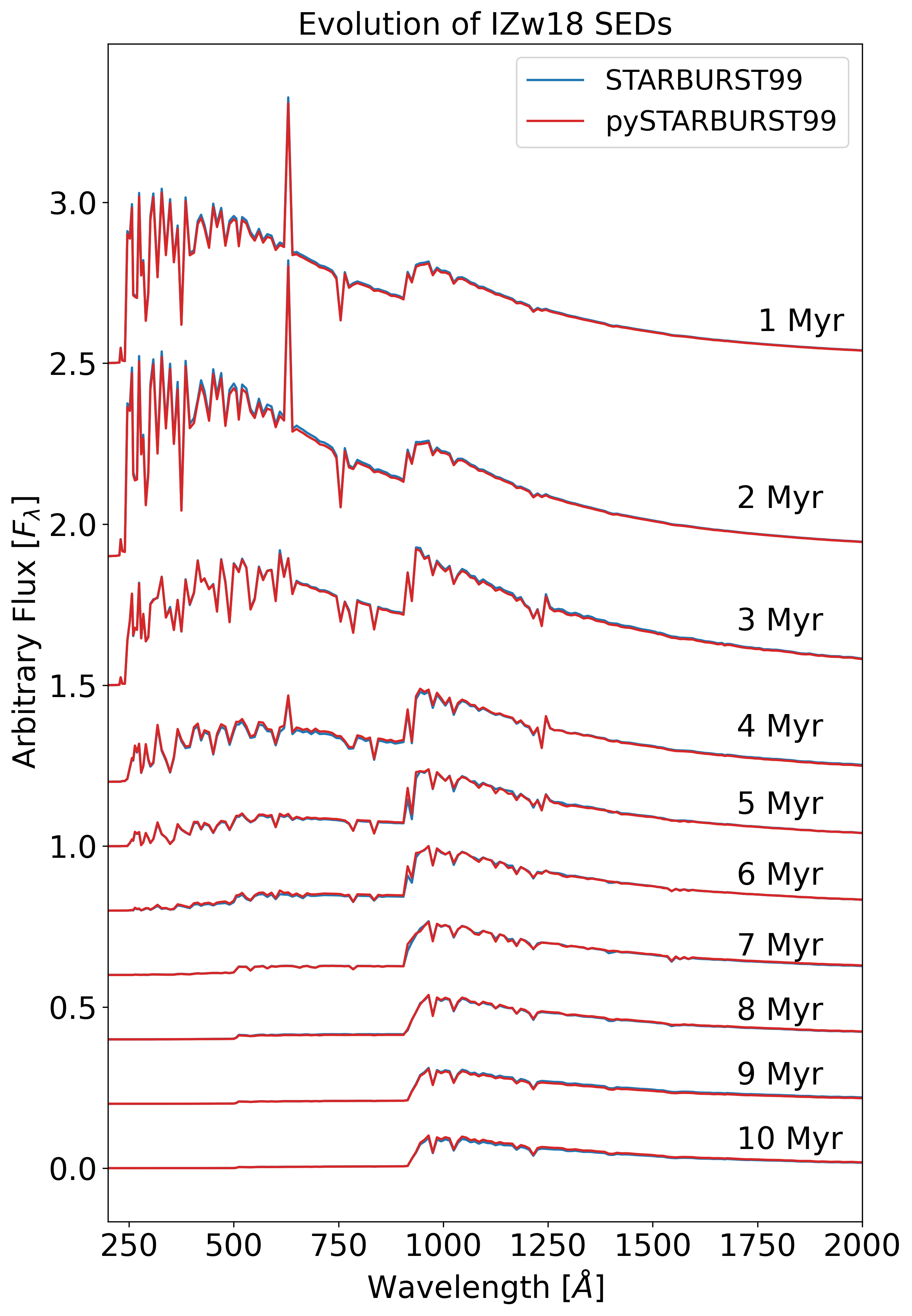}
    \caption{Synthetic FUV SEDs from \starburst\, compared to those produced with \nstarburst, in both cases utilising the \cite{Groh2019} stellar evolutionary models at $Z=0.0004$ and \wmbasic\, spectra at $Z=0.001$. The evolution with time is shown from 1 Myr to 10 Myr at intervals of 1 Myr.}
    \label{fig: py_fort_comp_izw18}
\end{figure}

\begin{figure}[t!]
    \includegraphics[scale=0.7]{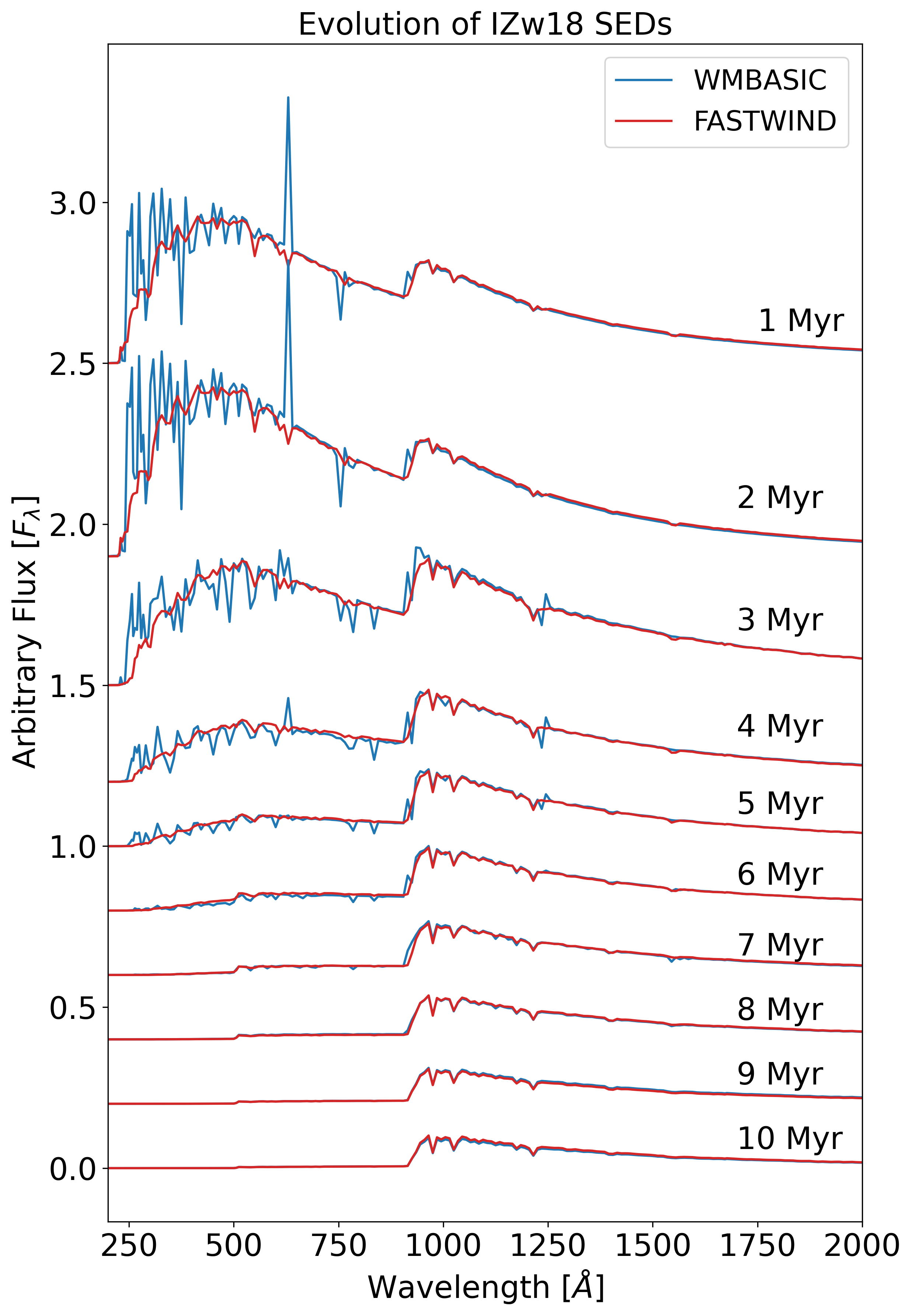}
    \caption{Synthetic FUV SEDs, utilising the \cite{Groh2019} stellar evolutionary models at $Z=0.0004$ for both, with \wmbasic\, spectra at $Z=0.001$ compared with \fastwind\, spectra at $Z=0.004$. The evolution with time is shown from 1 Myr to 10 Myr at intervals of 1 Myr.}
    \label{fig: wm_fw_comp_izw18}
\end{figure}

\begin{figure}[t!]
    \includegraphics[scale=0.7]{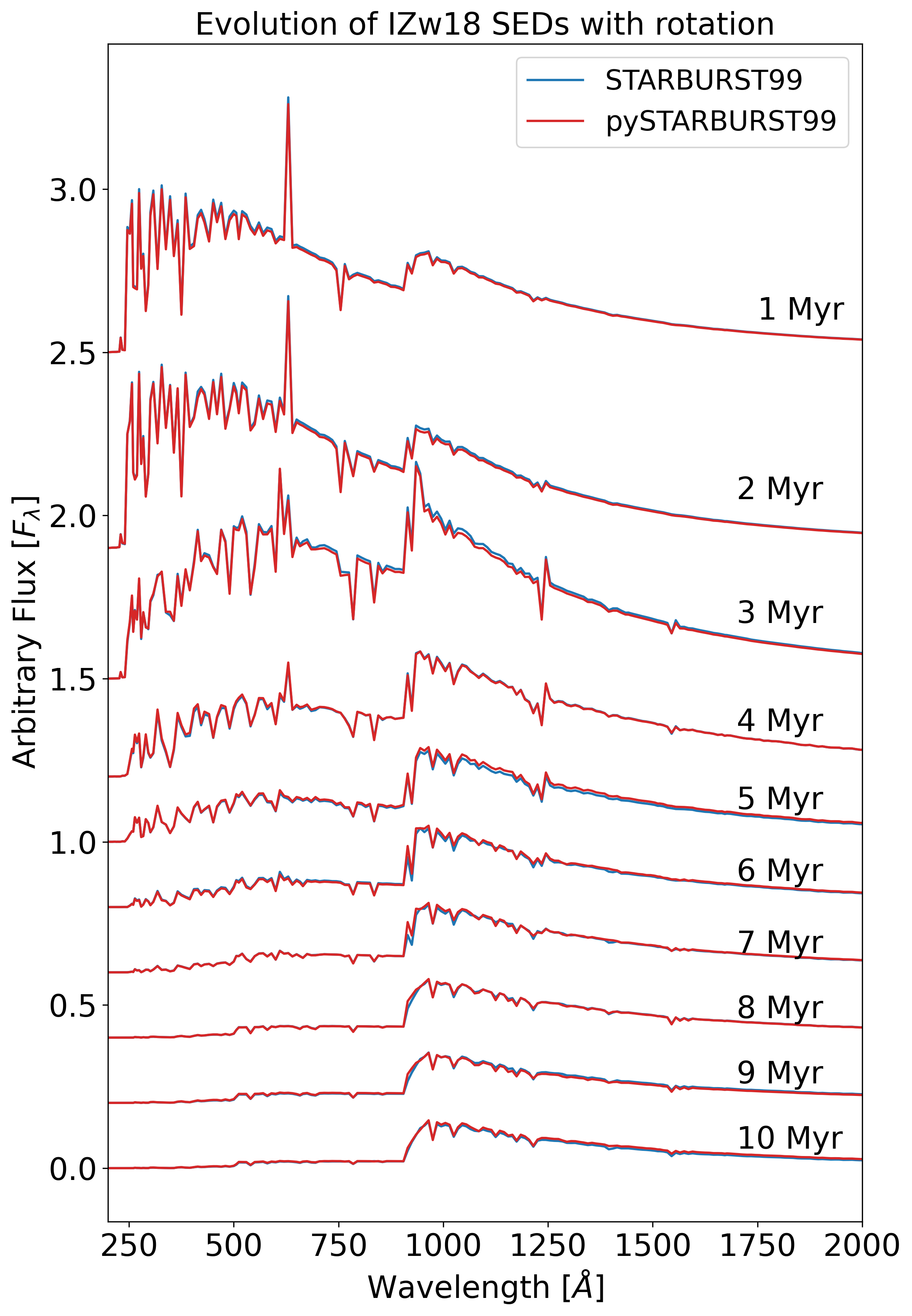}
    \caption{Synthetic FUV SEDs from \starburst\, compared to those produced with \nstarburst, in both cases utilising the \cite{Groh2019} stellar evolutionary models at $Z=0.0004$ with rotation and \wmbasic\, spectra at $Z=0.001$. The evolution with time is shown from 1 Myr to 10 Myr at intervals of 1 Myr.}
    \label{fig: py_fort_comp_izw18_rot}
\end{figure}

\begin{figure}[t!]
    \includegraphics[scale=0.7]{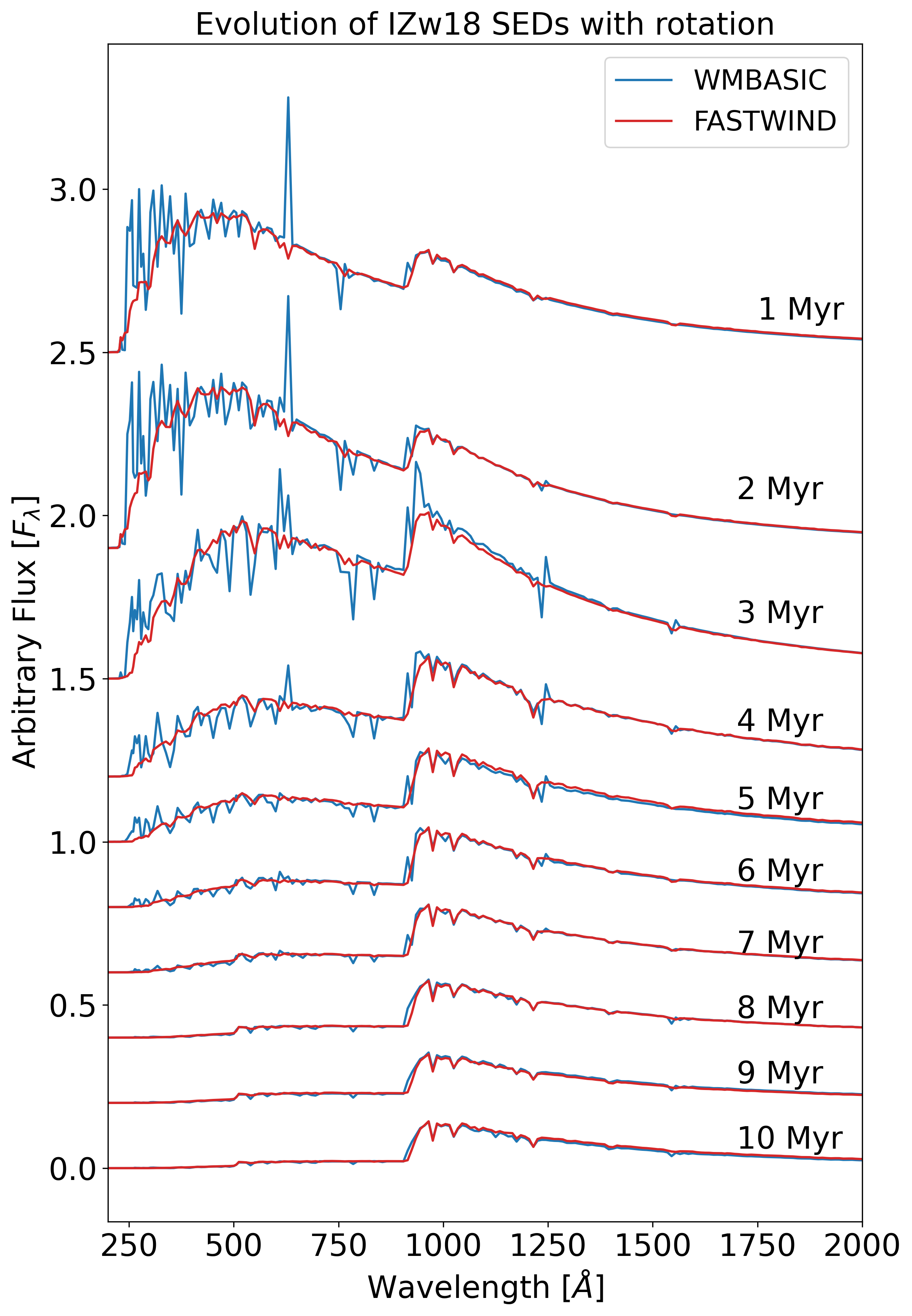}
    \caption{Synthetic FUV SEDs, utilising the \cite{Groh2019} stellar evolutionary models at $Z=0.0004$ with rotation for both, with \wmbasic\, spectra at $Z=0.001$ compared with \fastwind\, spectra at $Z=0.004$. The evolution with time is shown from 1 Myr to 10 Myr at intervals of 1 Myr.}
    \label{fig: wm_fw_comp_izw18_rot}
\end{figure}

\begin{figure}[t!]
    \includegraphics[scale=0.7]{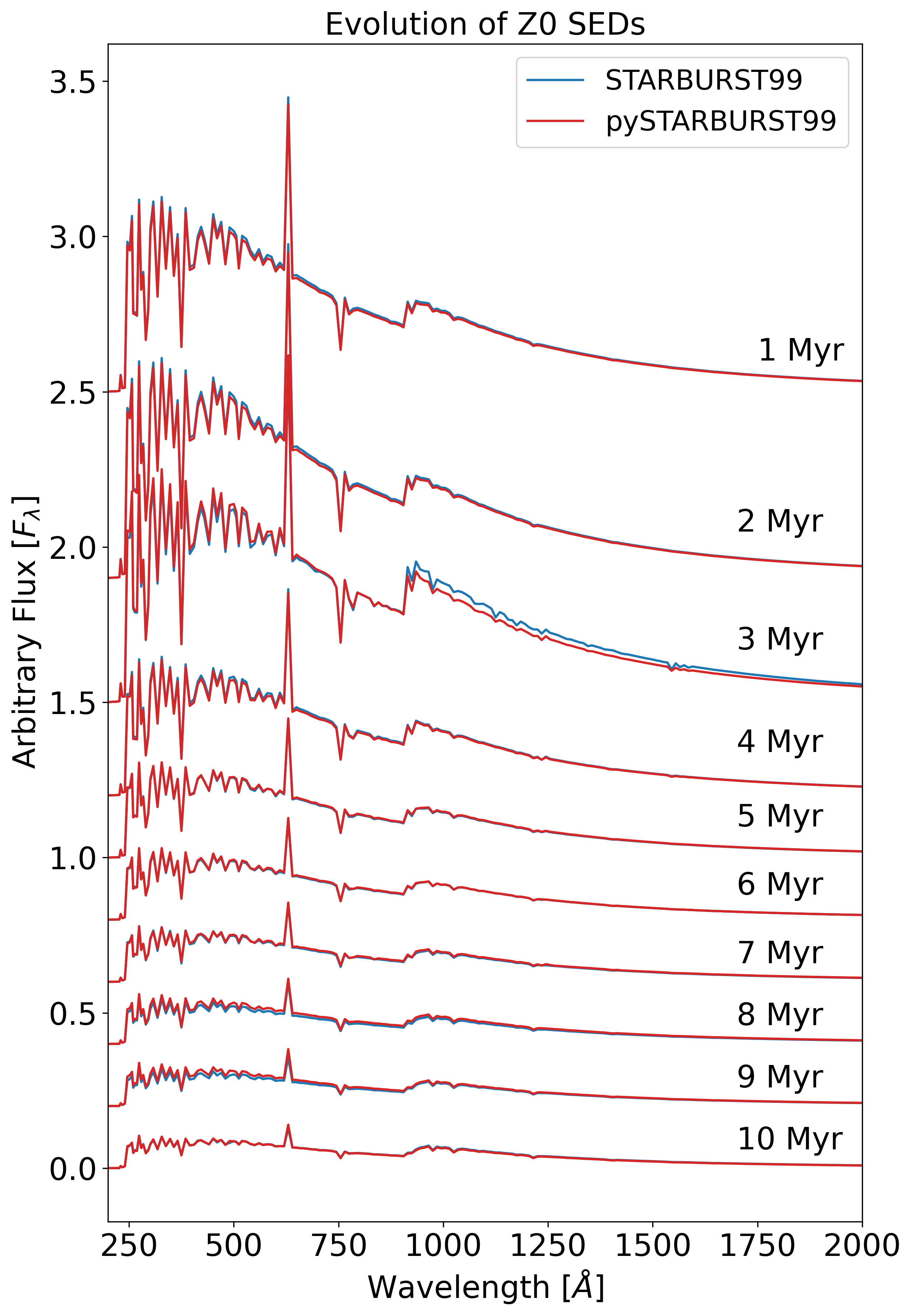}
    \caption{Synthetic FUV SEDs from \starburst\, compared to those produced with \nstarburst, in both cases utilising the \cite{Murphy2021}
    stellar evolutionary models at $Z=0.0$ and \wmbasic\, spectra at $Z=0.001$. The evolution with time is shown from 1 Myr to 10 Myr at intervals of 1 Myr.}
    \label{fig: py_fort_comp_z0}
\end{figure}

\begin{figure}[t!]
    \includegraphics[scale=0.7]{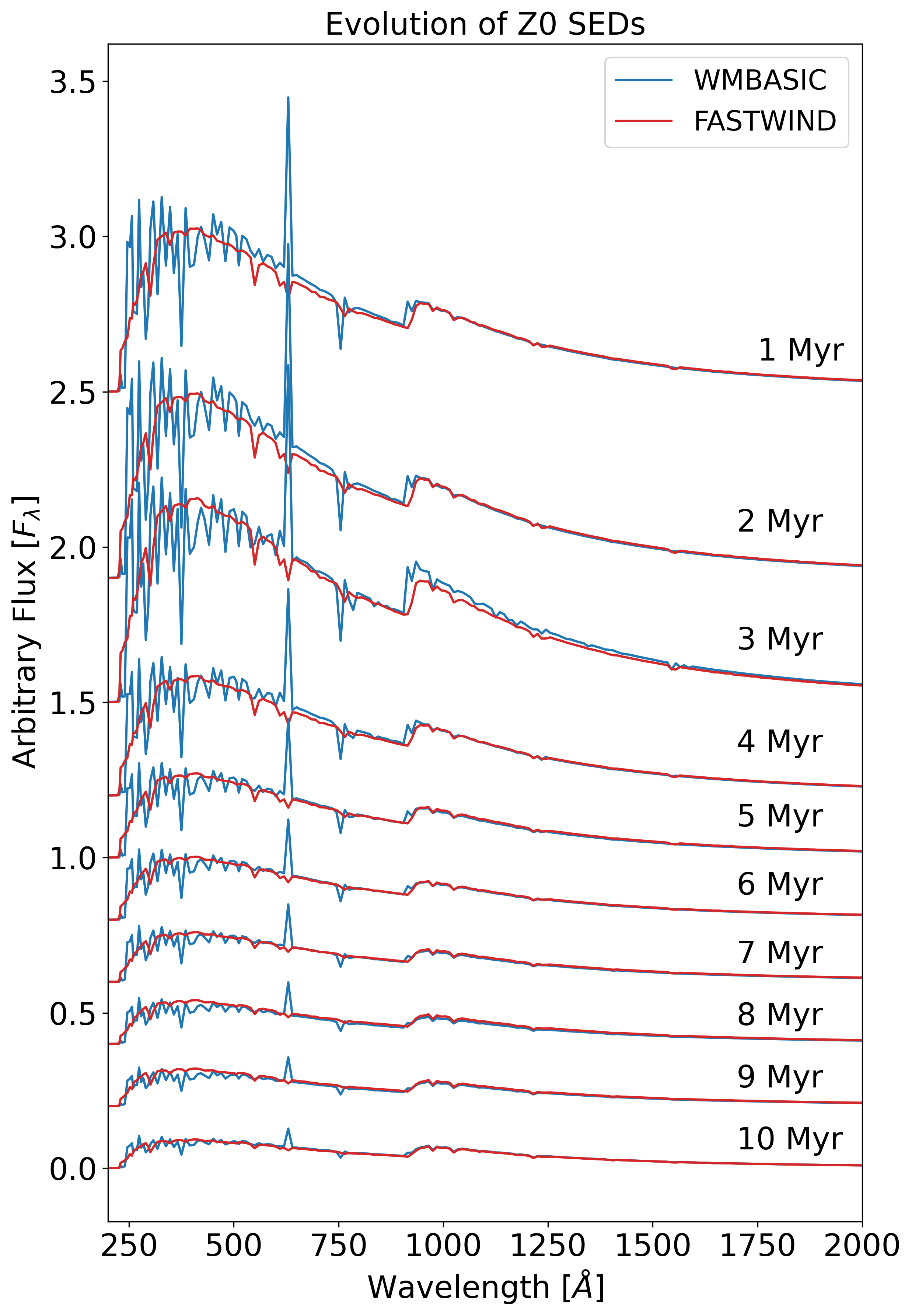}
    \caption{Synthetic FUV SEDs, utilising the \cite{Murphy2021} stellar evolutionary models at $Z=0.0$ for both, with \wmbasic\, spectra at $Z=0.001$ compared with \fastwind\, spectra at $Z=1E-6$. The evolution with time is shown from 1 Myr to 10 Myr at intervals of 1 Myr.}
    \label{fig: wm_fw_comp_z0}
\end{figure}

\begin{figure}[t!]
    \includegraphics[scale=0.7]{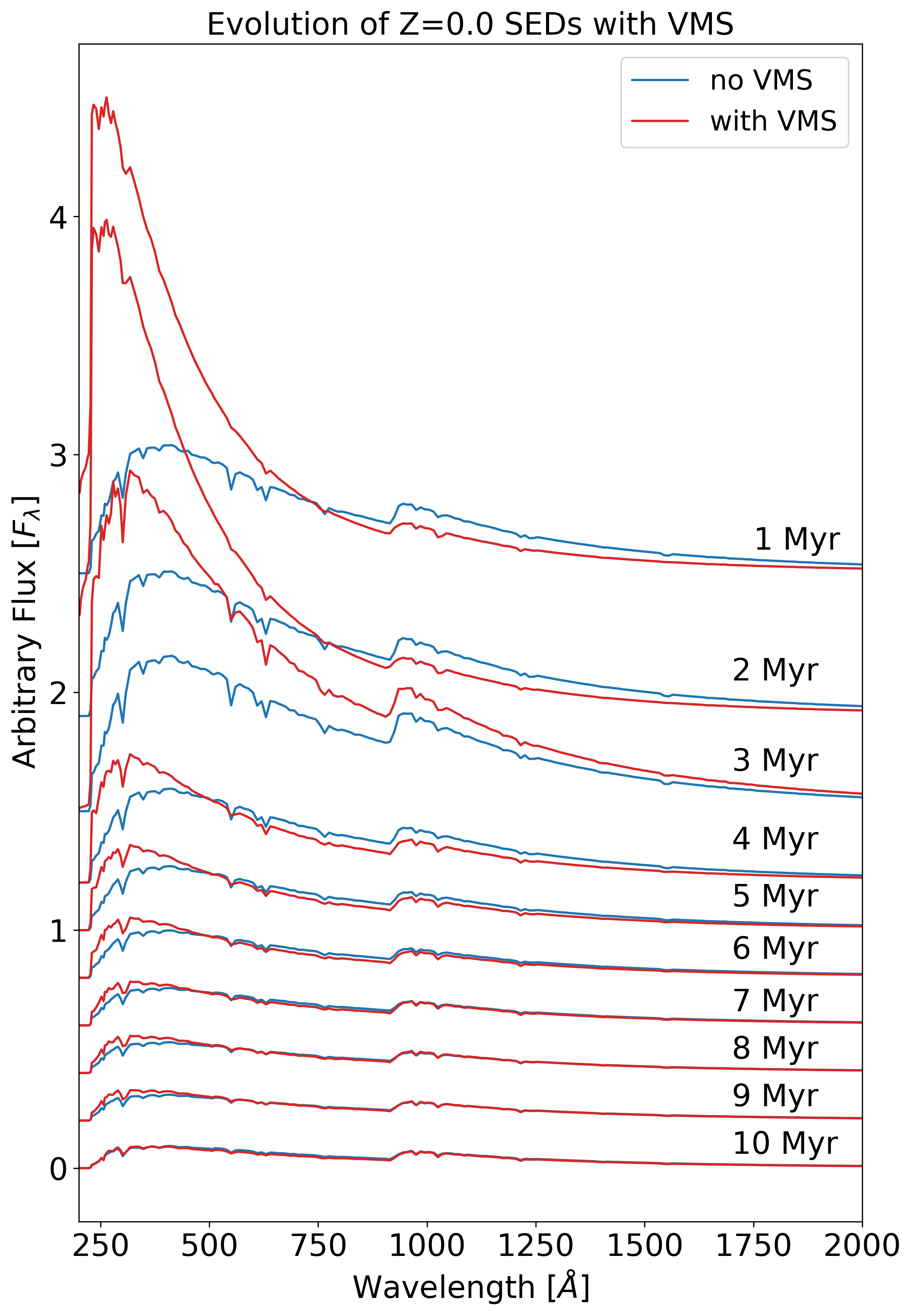}
    \caption{Synthetic FUV SEDs from \nstarburst, utilising the \cite{Murphy2021} stellar evolutionary models at $Z=0.0$ and \fastwind\, spectra at $Z=1E-6$. This is compared with the addition of VMS evolutionary tracks up to $300M_{\odot}$ from \cite{Martinet2023} and \fastwind\, models tailored to match the extended parameter space coverage from VMS evolutionary tracks. The evolution with time is shown from 1 Myr to 10 Myr at intervals of 1 Myr.}
    \label{fig: sed_z0_VMS_norot}
\end{figure}

\begin{figure}[t!]
    \includegraphics[scale=0.7]{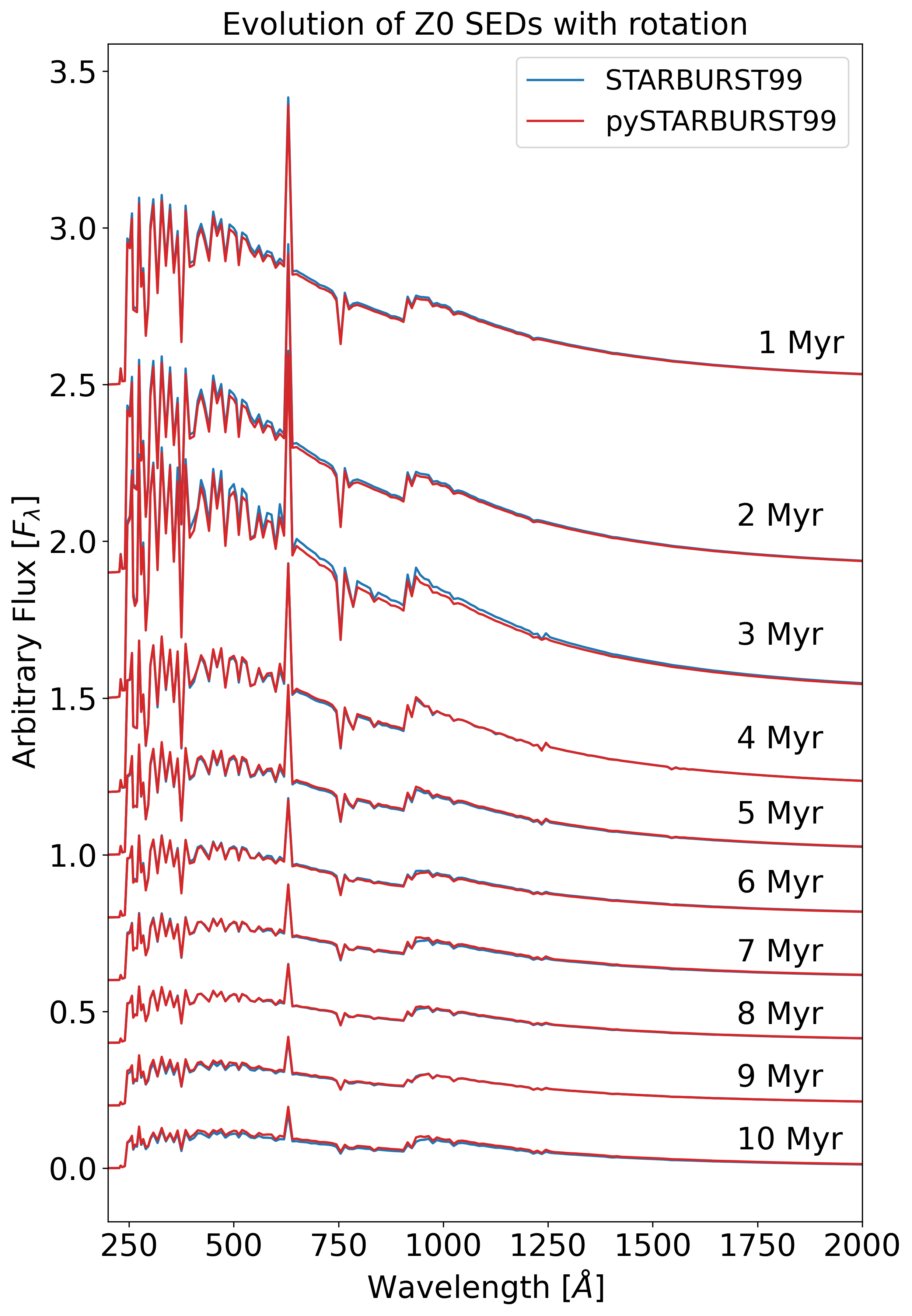}
    \caption{Synthetic FUV SEDs from \starburst\, compared to those produced with \nstarburst, in both cases utilising the \cite{Murphy2021}
    stellar evolutionary models with rotation at Z=0.0 and \wmbasic\, spectra at $Z=0.001$. The evolution with time is shown from 1 Myr to 10 Myr at intervals of 1 Myr.}
    \label{fig: py_fort_comp_z0_rot}
\end{figure}

\begin{figure}[t!]
   \includegraphics[scale=0.7]{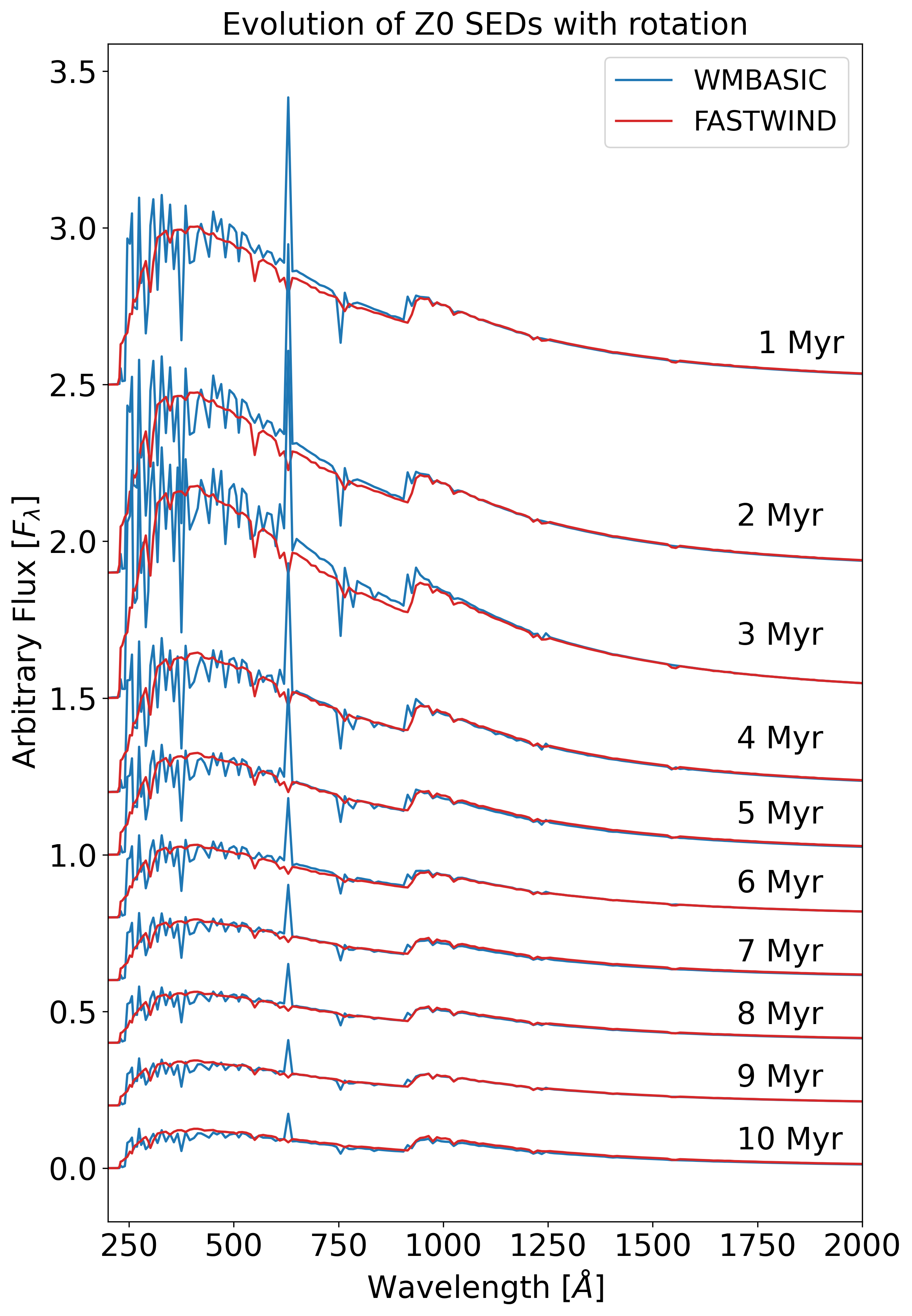}
    \caption{Synthetic FUV SEDs, utilising the \cite{Murphy2021} stellar evolutionary models with rotation at $Z=0.0$ for both, with \wmbasic\, spectra at $Z=0.001$ compared with \fastwind\, spectra at $Z=1E-6$. The evolution with time is shown from 1 Myr to 10 Myr at intervals of 1 Myr.}
    \label{fig: wm_fw_comp_z0_rot}
\end{figure}


\end{appendix}

\end{document}